\documentclass[11pt,a4paper]{article}
\usepackage{enumitem}
\usepackage{jcappub}
\usepackage{float}
\usepackage{jcappub}
\usepackage{amsmath}
\usepackage{hyperref}
\hypersetup{colorlinks=True,citecolor=blue}
\usepackage{amsfonts}
\usepackage{amssymb}
\usepackage{caption}
\usepackage{float}
\usepackage{bm}
\usepackage{graphicx}
\usepackage{color}
\usepackage{verbatim}
\usepackage{subfigure}
\usepackage{multirow}
\newcommand\Tstrut{\rule{0pt}{2.9ex}}

\def\beq{\begin{equation}}
\def\eeq{\end{equation}}
\def\ber{\begin{eqnarray}}
\def\eer{\end{eqnarray}}
\def\benu{\begin{enumerate}}
\def\eenu{\end{enumerate}}

\def\l{\left}
\def\r{\right}

\def\n{\nabla}

\def\f{\frac}
\def\mpl{m_{p}}

\def\ns{n_{_S}}
\def\nt{n_{_T}}

\def\Tre{T_{_{\rm re}}}

\def\wre{w_{_{\rm re}}}
\def\Nre{N_{_{\rm re}}}
\def\Rre{\rho_{_{\rm re}}}
\def\Nrd{N_{_{\rm RD}}}
\def\ng{n_{_{\rm GW}}}
\def\Og{\Omega_{_{\rm GW}}}

\def\sq{\lower.25ex\hbox{\large$\Box$}}
\def \lleq {\lower0.9ex\hbox{ $\buildrel < \over \sim$} ~}
\def \ggeq {\lower0.9ex\hbox{ $\buildrel > \over \sim$} ~}

\def\prl{{Phys.\@ Rev.\@ Lett.\ }}
\def\prd{{Phys.\@ Rev.\@ D\ }}

\def\plb {{Phys.\@ Lett.\@ B\ }}

\def\etal{{\it et al.}}

\def\n {\noindent}

\title{Curing inflationary degeneracies using reheating predictions and relic
gravitational waves}

\author[a]{Swagat S. Mishra,}
\author[a]{ Varun Sahni}
\author[b]{and Alexei A. Starobinsky}

\affiliation[a]{Inter-University Centre for Astronomy and Astrophysics,
Post Bag 4, Ganeshkhind, Pune 411~007, India}

\affiliation[b]{ L.~D.~Landau Institute for Theoretical Physics RAS, 119334 Moscow, Russia}

\emailAdd{swagat@iucaa.in}
\emailAdd{varun@iucaa.in}
\emailAdd{alstar@landau.ac.ru}

\date{\today}

\abstract{It is well known that the  inflationary scenario often displays different sets of
degeneracies in its predictions for CMB observables. 
These degeneracies usually arise either because multiple inflationary models
predict similar values for the scalar spectral index $n_{_S}$ and the tensor-to-scalar
ratio $r$, or because within the same model, the values of $\lbrace n_{_S}, r \rbrace$
are insensitive to some of the model parameters, 
making it difficult for CMB observations alone to constitute a unique probe of inflationary cosmology. 
We demonstrate that by taking into account constraints on the post-inflationary reheating parameters such as 
the duration of reheating $N_{_{\rm re}}$, its temperature $T_{_{\rm re}}$ and especially its  equation
of state (EOS), 
$w_{_{\rm re}}$, it is possible to break this degeneracy  in certain classes of inflationary models 
where identical values of $\lbrace n_{_S}, r \rbrace$ can correspond to different reheating $w_{_{\rm re}}$. 
In particular, we show how reheating constraints can break inflationary degeneracies in the T-model and the 
E-model $\alpha$-attractors. Non-canonical inflation is also studied. 
The relic gravitational wave (GW) 
 spectrum provides us with another tool to break inflationary degeneracies.
This is because the GW spectrum
is sensitive
to the post-inflationary EOS of the universe. Indeed
 a stiff EOS during reheating $(w_{_{\rm re}} > 1/3)$ gives rise to a small scale blue tilt in the  spectral index
$n_{_{\rm GW}} = \frac{d\log{\Omega_{_{\rm GW}}}}{d\log{k}} > 0$, while a soft EOS $(w_{_{\rm re}} < 1/3)$ 
results in a red tilt.
Relic GWs therefore provide us with valuable information about the post-inflationary epoch,  
and their spectrum
can be used to cure inflationary degeneracies in $\lbrace n_{_S}, r\rbrace$.
} 


\begin{document}
\maketitle

\section{Introduction}
\label{sec:intro}

There has been a remarkable  progress  in our understanding  of the early universe over the past three decades fostered by new theoretical insights and reinforced by a plethora of precision cosmological missions, ranging from  Cosmic Microwave Background (CMB)  to large scale structure (LSS) observations.  
As a result, the inflationary paradigm 
\cite{inf_star80,inf_guth81,inf_linde82,inf_alstein82,linde90,baumann07} has emerged as a key
scenario for describing the early universe and for setting initial conditions for the hot Big Bang phase 
of expansion.  One of the key predictions of the inflationary scenario is the quantum-mechanical 
production of primordial tensor fluctuations which give rise to a stochastic background of relic gravitational 
waves with an approximately scale-invariant primordial power spectrum, $|n_T|\ll 1$, at large scales 
including the cosmological ones~\cite{star79}. The reason for this lies in the fact that unlike other massless
fields such as photons and massless neutrinos which couple conformally to gravity and whose production 
is therefore suppressed in the conformally flat Friedmann-Lemaitre-Robertson-Walker (FLRW) universe, 
gravitational waves in General Relativity (GR) couple minimally to gravity \cite{grish75} that results in 
their non-adiabatic production in an expanding isotropic universe if the Ricci scalar $R$ is 
non-zero~\footnote{More complicated situation may occur in modifed gravity. In particular, in $f(R)$ 
gravity small oscillations of the gravitational scalar degree of freedom with a non-zero $R$ do not create 
gravitons; in quantum language, decay of scalarons into pairs of gravitons is suppressed~\cite{st81}.}. 
While several  distinct predictions of the single field slow-roll scenario of inflation have received spectacular 
observational confirmation, both from CMB as well as LSS observations,  the detection of primordial tensor 
fluctuations, both in the form of CMB B-mode polarization on large angular scales as well as a spectrum of  
relic gravitational waves (GWs), remains one of the major challenges confronting observational cosmology 
in the coming decade.

It is well known that the GW primordial power spectrum at large scales provides us with important 
information about the nature of an inflaton field due to its direct relation to the inflaton potential in the case 
of the minimal inflaton coupling to gravity (or, in the Einstein frame).
Of equal importance is the fact that their spectrum, $\Og(k)$, and spectral index 
$\ng = \frac{d\log{\Omega_{\rm GW}}}{d\log{k}}$ at sufficiently small scales can serve
as a key probe to physical processes occurring after inflation. As originally shown in \cite{sahni90},
the primordial spectrum of relic gravitational radiation at small scales is exceedingly sensitive to the post-inflationary
equation of state (EOS), $w$. In fact the GW spectrum has distinctly different properties
for stiff/soft equations of state. For a stiff EOS, $w > 1/3$, the GW spectrum shows a blue tilt:
$\ng > 0$, that increases the GW amplitude on small scales. Softer equations of state, $w < 1/3$, on
the other hand, lead to a red tilt, whereas the radiation EOS, $w = 1/3$, results in a flat
spectrum with $\ng \simeq 0$.

Another key aspect of inflationary cosmology, namely the epoch of `reheating', also
remains observationally inaccessible at present, despite a profusion of theoretical progress in this direction. 
It is well known that the post-inflationary universe passed through a series of physical epochs
each of which can be characterized by an EOS, $w_i$. Of these, the most recent ones are:
the radiation dominated stage with $w_r \simeq 1/3$, the matter dominated stage with $w_m \simeq 0$
and the present stage of accelerated expansion with $w _{\rm de} < -1/3$.
However, after the end of inflation and before the commencement of the radiation dominated stage,
the universe went through the epoch of reheating during which the energy contained in the inflaton
field was transferred to the other matter/radiation degrees of freedom present in the universe.

The nature of the reheating epoch, including its duration $\Nre$ and EOS $\wre$, depends crucially upon how
the inflaton couples with (and hence releases its energy into) other matter fields in the universe.
If this process is slow then reheating takes place perturbatively and the inflaton scalar field oscillates
for a very long time, gradually releasing its energy into matter/radiation. 
In this case, the EOS during the oscillatory regime is determined primarily by the shape of the
inflaton potential near its minimum value, about which the inflaton oscillates.
Perturbative reheating in GR is expected to occur if the inflaton $\phi$ decays primarily into fermions
(which soon decay into the standard model fields), its decay into bosons being strongly suppressed
in the absence of the trilinear $\phi\chi^2$ interaction \cite{kofman96a}~\footnote{The situation is changing 
dramatically in the case of strong non-minimal coupling of bosons to gravity~\cite{ema17,he19}.}.

On the other hand, if the inflaton decays into bosons, $\chi$, through a coupling $g^2\phi^2\chi^2$
with $g \gg 10^{-3}$, then oscillations of $\phi$ can lead to a parametric resonance during which
quanta of the field $\chi$ are produced in copious amounts. This stage is usually referred to as
preheating \cite{kofman96a,kofman94,yuri95,kofman96,kofman97}.
The backreaction of $\chi$ on $\phi$ ends the resonance and the subsequent decay of excitations
of the $\phi$ and $\chi$ fields into standard model (SM) fields gives rise to reheating
and the subsequent thermalization of the universe at a temperature $\Tre$.
The duration of the pre-radiative epoch, which includes the end of inflation, the parametric
resonance, the decay of the inflaton into bosons ($\phi \to \chi\,\chi$) and fermions
($\phi \to \psi\,\psi$) and thermalization can be quite long,
and it is convenient to encode its physics by means of an effective EOS parameter $\wre$.
Since $\wre$ influences the spectrum of relic gravitational waves, observations of the GW
spectrum can shed light on the complex, non-linear and out of equilibrium physics
which operates during the reheating epoch.

In addition to primordial tensor fluctuations which result in the GW background, 
a key prediction of inflationary cosmology is the generation
of primordial scalar fluctuations which later grow to form the LSS of the universe.
Scalar and tensor perturbations generated during inflation create an imprint in the cosmic
microwave background (CMB) which can be used to deduce the scalar spectral index $n_{_S}$ and the
tensor to scalar ratio $r$ -- two important observables which can be used to rule out competing
inflationary models \cite{baumann07,liddle_lyth}.
However it is well known that inflationary models often display degeneracies, with two (or more) models
predicting essentially the same values of $\lbrace n_{_S}, r \rbrace$.
This leads to so called `cosmological attractors' or `universality classes' of the inflationary scenario \cite{linde1,linde2,Roest1}.
This degeneracy
makes it difficult for CMB observations alone to constitute a unique probe of inflationary cosmology
\cite{Planck1,bicep2}.

In this paper we show that reheating predictions
 including the reheating duration  $\Nre$, temperature $\Tre$ and particularly the EOS $\wre$, 
can help break degeneracies  in inflationary scenario's in which 
identical values of $\lbrace n_{_S}, r \rbrace$ can correspond to different $\wre$; also see \cite{Martin:2010kz, 
Easson:2010zy,kamion_1,Creminelli,Martin:2014nya,kamion_2,cook15,German:2020cbw,German:2020iwg}.
Since the gravitational wave spectrum $\Og(k)$ is sensitive to the value of $\wre$,
observations of $\Og(k)$ by space-based GW observatories can shed valuable light both on the
dynamics of reheating as well as on the parameters of the inflationary potential.

Our paper is organised as follows: section \ref{sec:inflation} demonstrates the existence of
inflationary degeneracies in the 
T-model and E-model $\alpha$-attractors and in non-canonical inflation. In section \ref{sec:reheat} 
we provide an  introduction to reheating and discuss 
implications of reheating predictions on inflationary degeneracies  in the aforementioned models.  
The spectrum of relic GWs for the  three inflationary models is  determined in section \ref{sec:GW}  and
a summary of  our results is presented in  section \ref{sec:dis}. 

\section{Inflationary degeneracies}
\label{sec:inflation}

It is well known that  the inflationary scenario displays different sets of
degeneracies in its predictions for the CMB observables. These usually arise either because multiple inflationary models
predict similar values for the scalar spectral index $\ns$ and the tensor-to-scalar
ratio $r$, or because within the same model, the values of $\lbrace \ns, r \rbrace$
are insensitive to some of the model parameters. We focus on three separate inflationary
models and show that the degeneracies in $\lbrace n_{_S}, r\rbrace$ which they
display can easily be broken by incorporating information obtained  from reheating predictions as well as using  the associated relic gravitational wave background. The models discussed in this paper are: the T-model and E-model  $\alpha$-attractors and the non-canonical $m^2\phi^2$ model.

\subsection{Inflationary degeneracies in $\alpha$-attractors}
\label{sec:alphainf}

\begin{enumerate}

\bigskip

\item {\large {\bf The T-model $\alpha$-attractor}} is associated with the potential

\beq
V (\phi) = V_0\tanh^{2p}{(\lambda\phi/\mpl)}\;;~~~p=1,2,3.... 
\label{eq:pot_Tmodel}
\eeq
Large absolute values of $\vert\lambda\phi\vert \gg \mpl$ lead to a
plateau-like, asymptotically flat potential with
$V (\phi) \simeq V_0$.
On the other hand small values, $|\lambda\phi| \ll \mpl$ describe the minimum of the potential
\beq
V (\phi) \simeq V_0\left (\frac{\lambda\phi}{\mpl}\right )^{2p}
\label{eq:Tmodel_min}
\eeq
around which the scalar field oscillates after inflation.
As originally shown in \cite{turner83}, a scalar field oscillating around the minimum of such
  a potential 
has the mean EOS~\footnote{For the particular case $p=1$, this was earlier derived in~\cite{st78}.}
\beq
\langle w_{\phi} \rangle = \frac{p-1}{p+1}~.
\label{eq:EOS}
\eeq

It is interesting that when $\lambda > 0.1$, the scalar ($\ns$) and tensor ($\nt$) spectral
indices and the tensor-to-scalar ratio ($r$) for the T-model (\ref{eq:pot_Tmodel}) acquire the form
\ber
\ns - 1 &\simeq& -\frac{2}{N_k}~, \\
\nt &\simeq& -\frac{1}{4 N_k^{2}\,\lambda^{2}}~,\\
r &\simeq& \frac{2}{N_k^{2}\,\lambda^{2}}~,
\label{eq:CMB_Tmodel}
\eer
which {\em does not depend} upon the value of $p$ in (\ref{eq:pot_Tmodel}).
This interesting degeneracy of the T-model is illustrated in figure \ref{fig:deg_Tmodel}. 

\begin{figure}[t]
\begin{center}
\includegraphics[width=0.496\textwidth]{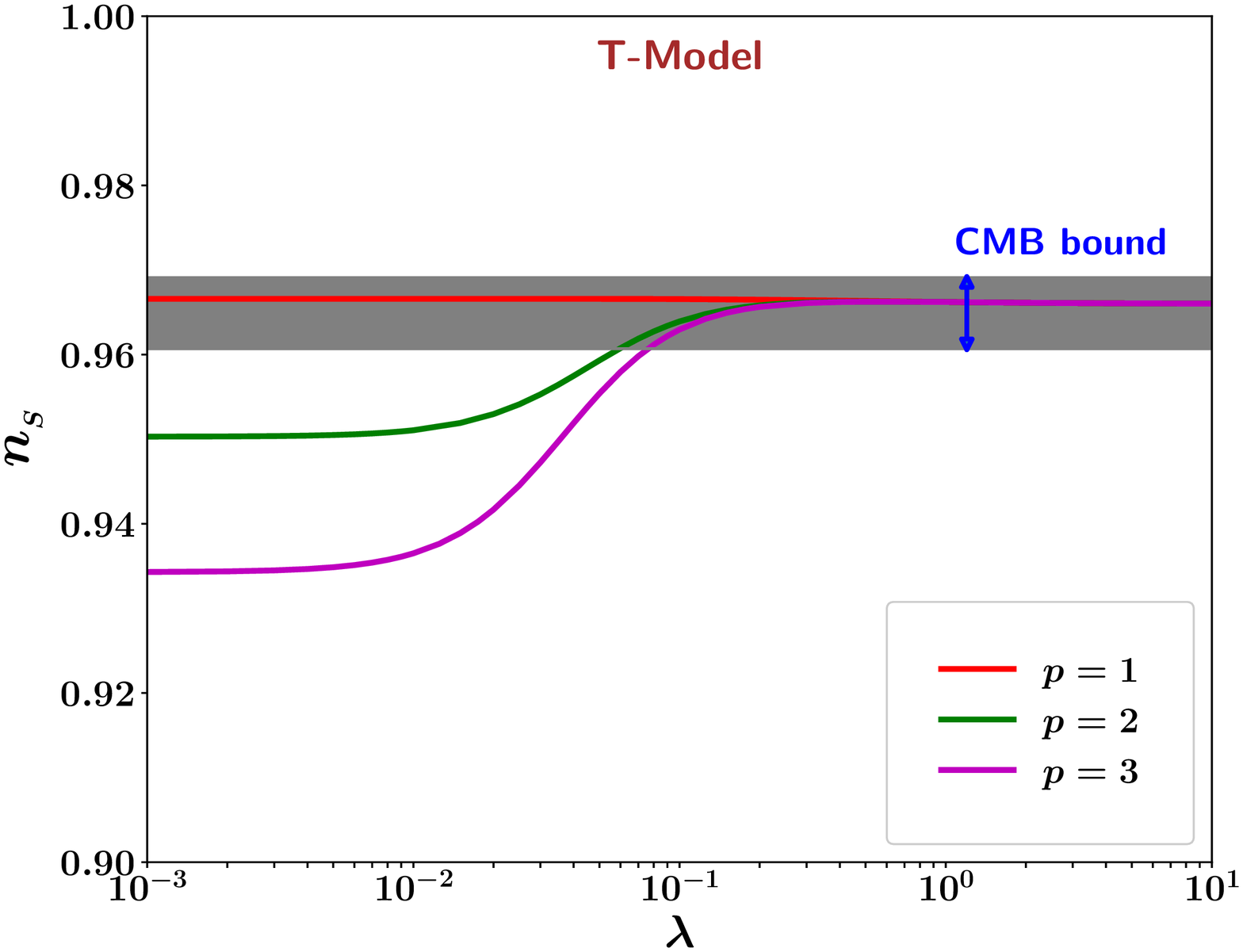}
\includegraphics[width=0.496\textwidth]{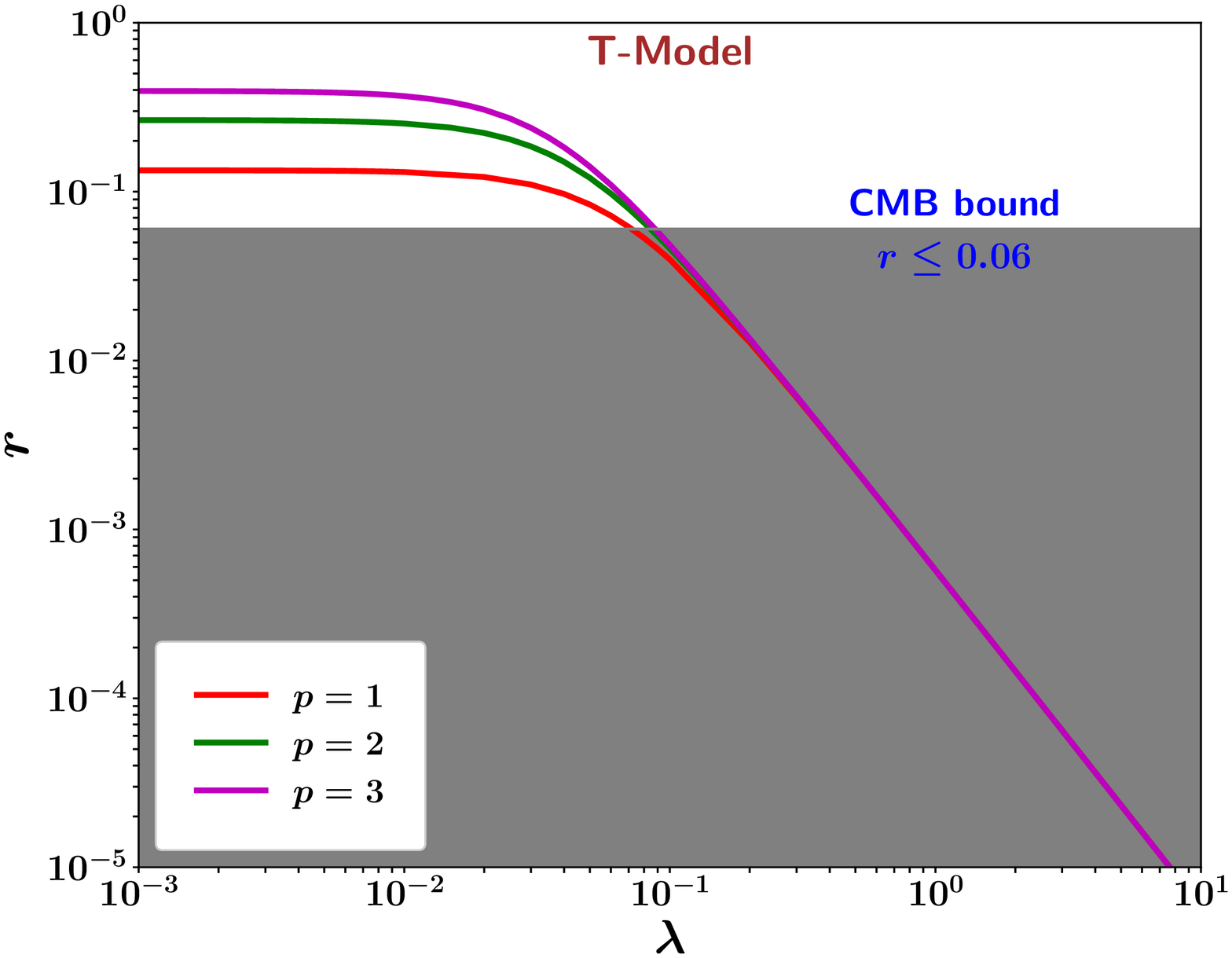}
\caption{This figure demonstrates the degeneracies of the T-model $\alpha$-attractor (\ref{eq:pot_Tmodel}).  In the left panel, the scalar spectral index $\ns$ is plotted as a
function of $\lambda$ for three different values of the parameter $p$ in the
potential~(\ref{eq:pot_Tmodel}). The right panel shows the tensor-to-scalar ratio $r$ 
as a function of $\lambda$  for the same set of values for $p$. All the curves  correspond to the number of $e$-folds $N_k = 60$. Note that when $\lambda > 0.1$, the scalar spectral index
approaches a constant value (e.g $\ns \simeq 0.967$ for $N_k=60$), whereas $r$ decreases as $\lambda^{-2}$. The shaded region refers to the CMB 1$\sigma$ limits on $\ns$ and $r$ as determined by Planck 2018 \cite{Planck1}, namely $\ns = 0.9649 \pm 0.0042$, $r\leq 0.06$.}
\label{fig:deg_Tmodel}
\end{center}
\end{figure}

\bigskip

\item {\large{\bf The generalized Starobinsky potential or the E-model $\alpha$-attractor }}
is described by
\beq
V (\phi) = V_0\left[1 \,-\, \exp\l(-\lambda\,\frac{\phi}{\mpl}\r) \right]^{2p}~ \,;~~~p=1,2,3....
\label{eq:pot_Emodel}
\eeq
This potential reduces to the Einstein
frame representation of Starobinsky's $R+R^2$ model of inflation
when $\lambda = \sqrt{2/3}$ and $p=1$.
For $\lambda \gg 0.5$
one finds
\ber
\ns - 1 &\simeq& -\frac{2}{N_k}\,\\
\nt  &\simeq& -\frac{1}{N_k^2\,\lambda^2}~, \\
r &\simeq& \frac{8}{N_k^2\,\lambda^2}~.
\label{eq:CMB_Emodel}
\eer
Note that these expressions {\em do not depend} upon the value of $p$ in
(\ref{eq:pot_Emodel}) as shown in figure  \ref{fig:deg_Emodel}.

\begin{figure}[t]
\begin{center}
\includegraphics[width=0.496\textwidth]{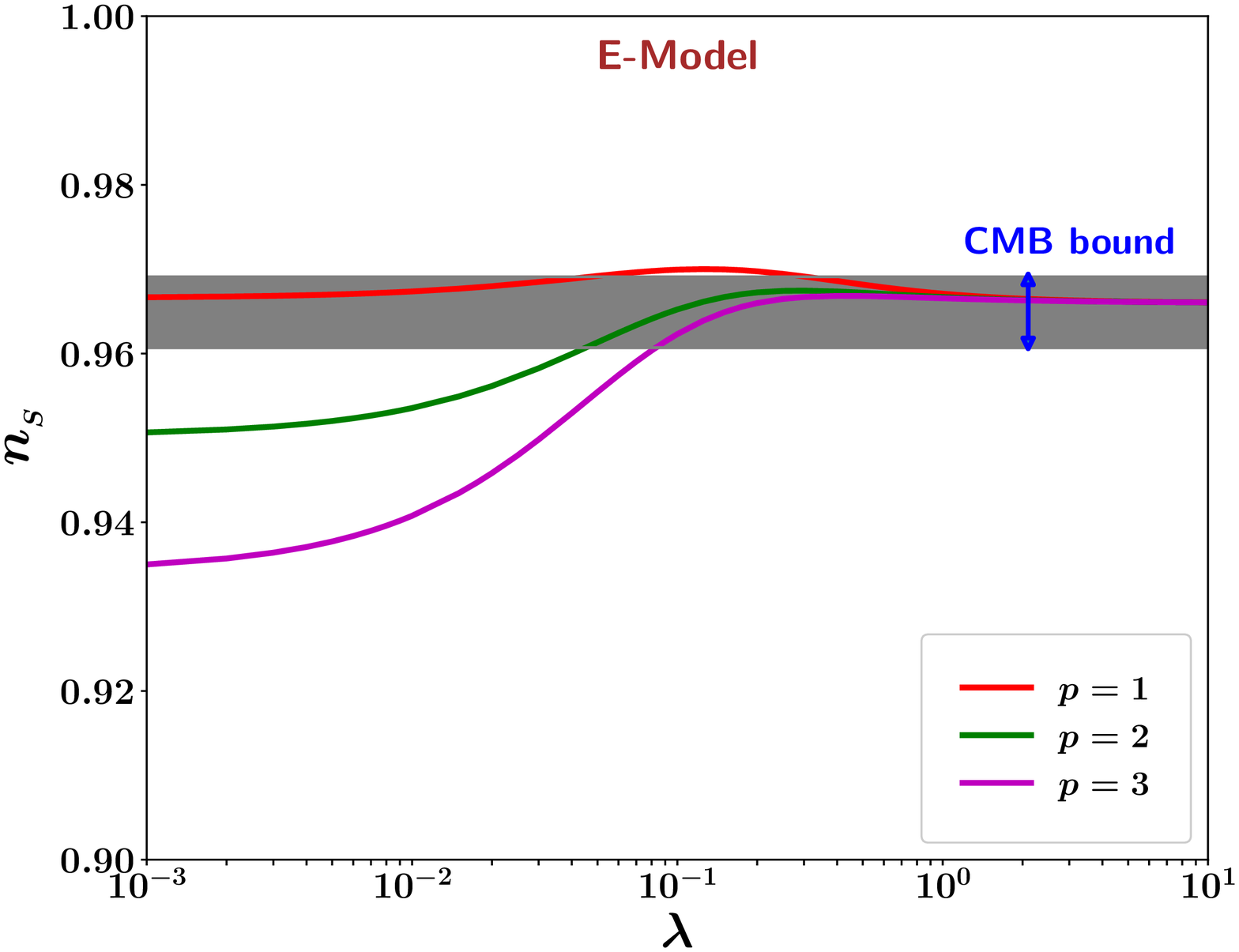}
\includegraphics[width=0.496\textwidth]{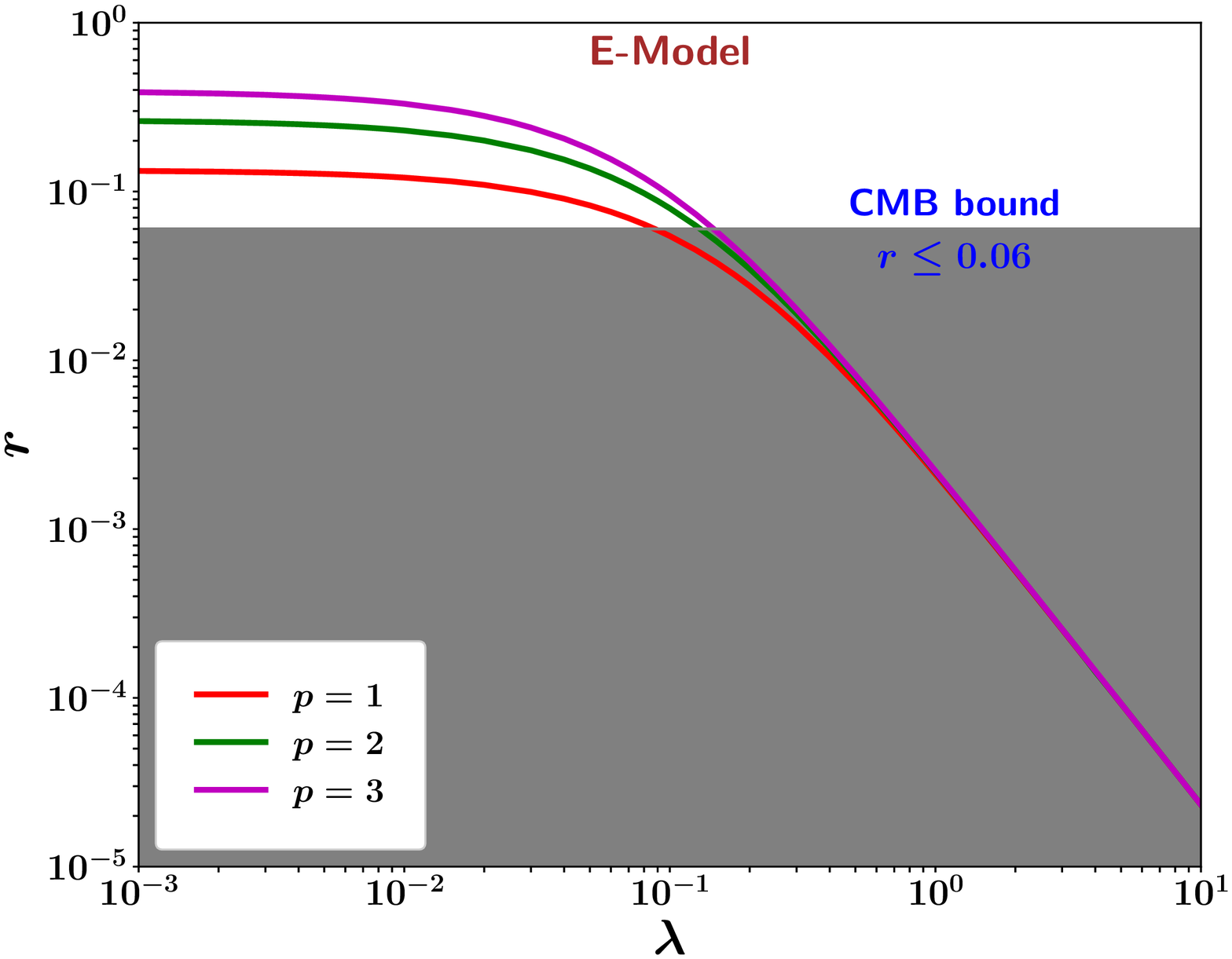}
\caption{This figure illustrates the degeneracies of the E-model $\alpha$-attractor (\ref{eq:pot_Emodel}).  In the left panel, the scalar spectral index $\ns$ is plotted as a
function of $\lambda$ for three different values of the parameter $p$ in the
potential~(\ref{eq:pot_Emodel}) while, in the right panel, the value of tensor-to-scalar $r$ is
plotted as a function of $\lambda$  for the same set of values for $p$. All the curves correspond to the number of $e$-folds $N_k = 60$. Note that when $\lambda > 0.5$, the scalar spectral index
approaches a constant value (e.g $\ns \simeq 0.967$ for $N_k=60$), whereas $r$ decreases as $\lambda^{-2}$.
The shaded region refers to the CMB 1$\sigma$ limits on $\ns$ and $r$ as determined by Planck 2018 \cite{Planck1}, namely $\ns = 0.9649 \pm 0.0042$, $r\leq 0.06$.}
\label{fig:deg_Emodel}
\end{center}
\end{figure}

\bigskip

One therefore finds that both the T-model (\ref{eq:pot_Tmodel}) as well as the E-model (\ref{eq:pot_Emodel}) display
degeneracies, since the same values of $\lbrace \ns, r\rbrace$
can correspond to different values of $p$.
Fortunately in both (\ref{eq:pot_Tmodel}) and (\ref{eq:pot_Emodel})
 this degeneracy can be broken once inflation
ends and the scalar field begins to oscillate. 
As we shall see later, $N_k$ -- the number of $e$-folds  between the Hubble exit of the CMB pivot scale and the end of inflation, is sensitive to
 the post-inflationary reheating regime. Indeed in both models, close to its minimum value, the potential has the form
$V \propto \phi^{2p}$, for which the mean EOS, $w_{_{\rm osc}} = \langle w_{\phi}\rangle$, is described by (\ref{eq:EOS}). In the perturbative reheating regime,  parameters such as the reheating duration $\Nre$ and temperature $\Tre$ are very sensitive to $\langle w_{\phi}\rangle = (p-1)/(p+1)$ and hence
to the value of $p$. Similarly the  primordial GW background is also sensitive 
to the value of $p$. Therefore a degeneracy in $\lbrace \ns, r\rbrace$
is easily broken if constraints on  the CMB observables  $\lbrace n_{_S}, r\rbrace$ are
 determined by taking into account the  reheating EOS.  This can be further supplemented by  the observations of the GW spectral density parameter $\Omega_g(k)$.

\end{enumerate}
\subsection{Non-canonical inflation}
\label{sec:non_can_inf}

Non-canonical scalars have the Lagrangian density \cite{non-can1}
\beq
{\cal L}(X,\phi) = X\l(\frac{X}{M^{4}}\r)^{\alpha-1} -\; V(\phi), ~~~~ X = \frac{1}{2}{\dot\phi}^2~,
\label{eq:lag_nc}
\eeq
where $M$ has dimensions
 of mass while $\alpha$ is dimensionless. When $\alpha = 1$
the Lagrangian (\ref{eq:lag_nc}) reduces to the usual canonical scalar field Lagrangian
${\cal L}(X,\phi) = X -\; V(\phi)$.

The energy density and pressure have the form
\ber
\rho_{_{\phi}} &=& \l(2\alpha-1\r)X\l(\frac{X}{M^{4}}\r)^{\alpha-1} +\;  V(\phi),\nonumber\\
p_{_{\phi}} &=& X\l(\frac{X}{M^{4}}\r)^{\alpha-1} -\; V(\phi), ~~
X \equiv \frac{1}{2} {\dot \phi}^{2}~,
\label{eq:rho_p_nc}
\eer
which reduces to the canonical expression
$\rho_{_{\phi}} = X + V$, ~$p_{_{\phi}} = X - V$ when $\alpha = 1$.

One should note that the
equation of motion
\beq
{\ddot \phi}+ \f{3\, H{\dot \phi}}{2\alpha -1} + \l(\f{V'(\phi)}{\alpha(2\alpha -1)}\r)\l(\f{2\,M^{4}}{{\dot \phi}^{2}}\r)^{\alpha - 1} =\; 0,
\label{eq:EOM_nc}
\eeq
is singular at $\dot{\phi} \rightarrow 0$ and needs to be
regularized so that
 the value of $\ddot{\phi}$ remains finite
in this limit. This can be done by modifying the Lagrangian (\ref{eq:lag_nc})
to \cite{non-can2,non-can3}
 \beq
{\cal L}_{_R}(X,\phi) = \l(\frac{X}{1 + \beta}\r)\l(1 + \beta\l(\frac{X}{M^{4}}\r)^{\alpha-1}\r) -\; V(\phi),
\label{eq:lag_nc_new}
\eeq
where $\beta$ is a dimensionless parameter.
In the limit when $\beta \gg 1$, equation (\ref{eq:EOM_nc})
 can be approximated as
\beq
{\ddot \phi}\,+\, \f{3\, H{\dot \phi}}{2\alpha -1}\, +\, \l(\f{V'(\phi)}{\epsilon\, +\, \alpha(2\alpha -1)\l(X/M^{4}\r)^{\alpha - 1}}\r)\, =\; 0, ~~ X = \frac{1}{2}{\dot \phi}^2~,
\label{eq:EOM_nc_new}
\eeq
where $\epsilon \equiv (1 + \beta)^{-1}$ is an infinitesimally small correction factor  when $\beta >> 1$.

As shown in \cite{non-can2} for potentials having the form
$V(\phi) = V_0\,\phi^n$ the average EOS during scalar field oscillations is
\beq
\l<w_\phi\r> = \frac{n\, -\, 2\alpha}{n\,(2\alpha -1)\, +\, 2\alpha}~.
\label{eq:EOS_nc}
\eeq
For $\alpha = 1$ the above expression reduces to the canonical result
(\ref{eq:EOS}).

Specializing to the quadratic potential with $n=2$ one gets
\beq
\l <w_\phi \r> = -\left (\frac{\alpha-1}{3\alpha-1}\right )
\label{eq:EOS_nc_quad}
\eeq
which informs us that the mean EOS during oscillations is
{\em negative} and lies in the interval $-1/3 < \l<w^{^{\rm NC}}_\phi\r> < 0$
for $\alpha > 1$.
This should be contrasted to the EOS of
oscillating canonical scalars
(\ref{eq:EOS})
which lies in the interval $0 \leq \l<w_\phi\r> < 1$.
One should note that the two ranges $-1/3 < \l<w^{^{\rm NC}}_\phi \r> < 0$ and
$0 \leq \l<w_\phi\r> < 1$ are complementary.
As a result the GW spectrum, which is sensitive to the precise value of
$\l<w_\phi \r>$, can easily distinguish between canonical
and non-canonical models of inflation.

\begin{figure}[t]
\begin{center}
\includegraphics[width=0.49\textwidth]{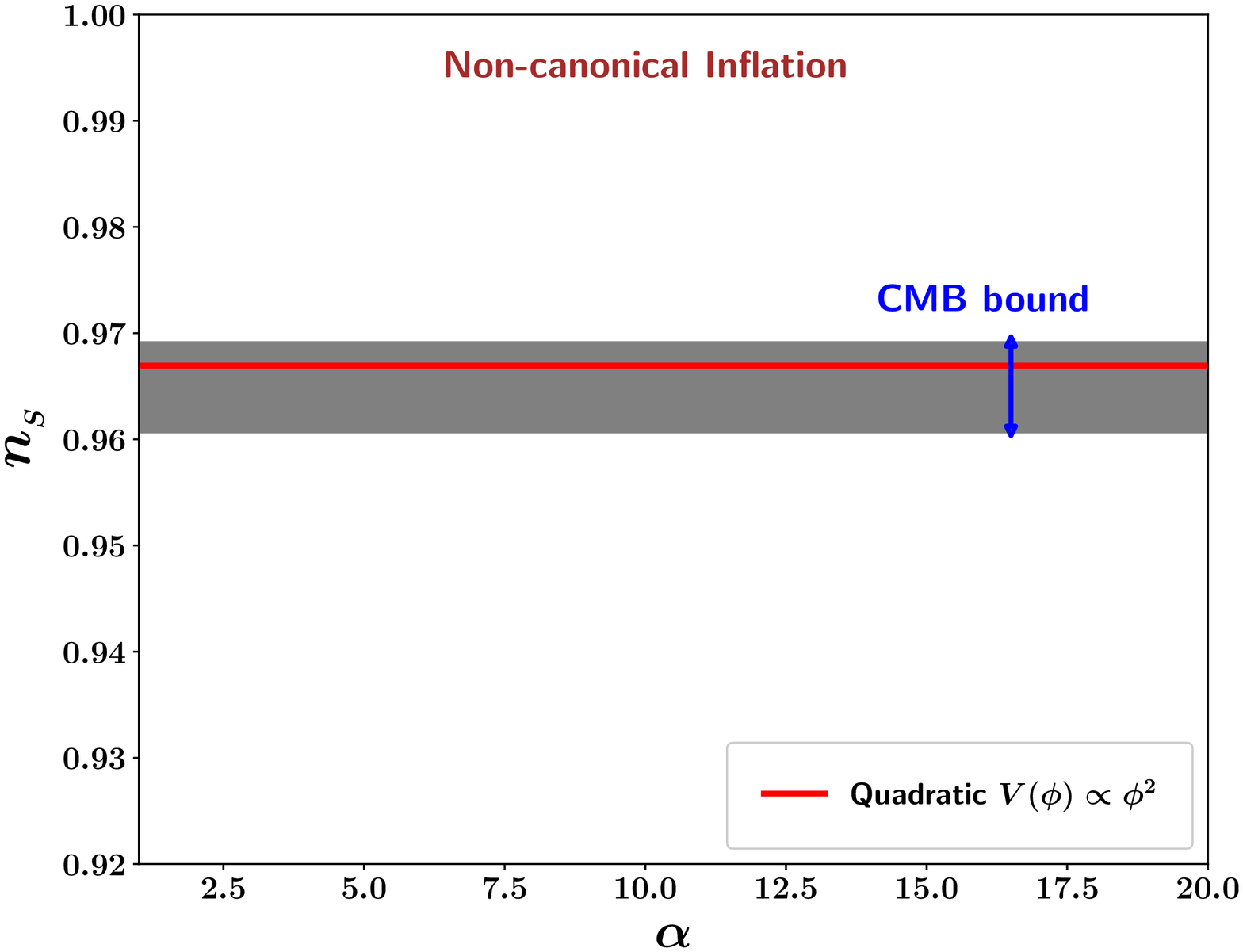}
\includegraphics[width=0.49\textwidth]{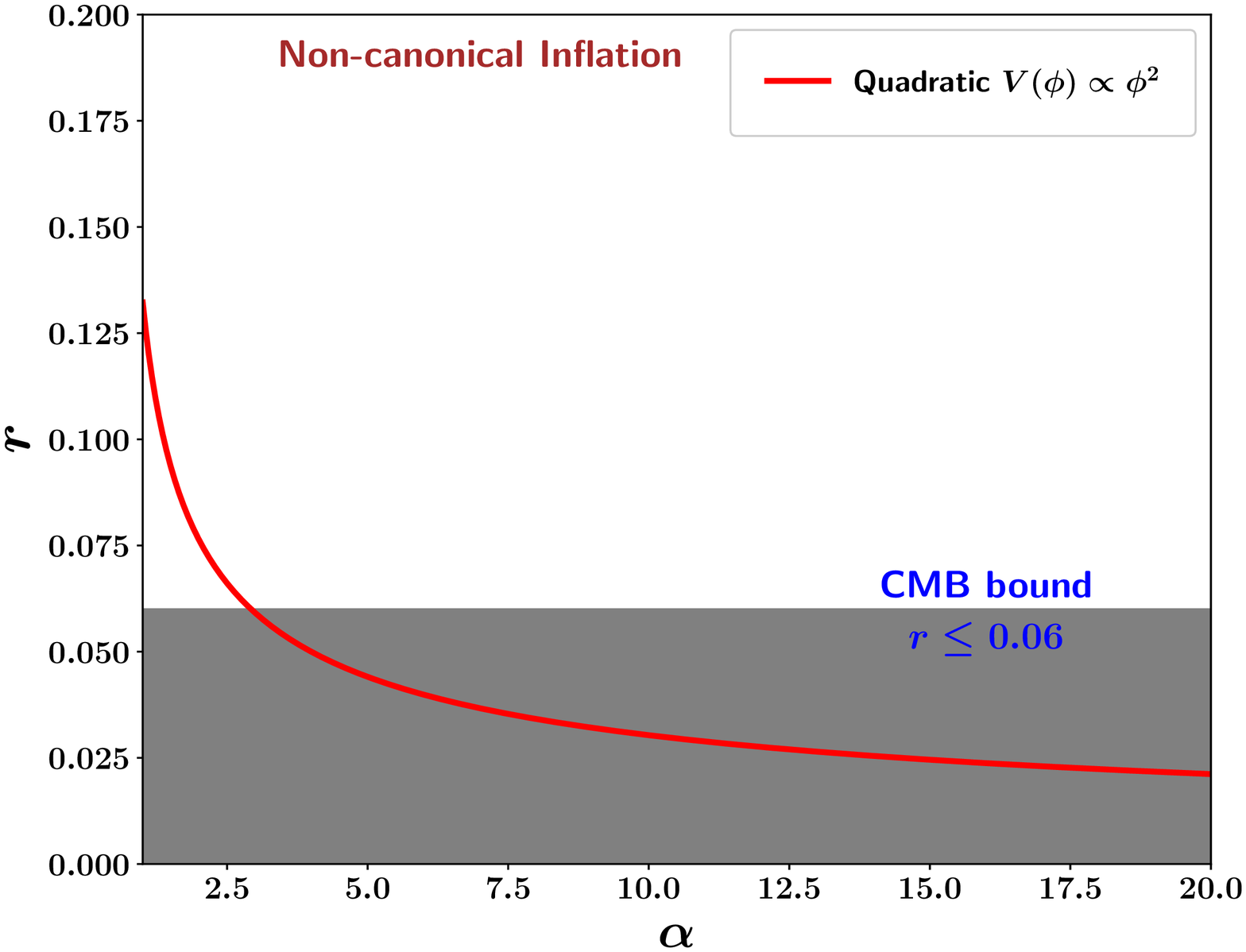}
\caption{The scalar spectral index $\ns$ (left panel)
and the tensor-to-scalar ratio $r$ (right panel) are determined for
the potential $V(\phi) = \f{1}{2}m^2\phi^{2}$ and shown as functions of
the non-canonical parameter $\alpha$ defined in
(\ref{eq:lag_nc}). 
Note that increasing $\alpha$ leads to a decrease in $r$ 
but has no effect on $\ns$ which is insensitive to this parameter; see (\ref{eq:ns_nc}). The shaded region refers to the CMB 1$\sigma$ limits on $\ns$ and $r$ as determined by Planck 2018 \cite{Planck1}, namely $\ns = 0.9649 \pm 0.0042$, $r\leq 0.06$.}
\label{fig:ns_r_nc}
\end{center}
\end{figure}

Turning next to the fluctuation parameters $\lbrace \ns, r\rbrace$ one finds
for  $V(\phi) = \frac{1}{2}m^2\phi^2$ the expression \cite{non-can2}
\beq
   \ns = 1 -\l(\f{4}{2N_k + 1}\r).
\label{eq:ns_nc}
\eeq
Surprisingly $\ns$ {\em does not} depend upon the value of $\alpha$
and coincides with the result
for canonical scalars.
Note that for $N_k \gg 1$ (\ref{eq:ns_nc}) reduces to the result
\beq
\ns- 1 \simeq -\frac{2}{N_k}
\eeq
obtained earlier for the T-model and the E-model potentials.

The value of $r$ is given by
\beq
r = \left (\frac{1}{\sqrt{2\alpha - 1}}\right )
\left (\frac{16}{2N_k+1}\right )
\label{eq:r_nc}
\eeq
which reduces to the standard result
\beq
r = \left (\frac{16}{2N_k+1}\right )
\eeq
for canonical scalars ($\alpha = 1$).
From (\ref{eq:r_nc}) we find that increasing the value of the
non-canonical parameter $\alpha$ reduces the scalar-to-tensor ratio $r$;
see figure  \ref{fig:ns_r_nc}.
Comparing (\ref{eq:r_nc}) with (\ref{eq:CMB_Tmodel})
\& (\ref{eq:CMB_Emodel}) we find that $\alpha$ plays the same role
as the parameter $\lambda$ in the potentials (\ref{eq:pot_Tmodel}), (\ref{eq:pot_Emodel}).
There therefore appears to be a close similarity, indeed a degeneracy,
between values of $\lbrace n_{_S}, r\rbrace$ in the non-canonical $m^2\phi^2$
model and in the T-model and Starobinsky model, respectively.
Interestingly, for a given value of $N_k$
all three models have the same value of $n_{_S}$.
The value of $r$
can also be identical in the three models by an
appropriate choice of $\lambda$ and $\alpha$.
(Note that $n_{_S}$ does not depend upon the free parameters
$\lambda$ and $\alpha$ present in the $\alpha$-attractor models and non-canonical models, respectively.)

The degeneracy in $\lbrace n_{_S}, r\rbrace$ in the three inflationary models
is easily broken by noting that
the equation of state during post-inflationary
oscillations of the scalar field is markedly different in canonical
(\ref{eq:EOS}) and
non-canonical (\ref{eq:EOS_nc}) models. This implies that the
 spectrum of the relic gravitational wave background
will differ in canonical and non-canonical models.
To summarize, the relic GW background carries an important imprint
of the post-inflationary universe which can be used to break degeneracies in
the inflationary parameters $\lbrace n_{_S}, r\rbrace$
probed by the CMB.

\section{Reheating}
\label{sec:reheat}

A key feature of inflationary cosmology is that it allows the universe to reheat\footnote{The term `reheating' is a misnomer carried over from early models of inflation
in which inflation began from a thermalized initial state. In later models, such as chaotic
inflation and those discussed in this paper,
 the universe commences inflating from a non-radiation state and heats up only once,
after inflation ends \cite{linde90}.} by transferring
the energy localized in the inflaton to the matter/radiative degrees of freedom present in the universe.

In potentials possessing a minimum,
reheating can occur in two distinct ways: (i) perturbatively (slowly),
(ii) rapidly -- via a parametric resonance.
Which of these two ways is realized depends upon the nature of the coupling between the inflaton
and bosons/fermions.
Below we provide a brief summary of perturbative and non-perturbative
 reheating in the context which is relevant for this paper.
The reader is referred to \cite{kofman94,yuri95,kofman96a,kofman96,kofman97} for more details on the subject of reheating.

\subsection{Perturbative reheating}
\label{sec:reheat_pert}

The perturbative theory of reheating after inflation corresponds to the case when the average number of particles created in each Fourier mode ${\bf k}$ is small. It was first used already in the same paper~\cite{inf_star80}, where the $R+R^2$ inflationary model was introduced, in order to obtain the transition from the quasi-de Sitter (inflationary) stage to radiation dominated one through an intermediate matter dominated stage. Its detailed presentation can be found in~\cite{Starobinsky:1981vz} and more recently in~\cite{DeFelice:2010aj}. In the context of the new inflationary scenario~\cite{inf_linde82,inf_alstein82} such theory was later developed in~\cite{albrecht82}. 
Phenomenologically it amounts to adding a friction term $\Gamma {\dot\phi}$ to the classical
equation of  motion of the scalar field oscillating around the minimum of  its potential \cite{albrecht82,Kolb_Turner}
\beq
{\ddot\phi} + 3H{\dot\phi} + \Gamma {\dot\phi} + V'(\phi) = 0~.
\label{eq:EOM_damp}
\eeq
The perturbative theory of reheating works well if either (a) the inflaton decays only into fermions 
$\psi$ through 
a $h\psi{\bar\psi}\phi$ coupling with $h^2 \ll m_\phi/\mpl$, or (b) 
the coupling of the inflaton to bosons, $\chi$, described by $\frac{1}{2}g^2\phi^2\chi^2$ is weak with
$g \ll 3\times 10^{-4}$, making particle production via parametric resonance ineffective \cite{kofman96a}.

As noted in the previous section, close to their minimum value, the potentials discussed in this paper
have the general form $V(\phi) \propto \phi^{2p}$.
Equation (\ref{eq:EOM_damp}) suggests that the amplitude of scalar field oscillations
around this minimum decreases as
\beq
\phi_{\rm max} \propto a^{-\left (\frac{3}{p+1}\right )} \exp{\left (-\frac{\Gamma\, t}{2p}\right )} {\rm}
\label{eq:oscillations}
\eeq
which reduces to the standard result \cite{kofman96a}
\beq
\phi_{\rm max} \propto a^{-{3}/{2}}\, e^{-\frac{1}{2}\Gamma\,t}
\eeq
for the chaotic potential $V \propto m^2\phi^2$.

Reheating in this perturbative scenario is complete when the (decreasing) expansion rate
becomes equal to the decay rate so that $H \simeq \Gamma$. Following thermalization, the reheating temperature
is given by $T_r \simeq 0.1\sqrt{\Gamma\mpl}$ which is independent of the duration of inflation and the
properties of $V(\phi)$.
The fact that the coupling between matter and the inflaton can alter, via radiative corrections,
the shape of $V(\phi)$, places strong constraints on the total decay rate: $\Gamma < 10^{-20}\mpl$.
This in turn implies that the reheating temperature in perturbative models can be relatively small
$T_r < 10^9$ GeV \cite{kofman96a,kofman97}.

It therefore follows that the post-inflationary oscillatory stage in models with perturbative (slow)
reheating can be quite long. It is important to note that during most of this stage
(while $H \gg \Gamma$) the EOS  of the oscillating scalar field is given by 
(\ref{eq:EOS}) namely
$\langle w_\phi \rangle = \frac{p-1}{p+1}$.

As pointed out in \cite{sahni90} and discussed in detail
  in  section \ref{sec:GW}, the spectrum of relic gravitational waves
 created during inflation is very sensitive to the post-inflationary
EOS,  $\wre$, and hence to
the value of the inflationary parameter $p$.
Observations of the GW spectrum can therefore
help in breaking the degeneracy between inflationary models which was pointed out in 
section \ref{sec:inflation}.

\subsection{Non-perturbative reheating}

For inflationary models in which the main source of reheating is through the decay of the inflaton
into bosons, the universe thermalizes and reheats through a sequence of successive stages.

\begin{enumerate}

\item
The first stage, sometimes called {\em preheating},
 sees the commencement of a parametric resonance brought about by coherent
oscillations of the inflaton $\phi$ around the minimum of its potential. The resonance can be
either narrow or broad depending upon (a) the value of coupling constant $g$ in the interaction
$\frac{1}{2}g^2\phi^2\chi^2$ between the inflaton and the bosonic field $\chi$, (b) the scalar field
amplitude $\Phi$, (c) its effective mass, $m_\phi^2 = V''$. 
If the resonance is broad ($g^2\Phi^2/m_\phi^2 \ggeq 1$)  then coherent oscillations of $\phi$
give rise to an exponentially large number of quanta of the field $\chi$ in a discrete
set of wave bands.
(The existence of self-interaction, such as the presence of
 a $\lambda \phi^4$ term in the Lagrangian, can also result in the creation
of quanta of the $\phi$ field during oscillations.)

\item
The second stage witnesses the backreaction of $\chi$ on $\phi$ (via scattering).
This effect can be quite significant
and can lead to the termination of the resonance. 

\item
During the third stage, which can be quite prolonged, quanta of the $\phi$ and $\chi$ fields
transfer their energy into other matter fields including radiation. The interaction between the
quanta of different fields leads to their thermalization and results in the universe acquiring
a reheating temperature $T_{\rm rh}$.

\end{enumerate}
 Thus the end of the third stage sees the commencement of
 the radiation dominated stage of expansion during which the EOS in the universe is
$p \simeq \epsilon/3$.
The dynamics of the three stages of reheating is quite complex and usually requires a numerical
treatment \cite{tkachev96}.
It is however quite instructive if one characterizes the pre-radiation stages
(1) - (3) by an effective EOS which, following \cite{cook15,kamion_1,kamion_2}, can be assumed to be a constant lying
in the interval $-1/3 < w \leq 1$.

As noted earlier, and discussed in detail in section \ref{sec:GW}, the pre-radiation EOS, $w$,
affects the spectrum of relic gravity waves produced during inflation.
Future space-based GW experiments might therefore shed light on this
important parameter, and through it on the physics of the reheating epoch.

Before moving forward, let us designate the  variables and parameters 
that are essential to describe the reheating kinematics by listing them down systematically at one place, for the convenience of the reader. We repeat the specifications in the text wherever necessary.
\begin{itemize}
\item
$a_k$, $H_k^{\rm inf}$, $\phi_k$ :  Scale factor, Hubble parameter and the inflaton field value, respectively, during the Hubble exit of a mode $k$, usually taken to be the CMB pivot scale $k\equiv k_*=0.05~{\rm Mpc}^{-1}$.
\item
$a_e$, $H_e$, $\rho_e$, $\phi_e$ : Scale factor, Hubble parameter, energy density and inflaton field value, respectively, at the end of inflation.
\item
$a_{\rm re}$, $H_{\rm re}$, $\Rre$, $\Tre$, $g_{\rm re}$, $g_{\rm re}^s$ : Scale factor, Hubble parameter, energy density, temperature, effective number of relativistic degrees of freedom in energy and entropy, respectively, at the end of reheating.
\item
$a_p$, $H_p$ : Scale factor and Hubble parameter, respectively, at the Hubble re-entry epoch of the pivot scale. 
\item
$a_{\rm BBN}$, $H_{\rm BBN}$, $T_{\rm BBN}$ : Scale factor, Hubble parameter and temperature, respectively, at the beginning of Big Bang Nucleosynthesis.
\item
$a_{\rm eq}$,  $H_{\rm eq}$, $T_{\rm eq}$, $g_{\rm eq}$, $g_{\rm eq}^s$ : Scale factor, Hubble parameter, temperature, effective number of relativistic degrees of freedom in energy and entropy, respectively, at the epoch of matter-radiation equality.
\item $a_0$, $H_0$, $T_0$, $g_0$, $g_0^s$ : Scale factor, Hubble parameter, temperature, effective number of relativistic degrees of freedom in energy and entropy, respectively, at the present epoch.
\item $N_k^{\rm inf}=\log{\l( a_e/a_k \r)}$ : Number of $e$-folds between the Hubble exit of scale $k$ and the end of inflation.
\item $\Nre = \log{\l(a_{\rm re}/a_e\r)}$ : Duration of reheating.

\item $\Nrd = \log{\l(a_{\rm eq}/a_{\rm re}\r)}$ : Duration of the radiation dominated epoch.
\item $\wre$ : Effective equation of state of the universe during the epoch of reheating.

\end{itemize}

\subsection{Essentials of reheating kinematics}

The main focus of this section will be on perturbative reheating. 
The epoch of reheating is usually characterized by a set of  three parameters $\lbrace \wre,\Nre,\Tre \rbrace$, namely the effective equation of state (EOS) during reheating $\wre$, the duration of reheating  $\Nre$ and the  temperature at the end of reheating $\Tre$, when the universe transits to a thermalized  radiation dominated  hot Big Bang phase (see \cite{kamion_1,Martin:2014nya, kamion_2,cook15}). The duration of reheating can be defined by the number of $e$-folds between the end of inflation $a_e$ and the end of reheating 
(commencement of the radiation dominated epoch) $a_{re}$, given by $\Nre=\log{\l(a_{\rm re}/a_e\r)}$. While $\Nre$ and $\Tre$ are interesting physical quantities in their own right describing the epoch of reheating, they are also potentially important for correctly interpreting the bounds on CMB observables such as the scalar spectral index $\ns$  and tensor-to-scalar ratio $r$, as we discuss next.

\begin{figure}[t]
\begin{center}
\includegraphics[width=0.9\textwidth]{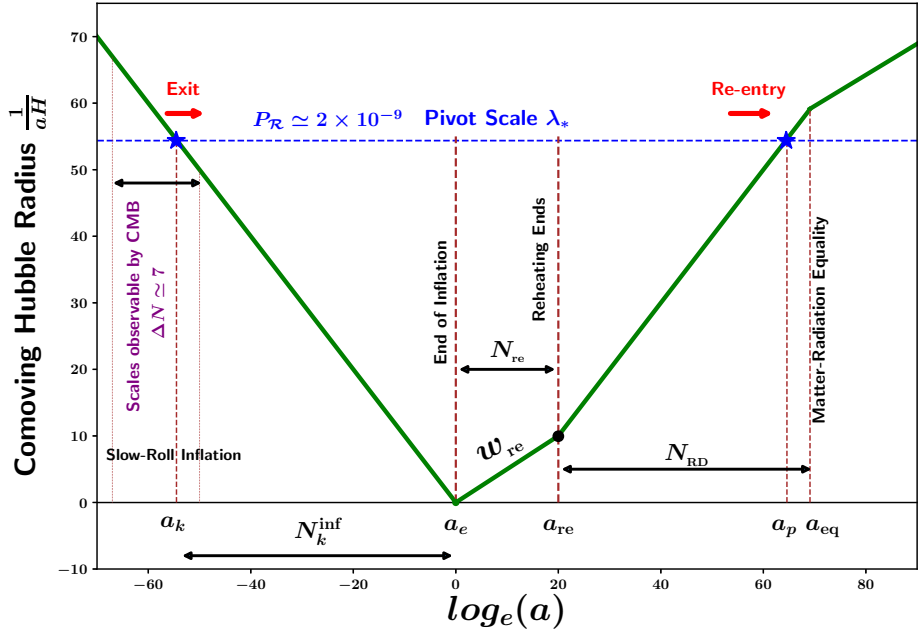}
\caption{This figure schematically illustrates the evolution of the comoving Hubble radius
$(aH)^{-1}$ with scale factor. During inflation $(aH)^{-1}$ decreases which causes physical scales to
exit the Hubble radius. After inflation ends $(aH)^{-1}$ increases, and physical scales begin to re-enter
the Hubble radius.
The CMB pivot scale, as used by the Planck mission, is set at $k_*=0.05~{\rm Mpc}^{-1}$.
It enters the 
Hubble radius during the radiation dominated epoch when $a_p \sim 4\times 10^{-5}\, a_0$. Note that the duration of reheating $\Nre$, and hence the duration of the radiation dominated epoch $\Nrd$, changes for different values of the reheating equation of state $\wre$. Note
 that $(aH)^{-1} \propto a$ during the radiation dominated regime and
$(aH)^{-1} \propto a^{-1}$ during
inflation.
}
\label{fig:RH_causal}
\end{center}
\end{figure}

Following the evolution of the comoving Hubble radius from the epoch of Hubble exit, at $a_k$, of scale $k$, 
until its late-time re-entry at $a_p$, one gets (see appendix \ref{app:A}) 
\beq
\log{\f{k}{a_0 H_0}} = -N_k^{\rm inf} - \Nre - \Nrd - \log{\l(1+z_{\rm eq}\r)} + \log{\f{H_k^{\rm inf}}{H_0}}~,
\label{eq:CMB_reheat_k}
\eeq
where $H_k^{\rm inf}$ is the Hubble parameter at the time of the Hubble exit of the scale $k$,  $N_k^{\rm inf}=\log{\l( a_e/a_k \r)}$ is 
the number of $e$-folds between the Hubble exit (of scale $k$) and the end of inflation, $\Nrd$ is the duration of the radiation dominated epoch and $z_{\rm eq}$ is the redshift at the epoch of  matter-radiation equality. In general, $k$ may correspond to any observable CMB scale in the range $k\in \l[0.0005,0.5\r]~{\rm Mpc}^{-1}$. However, in order to derive constraints on the inflationary observables $\lbrace \ns,r \rbrace$, we define $k$
 to be the CMB pivot scale, namely $k\equiv k_*=0.05~{\rm Mpc}^{-1}$, which makes its Hubble re-entry 
during the radiation dominated epoch at $a_p\sim 4\times 10^{-5}\, a_0$; see figure  \ref{fig:RH_causal}. 

Our main goal is to characterize the epoch of reheating between the end of inflation $a_e$ and the commencement of the radiation dominated epoch $a_{\rm re}$. Assuming the  effective equation of state $\wre$ during reheating to be a constant, 
allows one to match the density at the beginning of the radiation dominated epoch to the density at the end of inflation by 

\beq
\Rre = \rho_e \l(\f{a_e}{a_{\rm re}} \r)^{3(1+\wre)}~,
\label{eq:CMB_reheat_rho}
\eeq
which yields the following expression for the duration of reheating

\beq
\Nre \equiv \log{\l(\f{a_{\rm re}}{a_e}\r)} = \f{1}{3(1+\wre)}\log{\l(\f{\rho_e}{\Rre}\r)}~.
\label{eq:CMB_reheat_Nre}
\eeq

Expressing $\Rre$ in terms of the reheating temperature $\Tre$, one gets

\beq
\Nre = \f{1}{3(1+\wre)}\log{\l(\f{\rho_e}{\f{\pi^2}{30}g_{\rm re}\Tre^4}\r)}~,
\label{eq:CMB_reheat_Nre_Tre}
\eeq
 where $g_{\rm re} \equiv g(\Tre)$ is the effective number of relativistic degrees of freedom at the end of reheating. Applying entropy conservation to express $\Tre$ in terms of $a_{\rm re}$, one finds (appendix \ref{app:A})
 
\beq
\Tre = \l(\f{g_{\rm eq}^s}{g_{\rm re}^s}\r)^{\f{1}{3}} \l(\f{a_{\rm eq}}{a_{\rm re}}\r)T_{\rm eq}~,
\label{eq:CMB_reheat_Tre}
\eeq
where $g_{\rm eq}^s$ and $g_{\rm re}^s$ are the effective number of relativistic degrees of freedom in
the entropy at the epoch of matter-radiation equality and at the end of reheating respectively,
 while $T_{\rm eq}$  is the temperature at the matter-radiation equality. Incorporating (\ref{eq:CMB_reheat_Tre}) into (\ref{eq:CMB_reheat_Nre_Tre}), we obtain
\beq
\Nre = \f{4}{3(1+\wre)}\l[ \f{1}{4} \log{\l(\f{30}{\pi^2 g_{\rm re}}\r)} + \f{1}{3}\log{\l(\f{g_{\rm re}^s}{g_{\rm eq}^s}\r)} + \log{\l(\f{\rho_e^{\f{1}{4}}}{T_{\rm eq}}\r)}-\Nrd\r]~.
\label{eq:CMB_reheat_Nre_1}
\eeq
Substituting $\Nrd$ from (\ref{eq:CMB_reheat_k}) into (\ref{eq:CMB_reheat_Nre_1}), we arrive at an
 important expression for the duration of reheating, namely 
\beq
\Nre = -\f{4}{3(1+\wre)}\l[ \f{1}{4} \log{\l(\f{30}{\pi^2 g_{\rm re}}\r)} + \f{1}{3}\log{\l(\f{g_{\rm re}^s}{g_0^s}\r)} + \log{\l(\f{\rho_e^{\f{1}{4}}}{H_k^{\rm inf}}\r)} + \log{\l(\f{k}{a_0 T_0}\r)}+N_k^{\rm inf}+\Nre \r]~.
\label{eq:CMB_reheat_Nre_2}
\eeq
Note that if $\wre = 1/3$, then the term $\Nre$ cancels from both sides of (\ref{eq:CMB_reheat_Nre_2}), yielding the 
following expression for $N_k^{\rm inf}$
\beq
N_k^{\rm inf} = - \l[ \log{\l(\f{k}{a_0 T_0}\r)} + \log{\l(\f{\rho_e^{\f{1}{4}}}{H_k^{\rm inf}}\r)} + \f{1}{4} \log{\l(\f{30}{\pi^2 g_{\rm re}}\r)} + \f{1}{3}\log{\l(\f{g_{\rm re}^s}{g_0^s}\r)} \r] ~.
\label{eq:CMB_reheat_Nk}
\eeq
This arises because the end of reheating, and hence the beginning of the radiation dominated epoch, cannot be strictly defined 
within this framework if $\wre = 1/3$. However for
  $\wre\neq 1/3$ one obtains the following final expression 
for $\Nre$ from equation (\ref{eq:CMB_reheat_Nre_2})
\beq
\Nre = -\f{4}{1-3\wre}\l[N_k^{\rm inf} + \log{\l(\f{\rho_e^{\f{1}{4}}}{H_k^{\rm inf}}\r)}  + \log{\l(\f{k}{a_0 T_0}\r) + \f{1}{4} \log{\l(\f{30}{\pi^2 g_{\rm re}}\r)} + \f{1}{3}\log{\l(\f{g_{\rm re}^s}{g_0^s}\r)} } \r]~.
\label{eq:CMB_reheat_Nre_3}
\eeq
Accordingly the expression for the reheating temperature $\Tre$ in terms of the duration of reheating $\Nre$ and effective reheating EOS, $\wre$,
 follows from (\ref{eq:CMB_reheat_Nre_Tre}) to be 
\beq
\Tre = \l(\f{30 \rho_e}{\pi^2 g_{\rm re}}\r)^\f{1}{4} e^{-\f{3}{4}\l(1+\wre\r)\Nre}~.
\label{eq:CMB_reheat_Tre2}
\eeq
Having expressed the duration of reheating $\Nre$ and the 
reheating temperature $\Tre$ in terms of the effective reheating EOS, $\wre$,
in (\ref{eq:CMB_reheat_Nre_3}) and ({\ref{eq:CMB_reheat_Tre2}}) respectively, 
we now discuss how these two quantities can be used to obtain tighter constraints
 on the CMB observables $\lbrace \ns,r \rbrace$.  Note that  the expressions (\ref{eq:CMB_reheat_Nre_3}) and ({\ref{eq:CMB_reheat_Tre2}}) are valid only for $\wre\neq 1/3$. We return to the case $\wre = 1/3$ at the end of this subsection, for which the relevant final expression for $N_k^{\rm inf}$, following equation (\ref{eq:CMB_reheat_Nk}), is given in equation (\ref{eq:CMB_reheat_Nk_SR}).

In the context of single field slow-roll inflation with potential $V(\phi)$, the `potential slow-roll parameters' are defined by

\ber
\epsilon_{_V} = \f{\mpl^2}{2}\l(\f{V'}{V}\r)^2~,\label{eq:inf_eV} \\
\eta_{_V} = \mpl^2\l(\f{V''}{V}\r)~,\label{eq:inf_etaV}
\eer
and the slow-roll limit corresponds to $\epsilon_{_V},\eta_{_V} \ll 1 $. The value of the inflaton field at the end of inflation $\phi_e$ can be determined   from the condition
\beq
\epsilon_{_V}(\phi_e) = \f{\mpl^2}{2}\l(\f{V'}{V}\r)^2\bigg\vert_{\phi_e} \simeq 1~,
 \label{eq:inf_end}
\eeq
and the corresponding inflaton density at the end of inflation is given by (appendix \ref{app:B})
$$\rho_e \equiv \rho_\phi\bigg\vert_{\phi_e} = \f{1}{2}\dot{\phi}^2+V(\phi)\bigg\vert_{\phi_e} \simeq \f{3}{2}V(\phi_e) 
\equiv \f{3}{2}V_e~.$$
Substituting $\rho_e = \f{3}{2}V_e$ in
(\ref{eq:CMB_reheat_Tre2}) results in the following expression for the reheating temperature 
\beq
\Tre =  \l(\f{45}{\pi^2 g_{\rm re}}\r)^\f{1}{4} V_e^{\f{1}{4}}\, e^{-\f{3}{4}\l(1+\wre\r)\Nre}~.
\label{eq:CMB_reheat_Tre_SR}
\eeq
Assuming $k$ to be the CMB pivot scale, $k\equiv k_* = a_k H_k^{\rm inf} = a_p H_p = 0.05~{\rm Mpc}^{-1}$ in 
(\ref{eq:CMB_reheat_Nre_3}) and inserting the values of $T_0$, $g_0^s$, $g_{\rm re}$ and $g_{\rm re}^s$, one arrives at the following formula 
which expresses the duration of reheating $\Nre$  as a function of the reheating EOS, $\wre$,
on the  one hand, and parameters of the inflationary potential $V_e$, $H_k^{\rm inf}$, on the other  (see appendix \ref{app:A})
\beq
\Nre = \f{4}{1-3\wre}\l[61.55 -N_k^{\rm inf} - \log{\l(\f{V_e^{\f{1}{4}}}{H_k^{\rm inf}}\r)}  \r]~, ~~~~\wre \neq 1/3
\label{eq:CMB_reheat_Nre_SR}
\eeq
For a given slow-roll inflationary model with potential  
\beq
V(\phi) = V_0 \, f\l(\f{\phi}{\mpl}\r)~,
\label{eq:inf_pot}
\eeq
the number of inflationary $e$-folds $N_k^{\rm inf}$ is given by 
\beq
N_k^{\rm inf} = \f{1}{m_p}\int_{\phi_e}^{\phi_k} \f{d\tilde\phi}{\sqrt{2\epsilon_{_V}(\tilde\phi)}}~,
\label{eq:inf_N_k}
\eeq
where $\phi_k$ is the value of the inflaton field at the Hubble exit of the scale $k$ (which we take to be the CMB pivot scale $k\equiv k_* = 0.05~{\rm Mpc}^{-1}$). Note that $N_k^{\rm inf}$ does not depend upon the value of $V_0$ which is fixed by CMB normalization to be \cite{Planck1} 
\beq
A_{_S} \equiv \f{1}{24\pi^2}\l(\f{V_0}{\mpl^4}\r)\f{f(\phi_k)}{\epsilon_{_V}(\phi_k)} = 2.1 \times 10^{-9}~.
\label{eq:CMB_As}
\eeq 
The expressions for the scalar spectral index $\ns$ and tensor-to-scalar ratio $r$, in the slow-roll limit, are given by (appendix \ref{app:B})

\ber
\ns = 1 + 2 \, \eta_{_V}(\phi_k) - 6 \, \epsilon_{_V} (\phi_k) ~,\label{eq:inf_ns_SR}\\
r  = 16\, \epsilon_{_V}(\phi_k)~.
\label{eq:inf_r_SR}
\eer
Note that the  CMB observables $\lbrace \ns,r \rbrace$ depend upon the value of inflaton field $\phi_k$ at the Hubble exit of the pivot scale $k$. On the other hand equation (\ref{eq:inf_N_k}) 
informs us that $\phi_k$ depends upon the  number of $e$-folds $N_k^{\rm inf}$ between the Hubble exit of 
scale $k$ and the end of inflation. It therefore follows that
  $\lbrace \ns,r \rbrace$ ultimately depend upon $N_k^{\rm inf}$. This implies that constraints on $\ns$ and $r$, for a given inflationary potential, directly translate onto a constraint on $N_k^{\rm inf}$. 

The CMB $1\sigma$ constraints on the scalar spectral index $\ns$ and the tensor-to-scalar ratio $r$ from the recent CMB observations are given by\footnote{The constraint on the tensor-to-scalar ratio $r$ has been obtained by the combined observations of Planck 2018 and BICEP-II \cite{bicep2}}
\ber
\ns = 0.9649\pm 0.0042 \, \label{eq:Planck2018_ns}\\
r \leq 0.06
\label{eq:Planck2018_r}
\eer
The constraint on $\ns$ is especially strong and effectively restricts the scalar spectral index to
the interval $\ns \in \l[0.9607,0.9691\r]$. Next generation  CMB missions are expected to determine 
$\ns$ to within $0.1\%$  precision as discussed in \cite{cook15,kamion_1,kamion_2}. 
For a given inflationary model, the CMB constraint on $\ns$ effectively restricts the value $N_k^{\rm inf}$ as discussed above. 
For a given $N_k^{\rm inf}$, equation (\ref{eq:inf_N_k}) can be inverted numerically\footnote{It is also possible to invert (\ref{eq:inf_N_k}) to obtain $\phi_k(N_k^{\rm inf})$ by using Lambert functions as described in \cite{inf_encyclo}.} 
to obtain the value of $\phi_k$ 
(since $\phi_e$ is determined from (\ref{eq:inf_end})). A knowledge of $\phi_k$  can then be used to obtain the value of  $H_k^{\rm inf}$ given by
\beq
\l(H_k^{\rm inf}\r)^2 = \f{1}{3m_p^2}\l(\f{1}{2}\dot{\phi}^2 + V(\phi)\r)\bigg\vert_{\phi = \phi_k} \simeq \f{1}{3m_p^2}V(\phi_k) =  \f{1}{3m_p^2}V_0 f(\phi_k)~,
\label{eq:inf_H_SR}
\eeq
 with   $V_0$ determined from the CMB normalization (\ref{eq:CMB_As}).
 The latter can also be used to obtain the value of  $V_e \equiv V(\phi_e)=V_0\, f(\phi_e)$ by substituting the value of $\phi_e$ from (\ref{eq:inf_end}).

In the perturbative reheating scenario the effective reheating EOS  is obtained from the inflationary potential via
  $\wre = \langle w_\phi \rangle$. For inflationary models in which the inflaton oscillates around the minimum 
of a $V \propto \phi^{2p}$ potential the effective reheating EOS is given by (\ref{eq:EOS}), namely
$\wre \equiv \langle w_\phi \rangle = \f{p-1}{p+1}$.
 The tight constraint on $\ns$ from CMB observations translate into constraints on $N_k^{\rm inf}$, $H_k^{\rm inf}$ and $V_e$ for a given inflationary potential as discussed above. One can then
use (\ref{eq:CMB_reheat_Nre_SR}) and (\ref{eq:CMB_reheat_Tre_SR}) 
to obtain constraints on the duration of reheating $\Nre$ and 
the reheating temperature $\Tre$, respectively.

\begin{figure}[t]
\begin{center}
\includegraphics[width=0.7\textwidth]{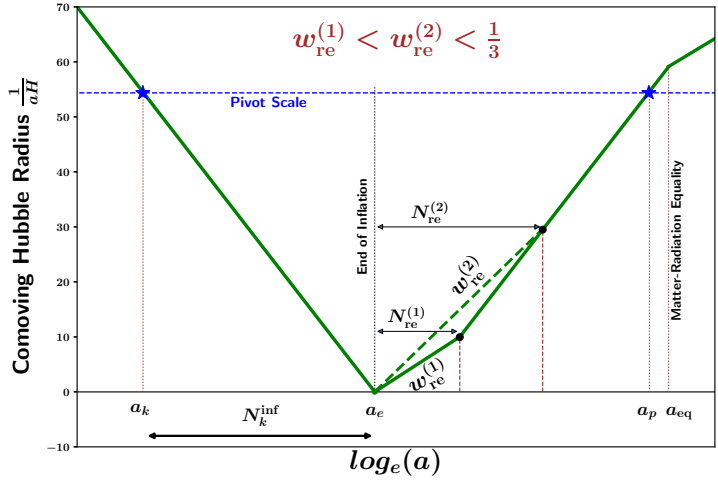}
\includegraphics[width=0.7\textwidth]{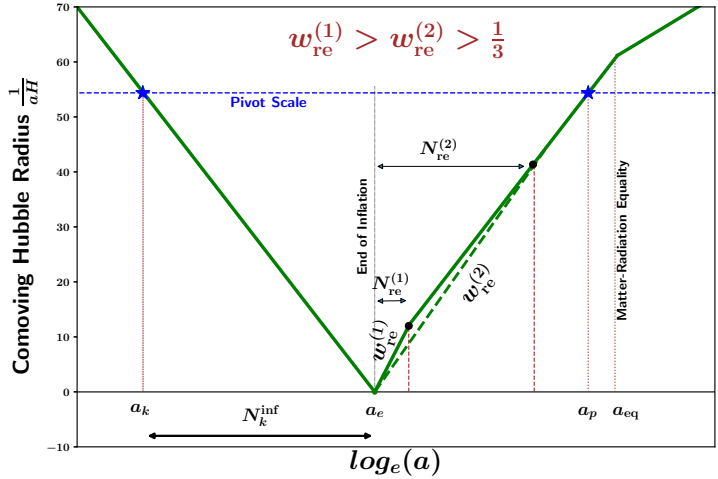}
\caption{This figure schematically illustrates the evolution of the comoving Hubble radius
$(aH)^{-1}$ with scale factor of the universe and explicitly depicts the dependence of the duration of reheating  on the reheating equation of state for a particular inflationary model with a given $N_k^{\rm inf}$. 
The {\bf top panel} shows  that for shallow reheating EOS $\wre<1/3$, the duration of reheating $\Nre$ is longer for a higher value of $\wre$, namely $\Nre^{(1)} < \Nre^{(2)}$ for
$\wre^{(1)}<\wre^{(2)}$. The {\bf bottom panel} demonstrates  that for a
 stiffer reheating EOS $\wre>1/3$, the duration of reheating $\Nre$ is shorter for a higher 
value of $\wre$, namely $\Nre^{(1)} < \Nre^{(2)}$ for
$\wre^{(1)}>\wre^{(2)}$, in accordance with equation (\ref{eq:CMB_reheat_Nre_SR}). 
Note that $(aH)^{-1} \propto a$ during the radiation dominated regime and
$(aH)^{-1} \propto a^{-1}$ during 
inflation.
For comparison see figure \ref{fig:RH_causal}.}
\label{fig:inf_causal_w_Nre}
\end{center}
\end{figure}

 Equations (\ref{eq:CMB_reheat_Nre_SR})  and (\ref{eq:CMB_reheat_Tre_SR})  capture some of the 
essential implications of reheating kinematics on CMB observables  and possess  important physical significance. For example, it is easy to see, from equation  (\ref{eq:CMB_reheat_Nre_SR}), that  
for a  softer reheating EOS with $\wre< 1/3$, a higher value of $N_k^{\rm inf}$ corresponds to a shorter reheating period $\Nre$, for a given model of inflation. 
Exactly the opposite is true for
a stiffer EOS with  $\wre > 1/3$. In this case the RHS of (\ref{eq:CMB_reheat_Nre_SR}) flips sign
so that a larger
 value of $N_k^{\rm inf}$ implies a larger $\Nre$ and hence a longer duration of reheating. Similarly 
equation (\ref{eq:CMB_reheat_Tre_SR}) implies that the longer is
 the duration of reheating $\Nre$, the lower will be the reheating temperature $\Tre$.
Moreover this result is independent of the value of $\wre$ simply because
 $1+\wre>0$ (since $\wre > -1/3$ by definition).  Another interesting aspect of equation (\ref{eq:CMB_reheat_Nre_SR}) is that,  given an inflationary potential with a fixed value of $N_k^{\rm inf}$ (which satisfies the CMB bound on $\ns\in \l[0.9607,0.9691\r]$), the duration of reheating $\Nre$ increases with an increase in the effective EOS $\wre$ as long as $\wre<1/3$. This is demonstrated in the left 
panel of figure  \ref{fig:inf_causal_w_Nre} in which $\Nre^{(1)} < \Nre^{(2)}$ for 
$\wre^{(1)}<\wre^{(2)}<1/3$.  Similarly  $\Nre$ increases with a {\em decrease}
 in the effective EOS, $\wre$, if $\wre>1/3$. This is shown
 in the right panel of figure  \ref{fig:inf_causal_w_Nre} where $\Nre^{(1)} < \Nre^{(2)}$
 for $\wre^{(1)}>\wre^{(2)}>1/3$.
These arguments also indicate that $w=1/3$ is a critical value of the EOS during reheating.

Turning our attention to $\Tre$, one notes that conservative  upper and lower bounds on this
quantity can be placed from the following considerations.
It is well known that the CMB upper bound on the tensor-to-scalar ratio, namely $r\leq 0.06$, 
translates into an upper bound on the inflationary Hubble scale $H_k^{\rm inf} \leq  6.1\times 10^{13}~{\rm GeV}$, which in turn sets an upper bound on the energy scale of inflation $T_{\inf} \leq 1.6\times 10^{16}~{\rm GeV}$, as described in appendix \ref{app:B}. Since reheating happens after the end of 
inflation, one gets  $\Tre \leq 1.6\times 10^{16}~{\rm GeV}$ as an absolute upper bound on the reheating temperature. Similarly, in order to preserve the success of the hot Big Bang phase, reheating must terminate before the beginning of Big Bang Nucleosynthesis (BBN) yielding the absolute lower bound $\Tre\geq 1~{\rm MeV}$. 
Hence the most conservative bounds on the reheating temperature are
\beq
1 \, {\rm MeV} \leq \Tre \leq 10^{16}\, {\rm GeV}~.
\label{eq:bound_Tre}
\eeq  
In accordance with the above discussion,  one can obtain interesting  reheating consistent bounds on the CMB observables for a given inflationary potential  by proceeding in the following  systematic way (also see \cite{cook15,kamion_1,kamion_2,Creminelli})
\begin{enumerate}
\item Given a slow-roll inflationary potential $V(\phi)$ with $V \propto \phi^{2p}$ as the asymptote  during reheating, the effective reheating equation of state is taken to be $\wre \equiv \langle w_\phi \rangle = \f{p-1}{p+1}$.
\item Given the strong  CMB constraint on the scalar spectral index, namely $\ns\in \l[0.9607,0.9691\r]$, we obtain a range of allowed values on $\phi_k$ from equation (\ref{eq:inf_ns_SR}) and hence on $N_k^{\rm inf}$ from equation (\ref{eq:inf_N_k}). It is important to stress that this bound on $N_k^{\rm inf}$ has been  obtained purely from CMB constraint on $\ns$, without taking into account the reheating constraints on $\Nre$ and $\Tre$ yet, which we shall  do in the next step.
\item We can then use equation (\ref{eq:CMB_reheat_Nre_SR}) to translate the bound on $N_k^{\rm inf}$ to a bound on $\Nre$ (as well as on $\Tre$ using equation (\ref{eq:CMB_reheat_Tre_SR})).  Note that $\Nre$ might turn out to be negative for some range of allowed values for $N_k^{\rm inf}$. Hence in the next step, we will discard  the corresponding range of $N_k^{\rm inf}$ that yields unphysical negative values of $\Nre$. This puts additional tighter constraint on $N_k^{\rm inf}$.
\item Next, as discussed above, imposing the condition  $\Nre>0$ as well as demanding  $\Tre \in [1 \, {\rm MeV}, 10^{16}\, {\rm GeV}]$, we obtain a tighter bound on $N_k^{\rm inf}$ and hence subsequently on $\lbrace \ns,r \rbrace$.
\item We tabulate the final allowed range of values for $N_k^{\rm inf}$, $\ns$, $r$, $\Nre$ and $\Tre$.
\end{enumerate}
The importance  of this procedure lies in the fact that it allows us to obtain reheating consistent 
constraints on the CMB observables $\lbrace \ns,r \rbrace$, which are {\em tighter}
 than those obtained in \cite{Planck1}. 
In the following subsection we apply this methodology
to determine reheating consistent 
values of $\lbrace \ns,r \rbrace$ in the T- and E- model $\alpha$-attractors and
 in the non-canonical quadratic potential.
This will help us to break the inflationary degeneracies in these models. Later in section \ref{sec:GW}, we will discuss the implications of the reheating constraints for the spectrum of relic gravitational waves background. 

Before moving on, one should clarify that the above strategy is only applicable to inflationary models 
with $\wre \neq 1/3$, which implies $p\neq 2$ in the potential $V \propto \phi^{2p}$. 
For $\wre = 1/3$, using equation (\ref{eq:CMB_reheat_Nk}), assuming $k$ to be the CMB pivot scale, i.e $k\equiv k_* = 0.05~{\rm Mpc}^{-1}$ and inserting the values of $T_0$, $g_0^s$, $g_{\rm re}$ and $g_{\rm re}^s$, as was previously done for the case $\wre\neq 1/3$,  one obtains the following strong prediction for $N_k^{\rm inf}$.

\beq
N_k^{\rm inf} =  61.55 - \log{\l(\f{V_e^{\f{1}{4}}}{H_k^{\rm inf}}\r)}~,
\label{eq:CMB_reheat_Nk_SR}
\eeq
which in turn  translates into predictions for  $\lbrace \ns,r \rbrace$.

\subsection{Reheating constraints on the T-model $\alpha$-attractor}
\label{sec:reheat_Tmodel}
\begin{figure}[t]
\centering
\subfigure[][]{
\includegraphics[width=0.4\textwidth]
{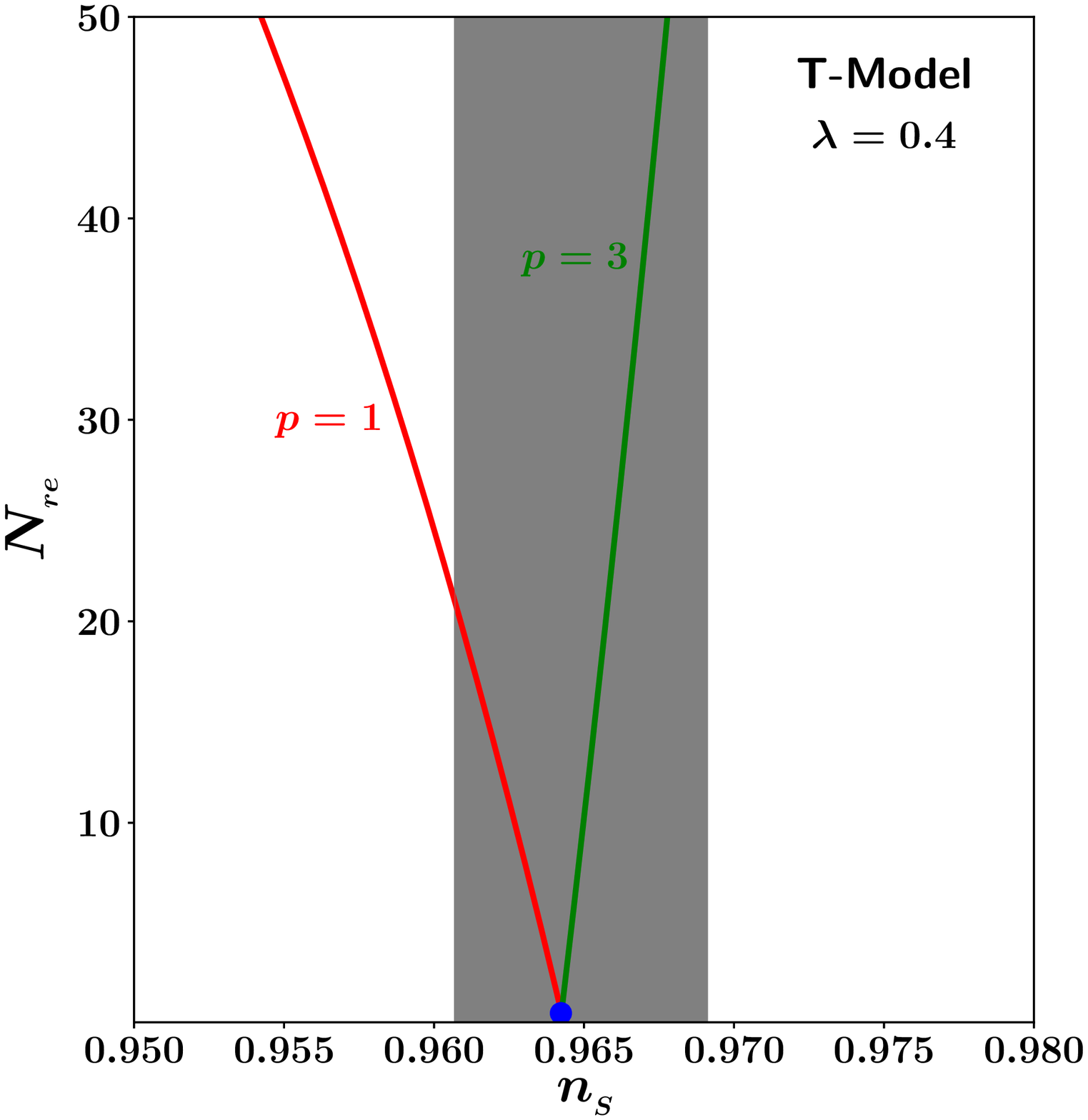}}
\subfigure[][]{
\includegraphics[width=0.4\textwidth]
{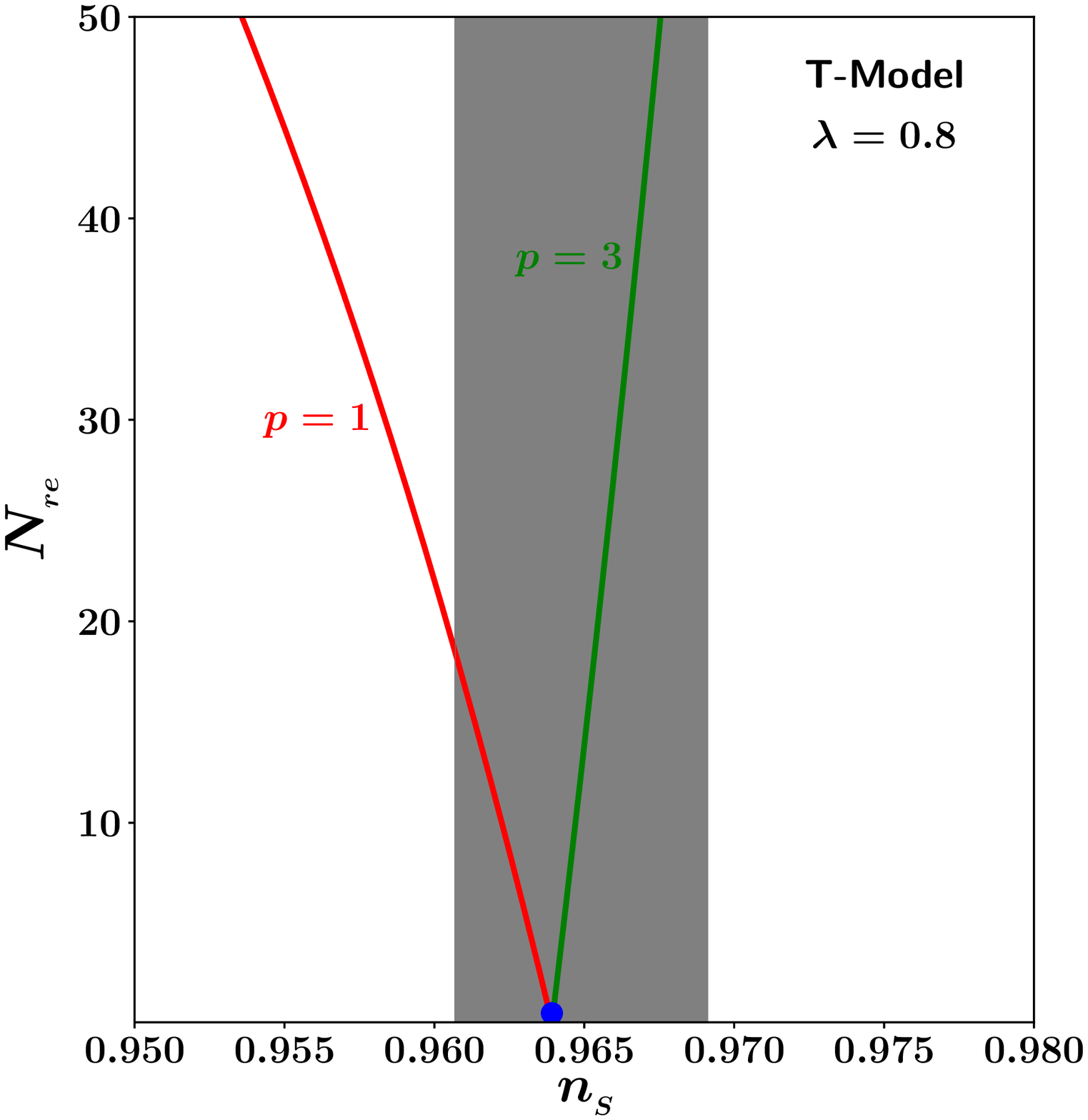}}
\subfigure[][]{
\includegraphics[width=0.4\textwidth]
{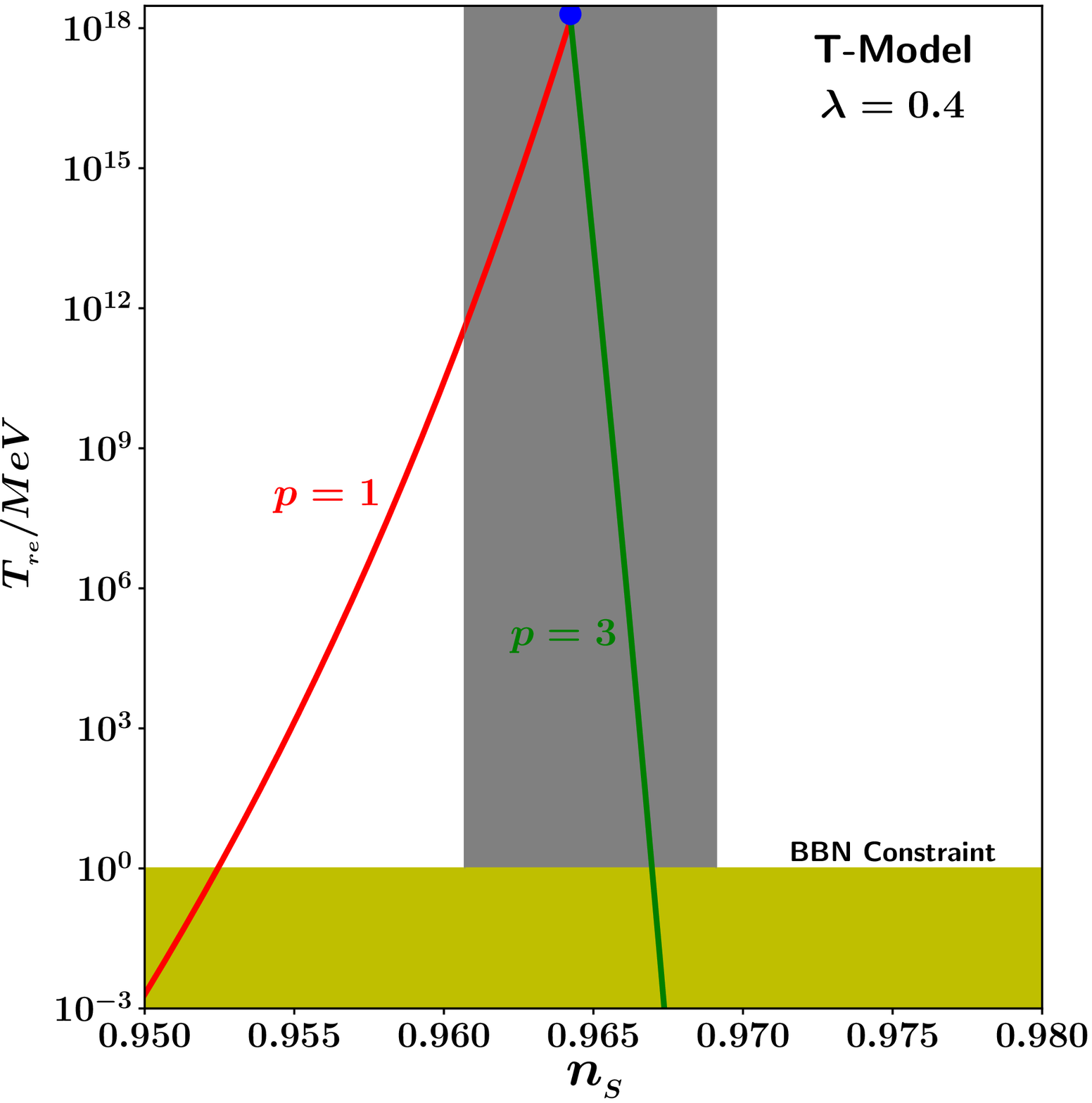}}
\subfigure[][]{
\includegraphics[width=0.4\textwidth]
{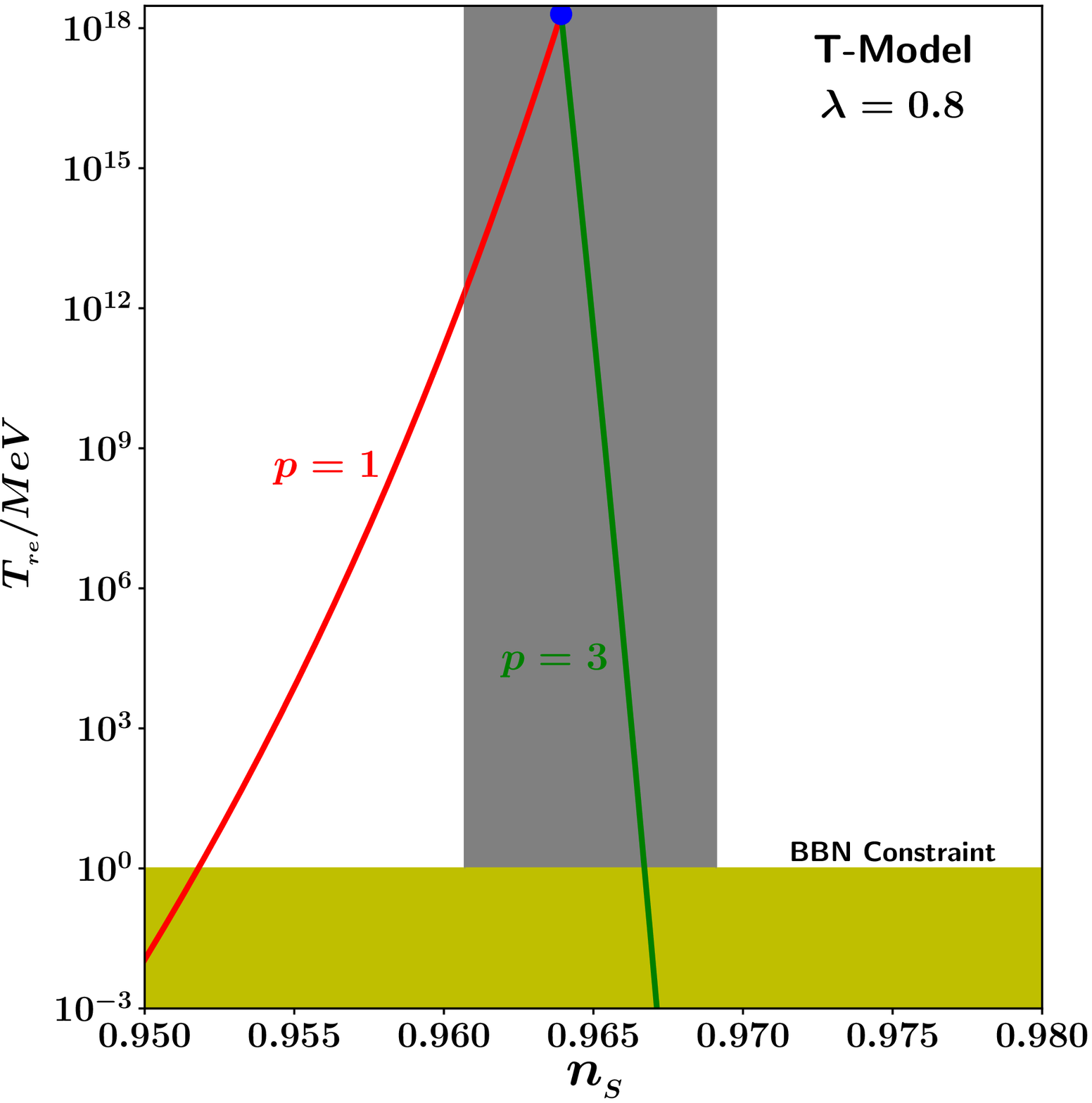}}
\caption{Constraints on the reheating parameters $\Nre$ ({\bf top row}) and $\Tre$ ({\bf bottom row}), given by equations (\ref{eq:CMB_reheat_Nre_SR}) and (\ref{eq:CMB_reheat_Tre_SR}) respectively, have been illustrated for the T-model $\alpha$-attractor (\ref{eq:pot_Tmodel}) for two different value of the inflationary parameter $\lambda$, namely  $\lambda = 0.4$ ({\bf left panel}) and  $\lambda=0.8$ ({\bf right panel}). The red curve in each diagram corresponds to $p=1$ and hence $\wre = 0$ in (\ref{eq:EOS}), while the green curve corresponds to $p=3$ and hence $\wre = 1/2$. The blue dot in each diagram, which corresponds to the case $p=2$ with $\wre = 1/3$, yields definitive  predictions for $\lbrace \ns,r \rbrace$ as can be inferred from (\ref{eq:CMB_reheat_Nk_SR}). We notice that the reheating temperature is typically higher for  $p=1$, notably $\Tre\geq 10^8$ GeV. 
The horizontal band in all figures corresponds to the Planck bound
$n_{_S} = 0.9649 \pm 0.0042$. Values of $\Nre$ and $\Tre$ lying outside of this bound are
disfavoured by CMB observations \cite{Planck1}.
We therefore conclude that different values of $p$ in the inflationary T-model potential
(\ref{eq:pot_Tmodel})
result in different relations for $\Nre(\ns)$ and $\Tre(\ns)$. This effectively breaks the CMB
degeneracy illustrated in figure \ref{fig:deg_Tmodel}.
}
\label{fig:T_model_deg_break}
\end{figure}

After the end of inflation in the T-model, 
the scalar field oscillates around the minimum of the T-model potential (\ref{eq:pot_Tmodel}) 
which acquires the form 
$$V\l(|\lambda\phi| \ll \mpl\r)\simeq V_0 \l(\lambda\f{\phi}{m_p}\r)^{2p}\;;~~~p=1,2,3...., $$
with the mean EOS during oscillations  given by (\ref{eq:EOS}). Taking the effective reheating EOS to be $\wre=\langle w_\phi \rangle = \frac{p-1}{p+1}$, in the context of perturbative reheating, the CMB bound  on $\ns$  places constraints on the reheating parameters $\Nre$ and $\Tre$. Additionally demanding $\Nre>0$ and $\Tre \in [1 \, {\rm MeV}, 10^{16}\, {\rm GeV}]$ allows us to
 obtain reheating-consistent constraints  on the CMB observables $\lbrace \ns,r \rbrace$, as discussed above. Our results are  illustrated  in figure  \ref{fig:T_model_deg_break} and tabulated in table \ref{table:1}.  

The red curve in each diagram of figure  \ref{fig:T_model_deg_break} corresponds to $p=1$ and hence $\wre = 0$ in (\ref{eq:EOS}), while the green curve corresponds to $p=3$ and hence $\wre = 1/2$. The blue dot in each diagram, which corresponds to the case $p=2$ with $\wre = 1/3$, yields definitive  predictions for $\lbrace \ns,r \rbrace$ which can be inferred from  (\ref{eq:CMB_reheat_Nk_SR}). 
Notice that $\Nre$ and $\Tre$ occupy different regions of space in the diagram for different values of $p$. Also note that the reheating temperature is higher for $p=1$ ($\Tre\geq 10^8$ GeV)
than for $p=3$ ($\Tre\geq 1$ MeV).  
As shown in table \ref{table:1}, reheating constraints segregate  the number of
inflationary $e$-foldings $N_k^{\rm inf}$, and hence the CMB observables $\lbrace \ns,r \rbrace$,
 into different ranges of parameter space for different values of $p$. This 
facilitates  the breaking of degeneracies associated with the T-model $\alpha$-attractor potential (\ref{eq:pot_Tmodel}), which was illustrated in figure  \ref{fig:deg_Tmodel}. Note that the 
tabulated constraints on the CMB  observables have been obtained by  taking into  account
 the conservative reheating constraints  $\Nre>0$ and $\Tre \in [1 \, {\rm MeV}, 10^{16}\, {\rm GeV}]$.

\begin{table}[htb]
\begin{center}
 \begin{tabular}{|c|c|c|c|c|c|c|}
 \hline
 \Tstrut
 $\lambda$ & Observables  & $p=1$, $\wre=0$ &  $p=2$, $\wre=1/3$ & $p=3$, $\wre=1/2$ \\ [1ex]
 \hline\hline \Tstrut
 \multirow{5}{*}{0.4}& $N_k^{\rm inf}$ & [50.4931, 55.686] & 55.72 & [55.724, 60.383]\\
  					 & $\ns$ & [0.9607, 0.9643] & 0.9643 & [0.96435, 0.967]\\
                      & $r$ & [0.003867, 0.004683] & 0.003899 & [0.003326, 0.003896]\\
  					 & $\Nre$ & [0, 20.9624] & 0 & [0, 37.5836]\\
& $\Tre$ & [$3.9\times 10^{8}$, $2.4\times 10^{15}$] GeV & $2.4\times 10^{15}$ GeV  & [$10^{-3}$, $2.4\times 10^{15}$] GeV\\[1ex]
 \hline  \Tstrut
 \multirow{5}{*}{0.8}	& $N_k^{\rm inf}$ & [50.6026, 55.17] & 55.1749  & [55.1761, 59.8198]\\
  					    & $\ns$ & [0.9607, 0.9639] & 0.9639 & [0.9639, 0.966706]\\
                       	& $r$ & [0.001009, 0.001198] & 0.00101 & [0.0008604, 0.00101]\\
  						& $\Nre$ & [0, 18.4401] & 0 & [0, 37.4696]\\
 						& $\Tre$ & [$2.2\times 10^{9}$, $2.1\times 10^{15}$] GeV & $2.1\times 10^{15}$ GeV & [$10^{-3}$, $2.1\times 10^{15}$] GeV\\[1ex]
 \hline
\end{tabular}
\captionsetup{
	justification=raggedright,
	singlelinecheck=false
}
\caption{This table demonstrates that by taking into account the reheating constraints $\Nre>0$ and $\Tre \in [1 \, {\rm MeV}, 10^{16}\, {\rm GeV}]$, the number of inflationary
 $e$-foldings $N_k^{\rm inf}$ and hence also the associated  CMB observables $\lbrace \ns,r \rbrace$, 
get segregated into different ranges of parameter space for different values of $p$. 
This allows one to break the degeneracies associated with the T-model $\alpha$-attractor potential 
(\ref{eq:pot_Tmodel}) illustrated in figure  \ref{fig:deg_Tmodel}.}
\label{table:1}
\end{center} 
\end{table}

\subsection{Reheating constraints on the E-model $\alpha$-attractor}
\label{sec:reheat_Emodel}
After inflation ends in the E-model, the inflaton begins to oscillate around the minimum of the 
E-model potential (\ref{eq:pot_Emodel})  which takes the form 
$$V(\phi)\simeq V_0 \l(\lambda\f{\phi}{m_p}\r)^{2p}\;;~~~p=1,2,3...., $$
\begin{figure}[t]
\centering
\subfigure[][]{
\includegraphics[width=0.4\textwidth]
{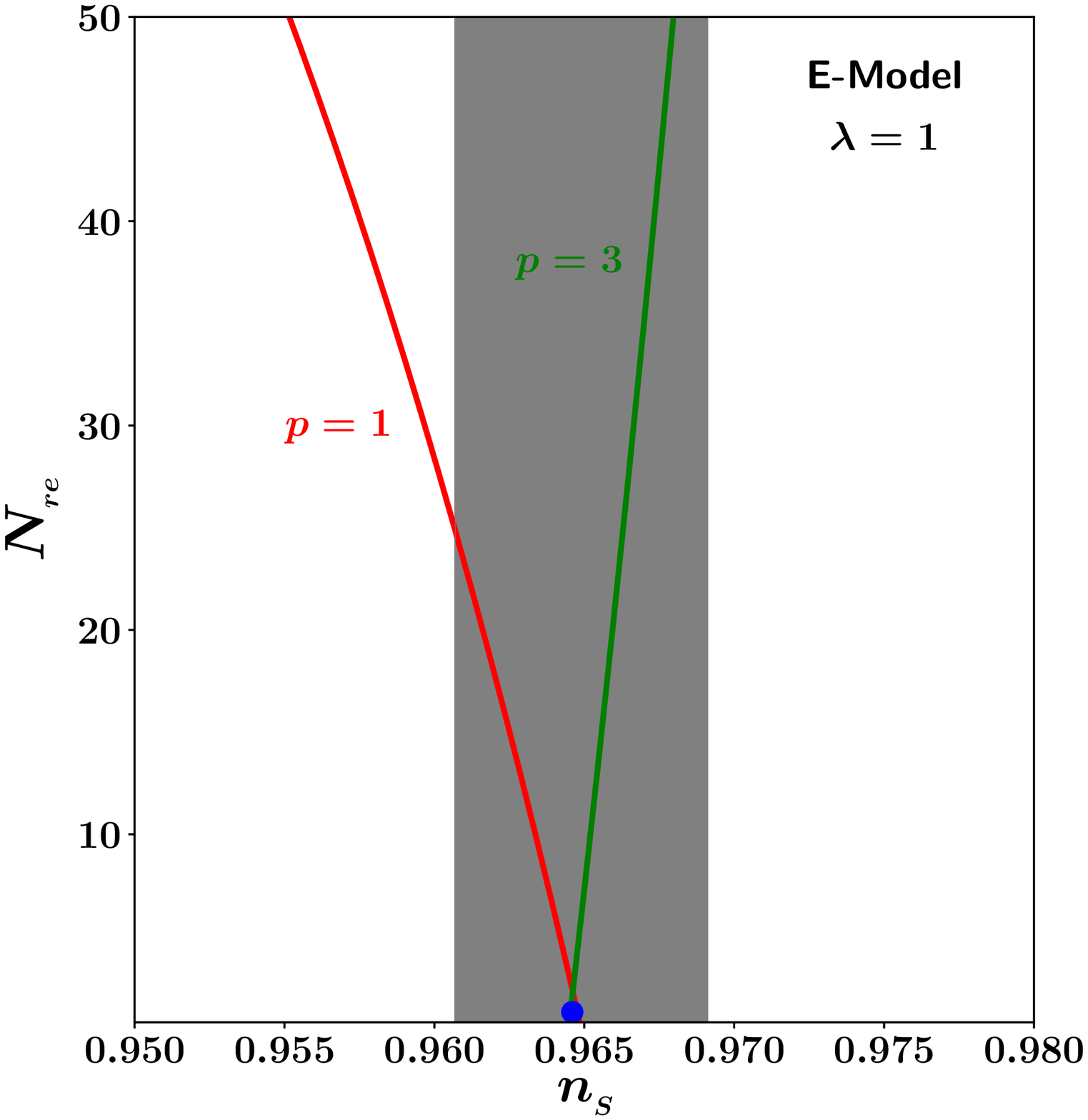}}
\subfigure[][]{
\includegraphics[width=0.4\textwidth]
{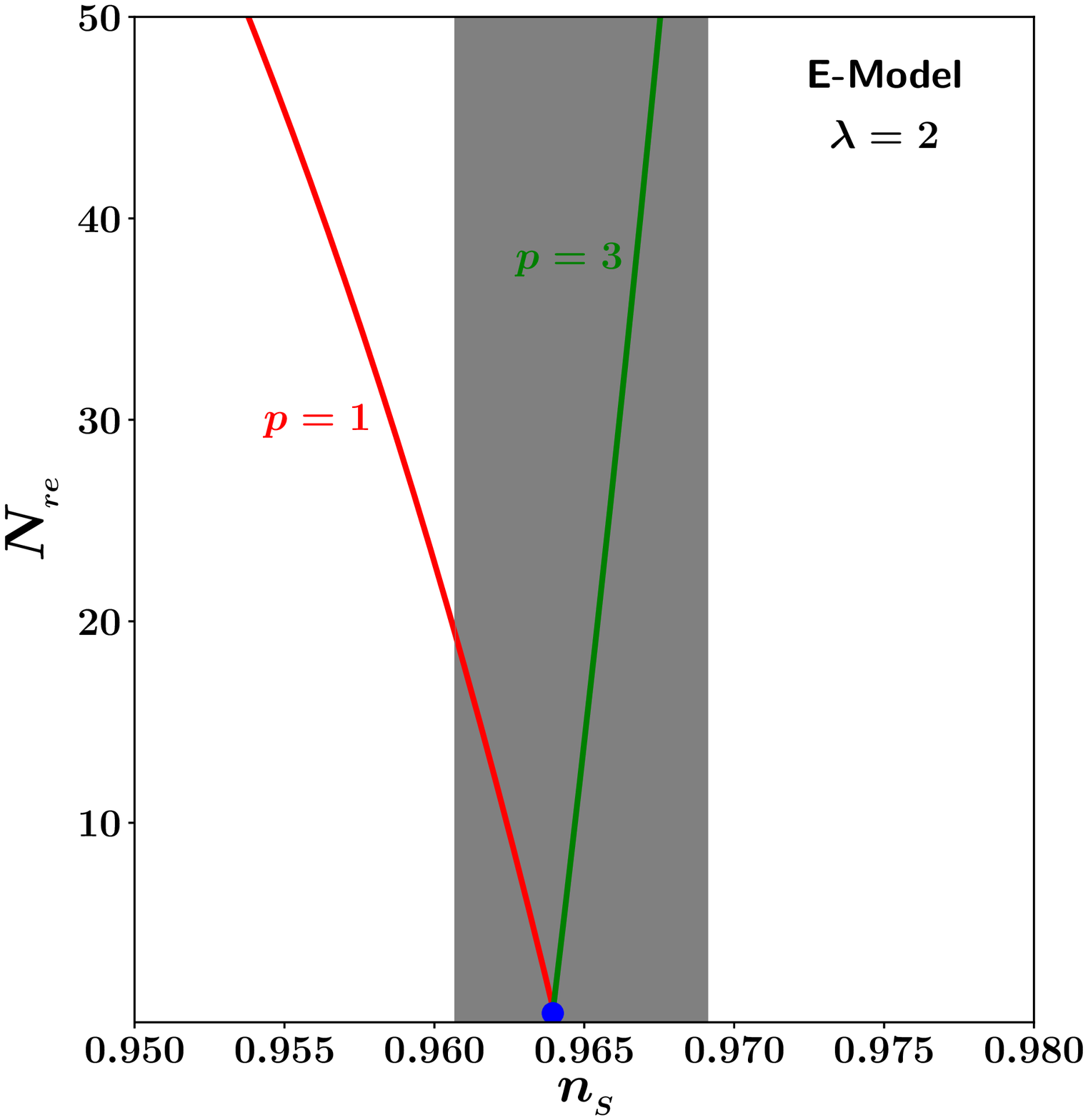}}
\subfigure[][]{
\includegraphics[width=0.4\textwidth]
{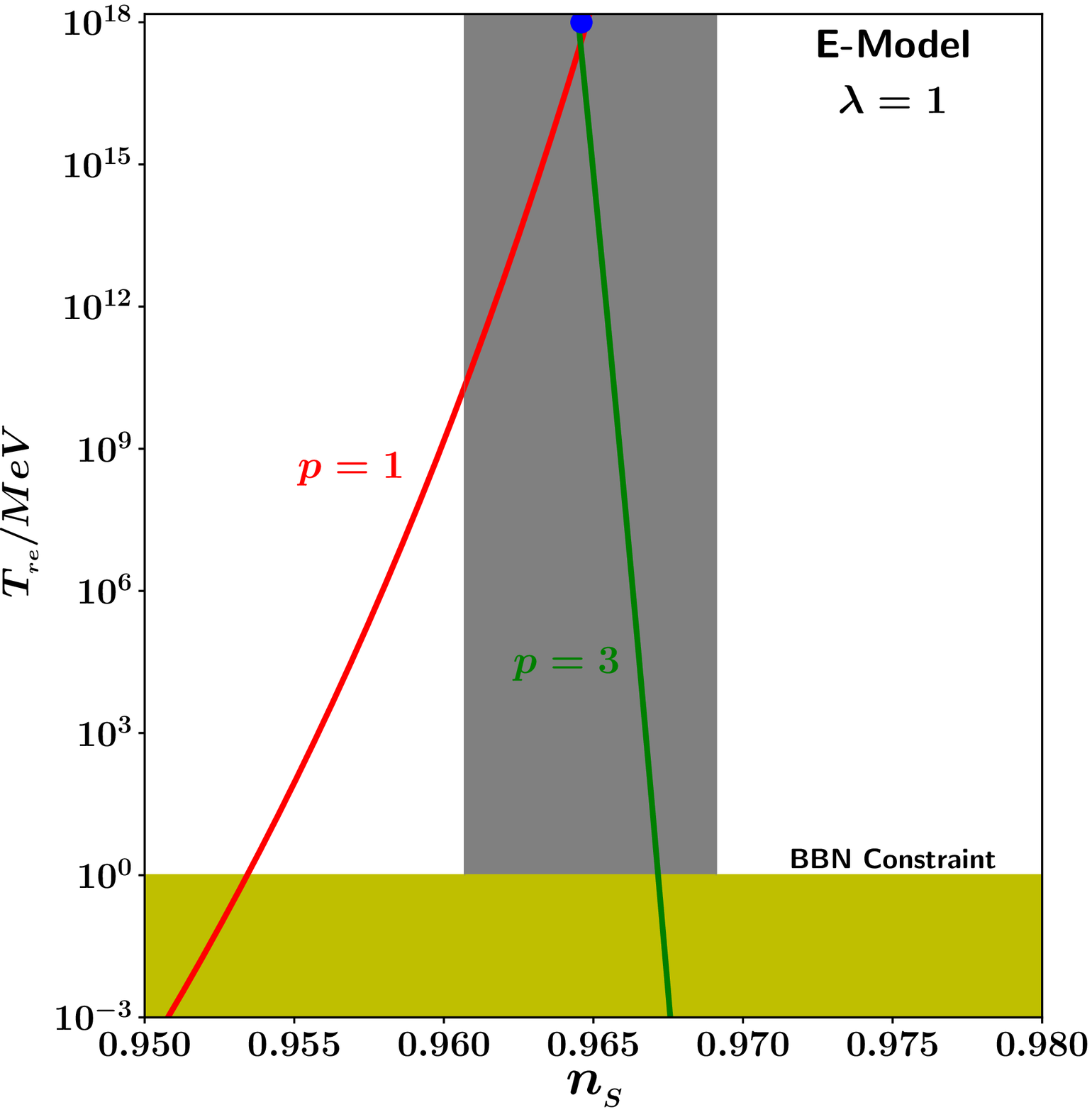}}
\subfigure[][]{
\includegraphics[width=0.4\textwidth]
{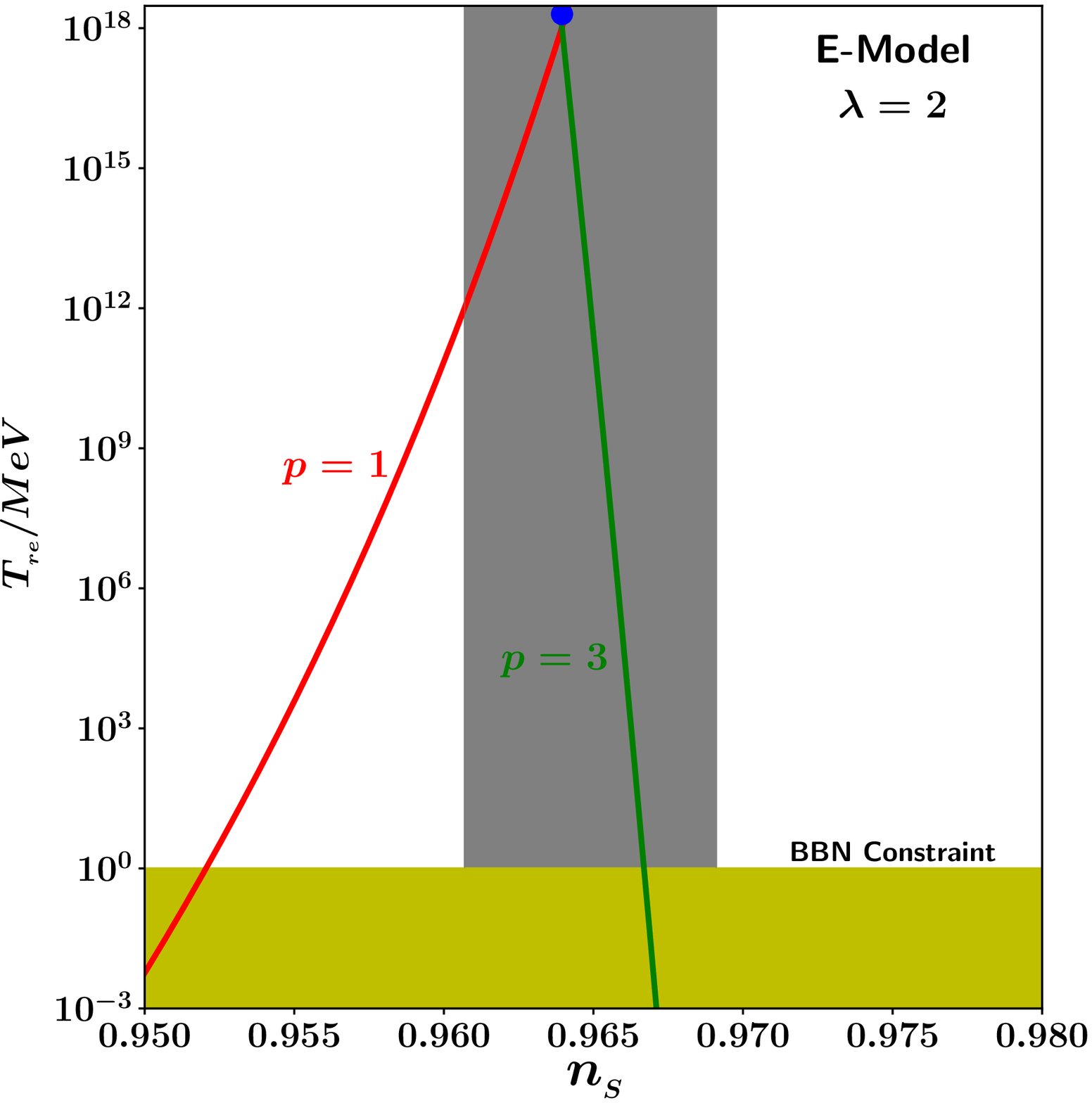}}
\caption{Constraints on the reheating parameters $\Nre$ ({\bf top row}) and $\Tre$ ({\bf bottom row}), given by equations (\ref{eq:CMB_reheat_Nre_SR}) and (\ref{eq:CMB_reheat_Tre_SR}) respectively, have been illustrated for the E-model $\alpha$-attractor (\ref{eq:pot_Emodel}) for two different value of the inflationary parameter $\lambda$, namely  $\lambda = 0.4$ ({\bf left panel}) and  $\lambda=0.8$ ({\bf right panel}). The red curve in each diagram corresponds to $p=1$ and hence $\wre = 0$ in (\ref{eq:EOS}), while the green curve corresponds to $p=3$ and hence $\wre = 1/2$. The blue dot in each diagram, which corresponds to the case $p=2$ with $\wre = 1/3$, yields definitive  predictions for $\lbrace \ns,r \rbrace$ as can be inferred from (\ref{eq:CMB_reheat_Nk_SR}). We notice that the reheating temperature is typically higher for  $p=1$, notably $\Tre\geq 10^7$ GeV. 
The horizontal band in all figures corresponds to the Planck bound
$n_{_S} = 0.9649 \pm 0.0042$. Values of $\Nre$ and $\Tre$ lying outside of this bound are
disfavoured by CMB observations \cite{Planck1}.
We therefore conclude that different values of $p$ in the inflationary E-model potential
(\ref{eq:pot_Emodel})
result in different relations for $\Nre(\ns)$ and $\Tre(\ns)$. This effectively breaks the CMB
degeneracy illustrated in figure \ref{fig:deg_Emodel}.
} \label{fig:E_model_deg_break}
\end{figure}

Assuming the effective reheating EOS to be $\wre=\langle w_\phi \rangle = \frac{p-1}{p+1}$, 
similar conclusions are drawn for the E-model as were obtained earlier for the T-model. Our results, described in figure   \ref{fig:E_model_deg_break}, and tabulated in table \ref{table:2}, indicate that the existing degeneracies in the E-model, illustrated in figure \ref{fig:deg_Emodel}, are
 easily broken by taking into consideration the  kinematics of reheating. 

\begin{table}[htb]
\begin{center}
 \begin{tabular}{|c|c|c|c|c|c|c|}
 \hline
 \Tstrut
 $\lambda$ & Observables  & $p=1$, $\wre=0$ &  $p=2$, $\wre=1/3$ & $p=3$, $\wre=1/2$ \\ [1ex]
 \hline\hline \Tstrut
 \multirow{5}{*}{1}& $N_k^{\rm inf}$ & [49.2795, 55.4335] & 55.6419 & [55.4944, 60.1534]\\
  					 & $\ns$ & [0.9607, 0.965] & 0.9646 & [0.9644, 0.9672]\\
                      & $r$ & [0.002372, 0.002873] & 0.002513 & [0.00211, 0.002471]\\
  					 & $\Nre$ & [0, 24.8427] & 0 & [0, 37.5851]\\
& $\Tre$ & [$2.1\times 10^{7}$, $2.5\times 10^{15}$] GeV & $2.4\times 10^{15}$ GeV  & [$10^{-3}$, $2.4\times 10^{15}$] GeV\\[1ex]
 \hline  \Tstrut
 \multirow{5}{*}{2}	& $N_k^{\rm inf}$ & [50.2202, 54.9935] & 55.0077  & [55.0125, 59.6497]\\
  					    & $\ns$ & [0.9607, 0.9641] & 0.964 & [0.9639, 0.9667]\\
                       	& $r$ & [0.000639, 0.000764] & 0.000645 & [0.000552, 0.000648]\\
  						& $\Nre$ & [0, 19.3462] & 0 & [0, 37.413]\\
 						& $\Tre$ & [$1.1\times 10^{9}$, $2\times 10^{15}$] GeV & $2\times 10^{15}$ GeV & [$10^{-3}$, $2\times 10^{15}$] GeV\\[1ex]
 \hline
\end{tabular}
\captionsetup{
	justification=raggedright,
	singlelinecheck=false
}
\caption{This table demonstrates that by taking into account the reheating constraints $\Nre>0$ and $\Tre \in [1 \, {\rm MeV}, 10^{16}\, {\rm GeV}]$, the number of inflationary
 $e$-foldings $N_k^{\rm inf}$ and hence also the associated  CMB observables $\lbrace \ns,r \rbrace$,
get segregated into different ranges of parameter space for different values of $p$.
This allows one to break the degeneracies associated with 
the E-model $\alpha$-attractor potential (\ref{eq:pot_Emodel}) which were
 illustrated in figure  \ref{fig:deg_Emodel}.}
\label{table:2}
\end{center} 
\end{table}

\subsection{Reheating constraints on non-canonical inflation}
\label{sec:reheat_nc}

\begin{figure}[t]
\begin{center}
\includegraphics[width=0.495\textwidth]{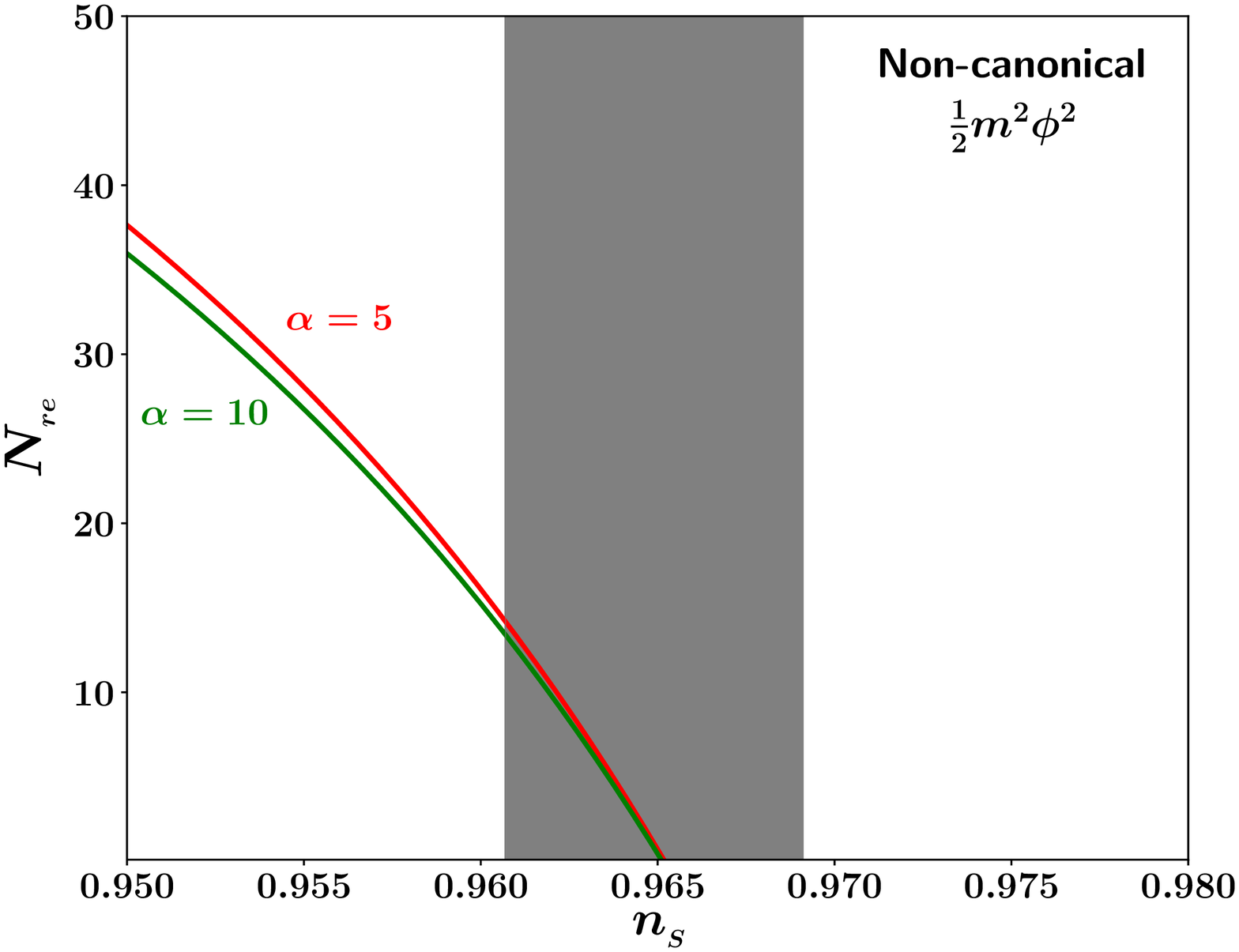}
\includegraphics[width=0.495\textwidth]{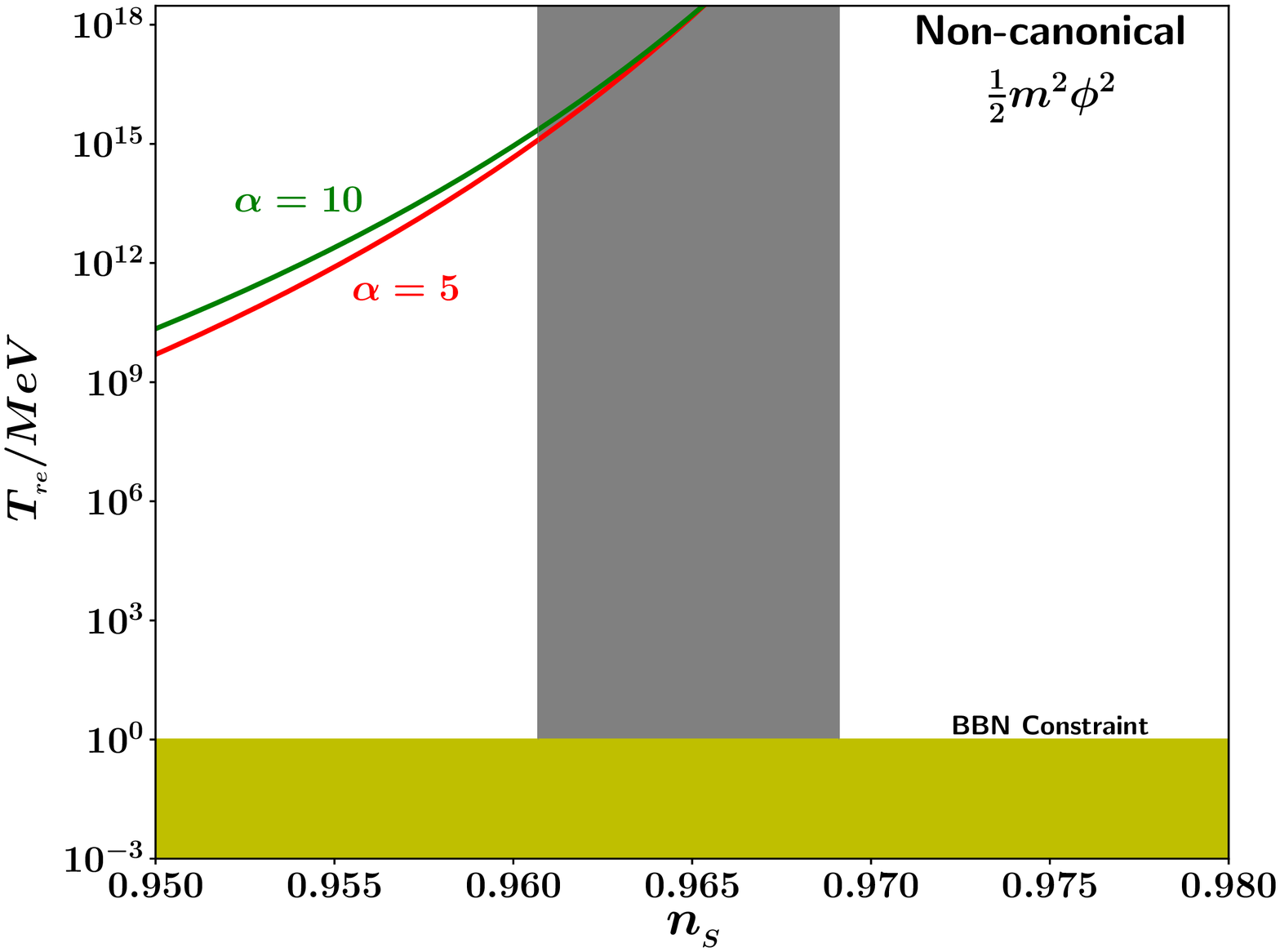}
\caption{Constraints on the duration of reheating $\Nre$ and the reheating temperature $\Tre$ are shown 
for the non-canonical quadratic potential discussed in section \ref{sec:non_can_inf}. One finds
 that the inflationary degeneracies shown in figure  \ref{fig:ns_r_nc} are not quite cured in this case, due
 to the fact that  $\l<w^{^{\rm NC}}_\phi\r> < 0$. Note that the reheating temperature is quite
 high, namely $\Tre \geq 10^{12}~{\rm GeV}$.
The horizontal band in both figures corresponds to the Planck bound
$n_{_S} = 0.9649 \pm 0.0042$.}
\label{fig:deg_nc}
\end{center}
\end{figure}

Next we apply the techniques developed in section \ref{sec:reheat_pert} to inflation in the 
non-canonical scenario discussed in section \ref{sec:non_can_inf}.
The unusual oscillatory EOS in the quadratic potential (\ref{eq:EOS_nc_quad}) namely
, $-1/3 < \l<w^{^{\rm NC}}_\phi\r> < 0$,
does not permit one to break the degeneracies which exist in this model,
and which arise because the scalar spectral index $\ns$ does not depend upon the non-canonical
parameter $\alpha$; see eqn. (\ref{eq:ns_nc}) and  figure  \ref{fig:ns_r_nc}.
However reheating considerations do
succeed in placing strong constraints on the reheating temperature in this model which is confined
to fairly high values $\Tre\geq 10^{12}~{\rm GeV}$, as shown in figure \ref{fig:deg_nc}.


Before moving forward, we would like to mention that our analysis in this work is based on perturbative reheating and has been   carried out in the framework similar to that developed in \cite{kamion_1,Creminelli,kamion_2,cook15}. However for  obtaining  reheating constraints in  a more general scenario, when $\wre$ is a priori unknown, a more suitable reheating  parameter (denoted as $R_{\rm reh}$) was introduced in \cite{Martin:2010kz}  as  a particular combination of the  reheating duration  $\Nre$ and  the average reheating EOS $\wre$ (also see \cite{Martin:2014nya}). For a given inflationary potential, CMB measurements constrain $R_{\rm reh}$ directly.

\section{\bf Relic gravitational waves from Inflation}
\label{sec:GW}

Relic gravitational waves  are a
generic prediction of the inflationary scenario \cite{star79}.
These  tensor fluctuations, which are created quantum mechanically,  get stretched to super-Hubble scales during inflation, where they remain frozen until their subsequent Hubble re-entry at late times after 
inflation ends. After becoming sub-Hubble, inflationary tensor fluctuations behave like a
stochastic relic gravitational waves background in the universe. The amplitude of relic GWs 
is sensitive to the value of the Hubble parameter
during inflation, while the GW spectrum encodes both the inflationary and post-inflationary
EOS of the universe \cite{star79,allen88,sahni90,sami2002,dany_18,dany_19,Bernal:2019lpc}. Since GWs interact minimally with other forms of matter/radiation
they constitute one of the cleanest probes of the physics of the very early universe.

In a homogeneous and isotropic universe
 gravitational waves satisfy the minimally coupled Klein-Gordon equation
$\sq h_{ij} = 0$
where $h_{ij} = \phi_k(\tau) e^{-i{\bf k}.{\bf x}}e_{ij}$ and $e_{ij}$ is  the
polarization tensor with $\tau$ being the conformal time defined by $\tau = \int dt/a(t)$.
As a result, each of the two polarization states of the graviton $h_{\times,+}(k) = 
\f{\phi_k(\tau)}{\mpl} e^{-i{\bf k}.{\bf x}}$,
satisfies the equation
\beq
\phi_k^{''} + 2\frac{a'}{a}\phi_k^{'} + k^2\phi_k = 0
\label{eq:GW_KG}
\eeq
where the derivative is with respect to  the conformal time $\tau$.
For near-exponential inflation,
$a = \tau_0/\tau$ $(\vert\tau\vert < \vert\tau_0\vert)$.

Eqn. (\ref{eq:GW_KG}) implies that the amplitude of a tensor Fourier mode freezes
to a constant value in the super-Hubble limit.
The corresponding dimensionless amplitude of a tensor  mode is related to
the inflationary Hubble parameter ${H_k^{\rm inf}}$ at Hubble exit by
\beq
P_{\rm GW}(k) \equiv h^2_{\times,+}(k) \simeq \frac{1}{2\pi^2}\left (\frac{H_k^{\rm inf}}{\mpl}\right )^2\bigg\vert_{k=aH}~.
\label{eq:GW_Ph}
\eeq
The inflationary tensor fluctuations are often defined in terms of a different normalization of the Fourier mode tensor amplitude $\phi_k$, which yields the following
 tensor power spectrum  (see appendix \ref{app:B} and \cite{star79,baumann07,dany_19,Ema_2020})
\beq
P_{T} = \f{2}{\pi^2}\l(\f{H_k^{\rm inf}}{m_p}\r)^2 = 4\times P_{\rm GW}(k)~,
\label{eq:GW_Pt}
\eeq
and the tensor-to-scalar ratio is defined, in terms of $P_T$, to be 
\beq
r = \f{P_T}{P_{\cal R}}~.
\label{eq:GW_r}
\eeq
The power spectrum $P_{\rm GW}(k)$ can be written as
\beq
P_{\rm GW}(k) = P_{\rm GW}(k_*)\left (\frac{k}{k_*}\right )^{\nt}~,
\label{eq:GW_Ph1}
\eeq
where the tensor power  at the CMB pivot scale $k_* = 0.05 ~{\rm Mpc}^{-1}$ is  given, in terms of the scalar power (see appendix \ref{app:B}),  by 
\beq
A_{_{\rm GW}} \equiv P_{\rm GW}(k_*) = \f{1}{4}\, r \, A_{_S} =r\times 5.25 \times 10^{-10}~,
\label{eq:GW_Ah}
\eeq
and the tensor tilt is found to be 
\beq
 n_T = \frac{d\log{P_{\rm GW}(k)}}{d\log{k}} = -\f{r}{8}~,
\label{eq:GW_nt}
\eeq
which satisfies the consistency relation.

The  quantum mechanically generated tensor modes discussed above, which become super-Hubble  during inflation, make their Hubble re-entry at late times when $k=aH$ and behave like stochastic GWs in the universe \cite{dany_18,dany_19}.  The physical frequency of these stochastic GWs at the present epoch is given by

\beq
f = \f{1}{2\pi}\l(\f{k}{a_0}\r)=\f{1}{2\pi}\l(\f{a}{a_0}\r)H~,
\label{eq:GW_f}
\eeq 
where $a$, $H$ correspond to the scale factor and Hubble parameter of the universe during the epoch when the corresponding    tensor mode makes its Hubble re-entry. In this work, we focus on the relic GWs that become sub-Hubble prior to the matter-radiation equality, so that their characteristic frequency is large enough to enable them to be detected by the GW observatories in the near future. Expressing $H$ in terms of  temperature, we get 
\beq
\f{H}{\mpl} = \l(\f{\rho}{3\mpl^4}\r)^{\f{1}{2}} = \l(\f{\f{\pi^2}{30}\,g_T\,T^4}{3\mpl^4}\r)^{\f{1}{2}} = \pi \, \l(\f{g_T}{90}\r)^{\f{1}{2}} \, \l(\f{T}{\mpl}\r)^2~.
\label{eq:GW_H_T}
\eeq
Using entropy conservation (discussed in appendix \ref{app:A}), we obtain
\beq
\f{a}{a_0} = \l(\f{a_{\rm eq}}{a_0}\r)\l(\f{g_{\rm eq}^s}{g_T^s}\r)^{1/3}\l(\f{T_{\rm eq}}{T}\r)~.
\label{eq:GW_a_T}
\eeq

Substituting $H$ from equation (\ref{eq:GW_H_T}) and  $a/a_0$ from equation (\ref{eq:GW_a_T}) in (\ref{eq:GW_f}), we obtain the following important  expression for the present day frequency of GWs in terms of their Hubble re-entry temperature.
\beq
f = 7.36\times 10^{-8}\, {\rm Hz}\,\l(\f{g_0^s}{g_T^s}\r)^{\f{1}{3}}\,\l(\f{g_T}{90}\r)^{\f{1}{2}}\,\l(\f{T}{{\rm GeV}}\r) ~.
\label{eq:GW_f_master}
\eeq

\begin{figure}[t]
\begin{center}
\includegraphics[width=0.75\textwidth]{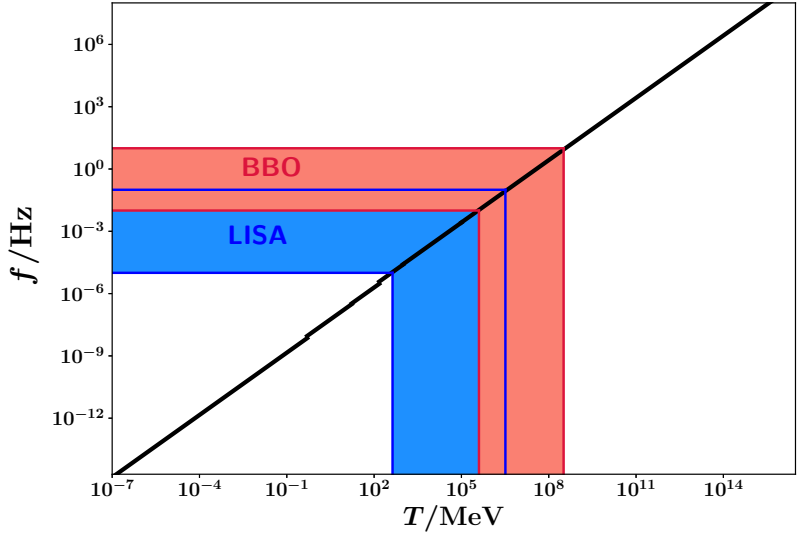}
\caption{The characteristic present day frequency of stochastic relic GWs is plotted as a function of the temperature of the universe at which the corresponding tensor modes became sub-Hubble. This figure also illustrates the frequency bands of future GWs detectors such as  LISA and BBO.}
\label{fig:GW_f_T}
\end{center}
\end{figure}

In figure  \ref{fig:GW_f_T}, we illustrate  the present day frequency of relic GWs as a function of their Hubble re-entry temperature along with the sensitivity bands of  future GWs observatories LISA and BBO \cite{LISA,BBO}. The values of $f$ corresponding to relic  GWs that became sub-Hubble at a number  of  important cosmic epochs are tabulated in table \ref{table:3}.

\begin{table}[h]
\begin{center}
 \begin{tabular}{||c|c|c|c|c||} 
 \hline\Tstrut
 \bf{Epoch} & \bf{Temperature} $T$ & \bf{GW Present day  {\large f} (in Hz)}\\ [1ex] 
 \hline\hline\Tstrut
 
 Matter-radiation equality & $\sim 1$~eV  & $1.7\times 10^{-17}$ \\ [1.2ex] 
 \hline\Tstrut
 CMB pivot scale re-entry & $\sim 5$~eV  & $8.5\times 10^{-17}$ \\ [1.2ex]
 \hline\Tstrut
 Big Bang Nucleosynthesis & $\sim 1$~MeV  & $1.8\times 10^{-11}$ \\ [1.2ex]
 \hline\Tstrut
 Electro-weak symmetry breaking & $\sim 100$~GeV  & $2.7\times 10^{-6}$ \\ [1.2ex]
 \hline
\end{tabular}
\captionsetup{
	justification=raggedright,
	singlelinecheck=false
}
\caption{Present day frequencies of relic GWs have been tabulated  for four different temperature  scales, associated with the Hubble re-entry of the respective primordial tensor modes. In order to probe
 the epoch of reheating using relic GWs, the physical frequency corresponding to tensor modes which 
become sub-Hubble during reheating must satisfy  $f>f_{_{\rm BBN}} \simeq 10^{-11}$ Hz in order to obey the BBN bound. }
\label{table:3}
\end{center}
\end{table}

The present day spectral density of stochastic GWs, defined in terms of the  critical density at the present epoch $\rho_{_{0c}}$  by \cite{dany_18,dany_19}

\beq
\Og (f) \equiv \f{1}{\rho_{_{0c}}}\f{d\rho_{\rm GW}^0(f)}{d\log{f}}~,
\label{GW_Omega_def}
\eeq
is given by the following set of equations \cite{sahni90,sami2002,dany_19}

\ber
\Og^{(\rm MD)}(f) &=& \frac{3}{32\pi^2}P_{\rm GW}(f)
\,\Omega_{0m}\left (\frac{f}{f_{h}}\right )^{-2},
~ f_h < f \leq f_{\rm eq}\label{eq:GW_spectrum_1a}\\
\Og^{(\rm RD)}(f) &=& \frac{1}{6}P_{\rm GW}(f)\,\Omega_{0r},
~ f_{\rm eq} < f \leq f_{\rm re}\label{eq:GW_spectrum_1b}\\
\Og^{(\rm re)} (f) &=& \Og^{(\rm RD)}
\left (\frac{f}{f_{\rm re}}\right )^{2\left (\frac{w-1/3}{w+1/3}\right )},
~ f_{\rm re} < f \leq f_{\rm e}
\label{eq:GW_spectrum_1c}
\eer

where $f_h$, $f_{\rm eq}$, $f_{\rm re}$, $f_e$ refer  to the present day frequency of relic GWs corresponding to tensor modes that became sub-Hubble at: the present epoch ($f_h$), 
the epoch of matter-radiation equality ($f_{\rm eq}$), at the end of reheating 
(commencement  of the radiation dominated epoch, $f_{\rm re}$) and at the end of inflation ($f_e$). 
The superscripts `MD', 'RD', `re' in $\Og$ refer to matter dominated epoch, radiation epoch and the epoch of reheating respectively. For $\Omega_{0r}=1$, Eq. (\ref{eq:GW_spectrum_1b}) just coincides with the result of the paper~\cite{star79}. 
Note that $f_{\rm re}>f_{\rm BBN} \simeq 10^{-11}~{\rm Hz}$ in order to 
satisfy the BBN bound on $\Tre$. Regarding the EOS $w = \wre$ during the epoch of reheating, it is important to keep in mind the following points.

\begin{figure}[t]
\begin{center}
\includegraphics[width=0.59\textwidth]{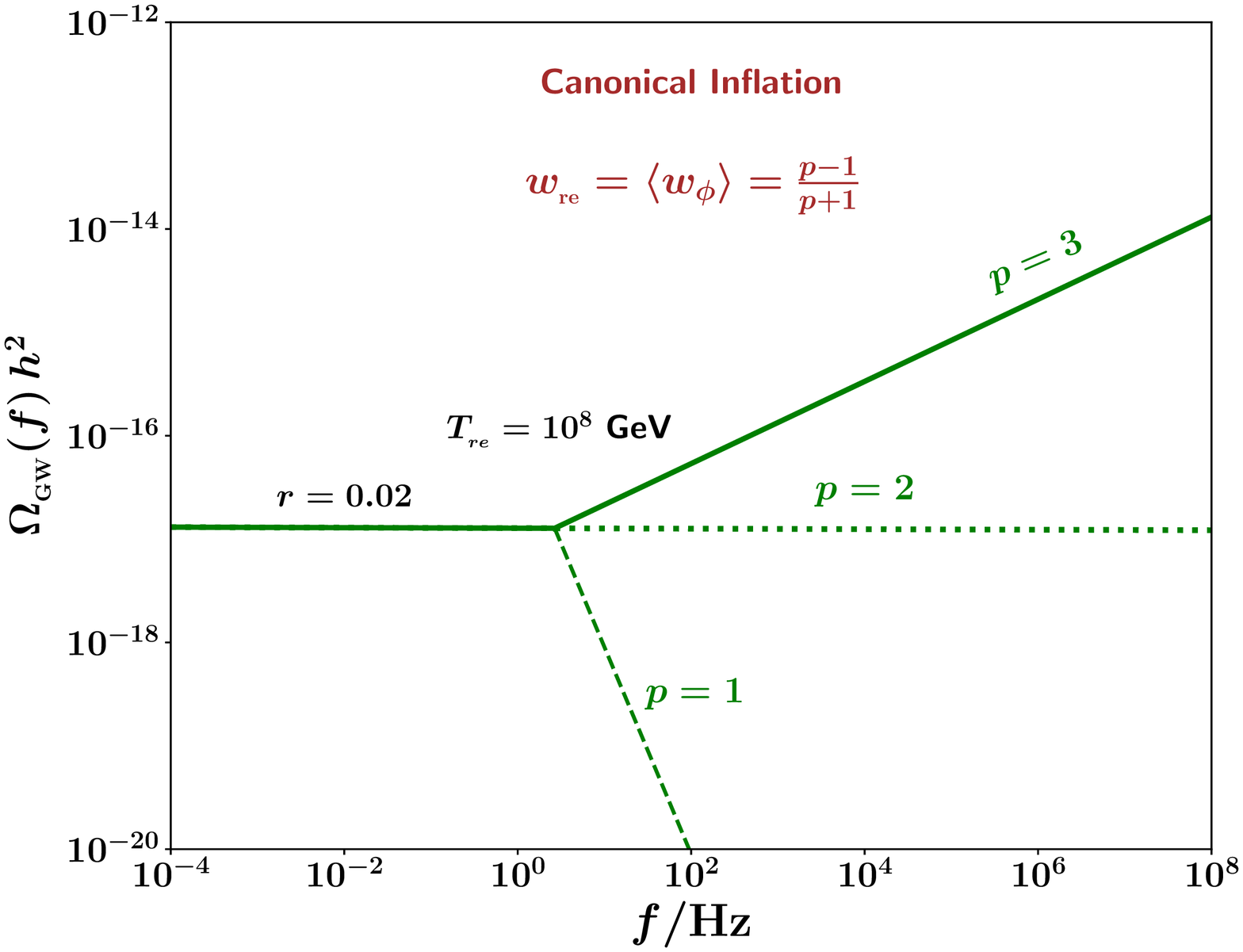}
\includegraphics[width=0.59\textwidth]{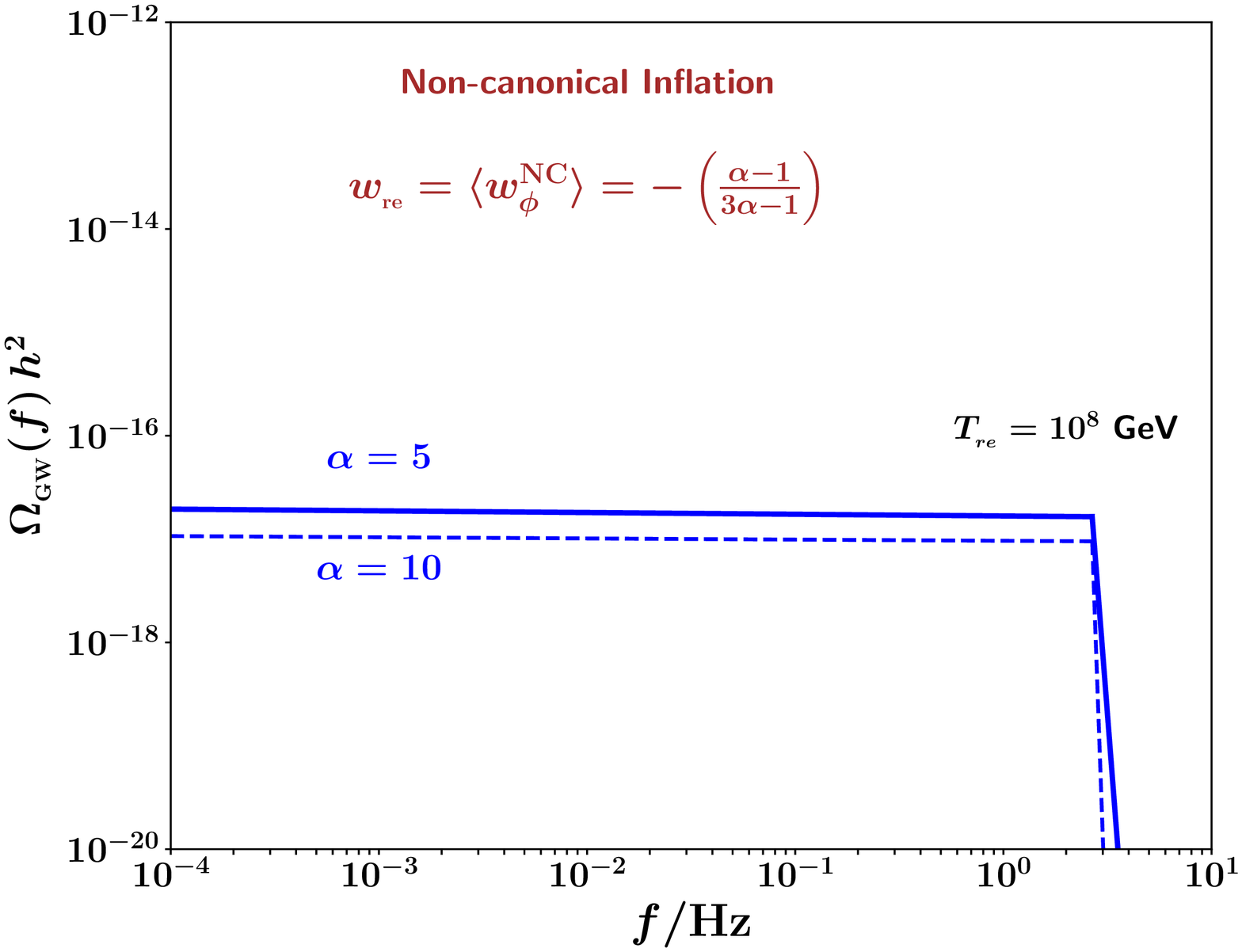}
\caption{This figure demonstrates that the spectra of relic GWs $\Og(f)$  can easily distinguish  
between canonical and non-canonical inflation. {\bf Top panel} shows the spectrum of relic GWs in the 
canonical case for which the post-inflationary EOS is described by (\ref{eq:GW_w_c}) 
with $p=1,\, 2 ~{\rm and}~ 3$,  plotted in dashed, dotted and solid green curves respectively. 
{\bf Bottom panel} shows the same for non-canonical inflation 
for which the post-inflationary EOS is described by (\ref{eq:GW_w_nc})
 with $\alpha = 5~{\rm and }~ 10$, plotted in solid and dashed blue curves respectively. }
\label{fig:GW_can_nc}
\end{center}
\end{figure}

\begin{itemize}

\item
In the case of perturbative reheating, the value of $\wre \equiv \l<w_{_{\phi}}\r>$
is given by (\ref{eq:EOS}) for canonical scalars, namely
\beq
\l<w_{_{\phi}}\r> = \frac{p-1}{p+1}~, ~~~ p \geq 1
\label{eq:GW_w_c}
\eeq
 and by
(\ref{eq:EOS_nc}) for non-canonical scalars (with quadratic potential), namely
\beq
\l<w^{^{\rm NC}}_{_{\phi}}\r> = -\left (\frac{\alpha-1}{3\alpha-1}\right )~, ~~~ \alpha \geq 1\,.
\label{eq:GW_w_nc}
\eeq
Note that the range permitted for non-canonical scalars
\beq -1/3 < \l<w^{^{\rm NC}}_{_{\phi}}\r> \leq 0\eeq
is complementary to that for canonical scalars
\beq 0 \leq \l<w_{_{\phi}}\r> < 1\,.\eeq Consequently,
the GW spectrum (\ref{eq:GW_spectrum_1c})
can easily distinguish between canonical
and non-canonical models of inflation as illustrated in figure  \ref{fig:GW_can_nc}.

\item In the case of non-perturbative reheating, the physics of the reheating
epoch can be quite complex. In this case $\wre$ is sometimes assumed 
to be a constant, for the sake of simplicity \cite{cook15,kamion_1,kamion_2}.

%

\end{itemize}

From the equations developed earlier in this section and
(\ref{eq:GW_spectrum_1b}),(\ref{eq:GW_spectrum_1c}),
  it follows that the 
spectral density of stochastic GWs corresponding to modes that became sub-Hubble prior to 
matter-radiation equality is
\ber
\mbox{{\bf Radiative epoch:}} ~~   \Og^{(\rm RD)}(f)  = \l(\frac{1}{24}\r)\,r\,A_{_S}\,\l(\f{f}{f_*}\r)^{\nt}\,\Omega_{0r} ~,  ~~ f_{\rm eq} < f \leq f_{\rm re}\, ,
\label{eq:GW_spectrum_2a}\\
\mbox{{\bf During reheating:}} ~~ \Og^{(\rm re)} (f)  = \Og^{(\rm RD)}(f)\, \left (\frac{f}{f_{\rm re}}\right )^{2\left (\frac{w-1/3}{w+1/3}\right )},~~ f_{\rm re} < f \leq f_{\rm e}\, ~,
\label{eq:GW_spectrum_2b}
\eer
where we have used $P_{\rm GW}(f) = P_{\rm GW}(f_*)\left (\frac{f}{f_*}\right )^{\nt}$, with $A_{_{\rm GW}} \equiv P_{\rm GW}(f_*) = \f{1}{4}\, r \, A_{_S}$ from equation (\ref{eq:GW_Ah}). Note that 
$f_*$ is the physical frequency (of GW) corresponding to the CMB pivot scale comoving wave number $k_*$.  

Equations (\ref{eq:GW_Ph1}), (\ref{eq:GW_spectrum_1a}), (\ref{eq:GW_spectrum_1b}), (\ref{eq:GW_spectrum_1c}) allow us to define a local post-inflationary
gravitational wave (tensor) spectral index as follows
\beq
\ng = \frac{d\log{\Og(k)}}{d\log{k}} = \frac{d\log{\Og(f)}}{d\log{f}}
\eeq
where
\beq
\ng =\nt + 2\,\left (\frac{w-1/3}{w+1/3}\right )
\label{eq:index1}
\eeq
which implies $\ng > \nt$ for $w > 1/3$, $\ng = \nt$ for $w = 1/3$ and
$\ng < \nt$ for $w < 1/3$,
where $w$ is the background EOS and is given by $w=0$ during matter domination,
$w=1/3$ during radiation domination and by $\l<w_{_{\phi}}\r>$ in (\ref{eq:EOS}) and (\ref{eq:EOS_nc})
during oscillations of canonical and non-canonical scalars respectively.

Note that since $\nt \simeq -2\epsilon_H$, 
CMB constraints on  the tensor-to-scalar ratio $r = 16\,\epsilon_H \leq 0.06$, imply 
$\vert \nt \vert \leq 0.0075$. Hence $\nt$  is a very small quantity that does not generate an appreciable change in $\Og(k)$ for more than 30 orders of magnitude variation in $k$ (and hence in $f$). Therefore (\ref{eq:index1}) effectively reduces to
\beq
\ng \simeq 2\,\left (\frac{w-1/3}{w+1/3}\right )~.
\label{eq:spectral_index}
\eeq
Thus the post-inflationary EOS has a direct bearing on the spectral index of relic gravitational
radiation with 
\ber
\ng \geq 0 ~~ {\rm for} ~~ w > 1/3\nonumber\\
 \ng \simeq 0 ~~ {\rm for} ~~ w = 1/3 \nonumber\\
\ng \lleq 0 ~~ {\rm for} ~~ w< 1/3
\label{eq:GW_ng_cases}
\eer
 which illustrates the extreme sensitivity of the 
GW spectral index to the background EOS in the universe.

\begin{figure}[t]
\begin{center}
\includegraphics[width=0.59\textwidth]{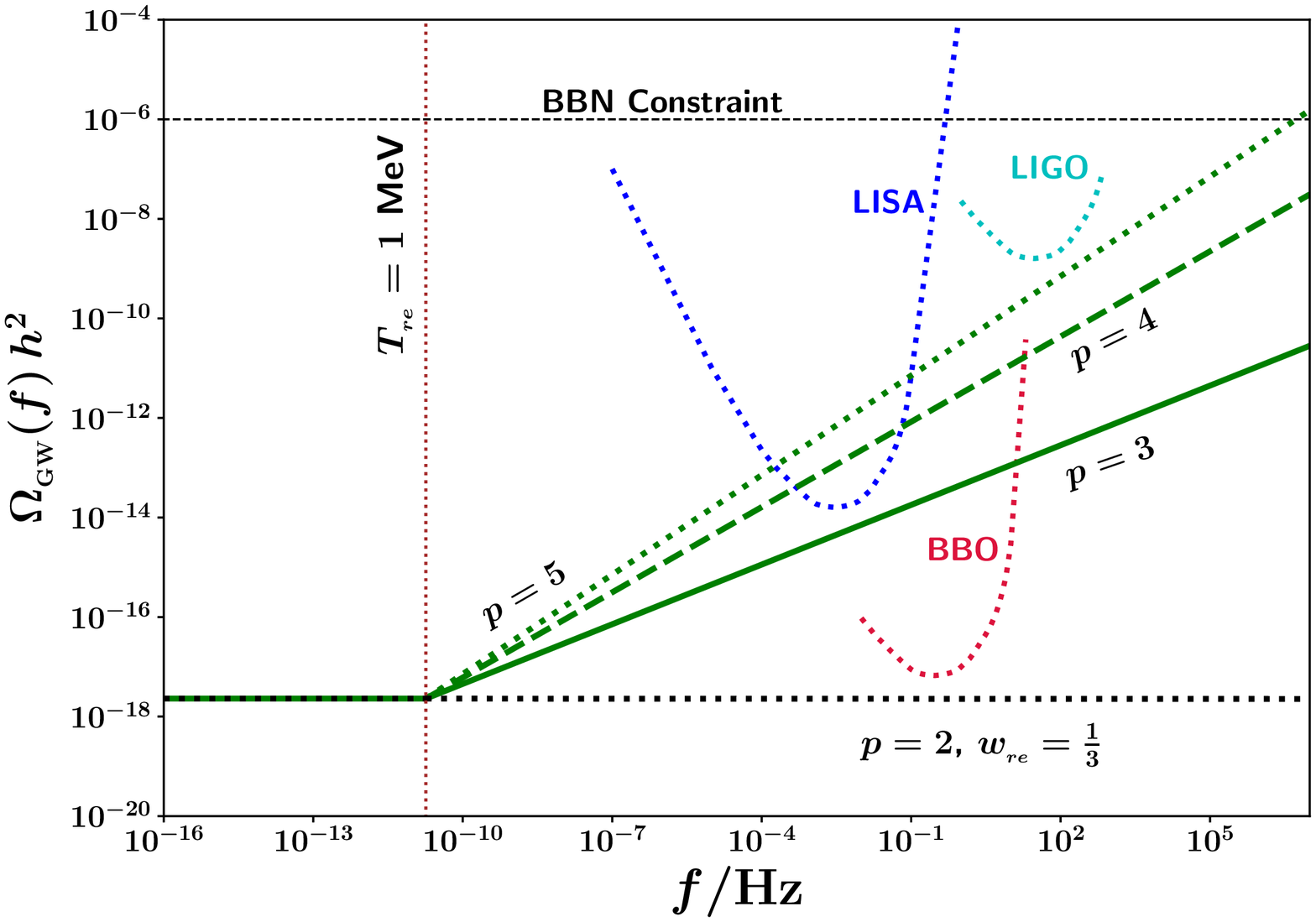}
\includegraphics[width=0.59\textwidth]{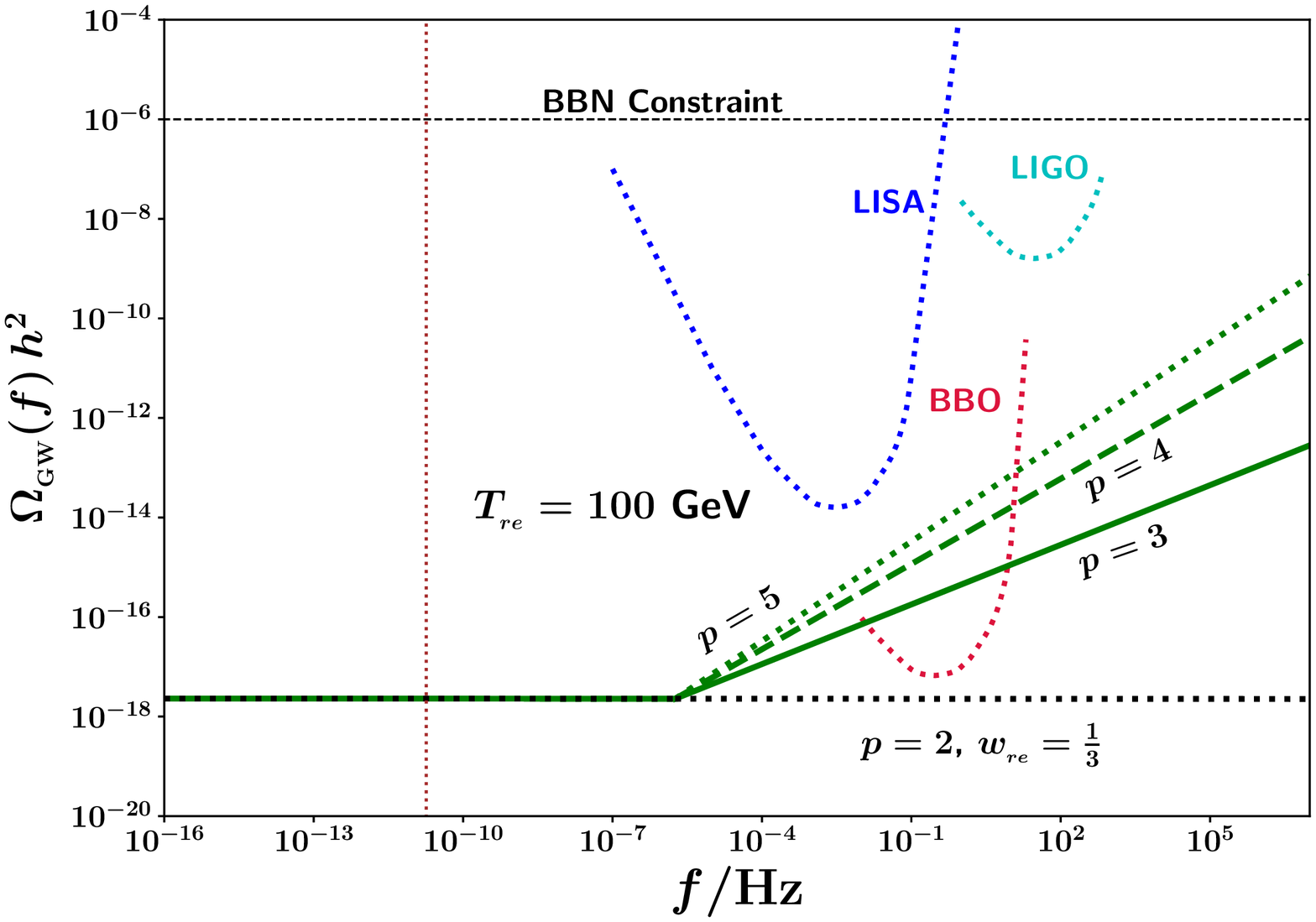}
\caption{This figure illustrates the potential implications of blue tilted relic GWs with $p>2$ in (\ref{eq:GW_w_c}) (and hence $\wre>1/3$) from  the perspective of near future GW observatories such as the advanced LIGO, LISA and BBO. {\bf Top panel} depicts the spectrum of blue tilted relic GWs corresponding  to $p=3,~4,~5$, plotted in solid, dashed and dotted green curves respectively, for  a fixed reheating temperature $\Tre=1$ MeV and tensor-to-scalar ratio $r\simeq 0.001$. 
The {\bf bottom panel} shows the same but with a higher reheating temperature  $\Tre=100$ GeV. The dotted green curve in the top panel indicates that relic GWs with a low enough reheating temperature $\Tre \leq 100$ MeV and EOS stiffer than $\wre=2/3$ (corresponding to $p>5$) would violate the BBN constraint $\Og\, h^2 \leq 10^{-6}$.
 Theories predicting such signals can therefore be regarded as being unphysical. }
\label{fig:GW_p}
\end{center}
\end{figure}

Setting $\nt = 0$ for simplicity, one gets, for the different cosmological epochs, the result
\ber
{\bullet} ~~~~ {\rm Matter ~ domination}~(w=0) ~~~~ \Rightarrow ~~ \ng(k)\bigg\vert_{\rm MD} &=& ~ - 2 \nonumber\\
{\bullet} ~~~~ {\rm Radiation ~ domination}~(w=1/3) ~~~~ \Rightarrow ~~ \ng(k)\bigg\vert_{\rm RD} &\simeq& ~~0
\eer

\begin{figure}[t]
\begin{center}
\includegraphics[width=0.85\textwidth]{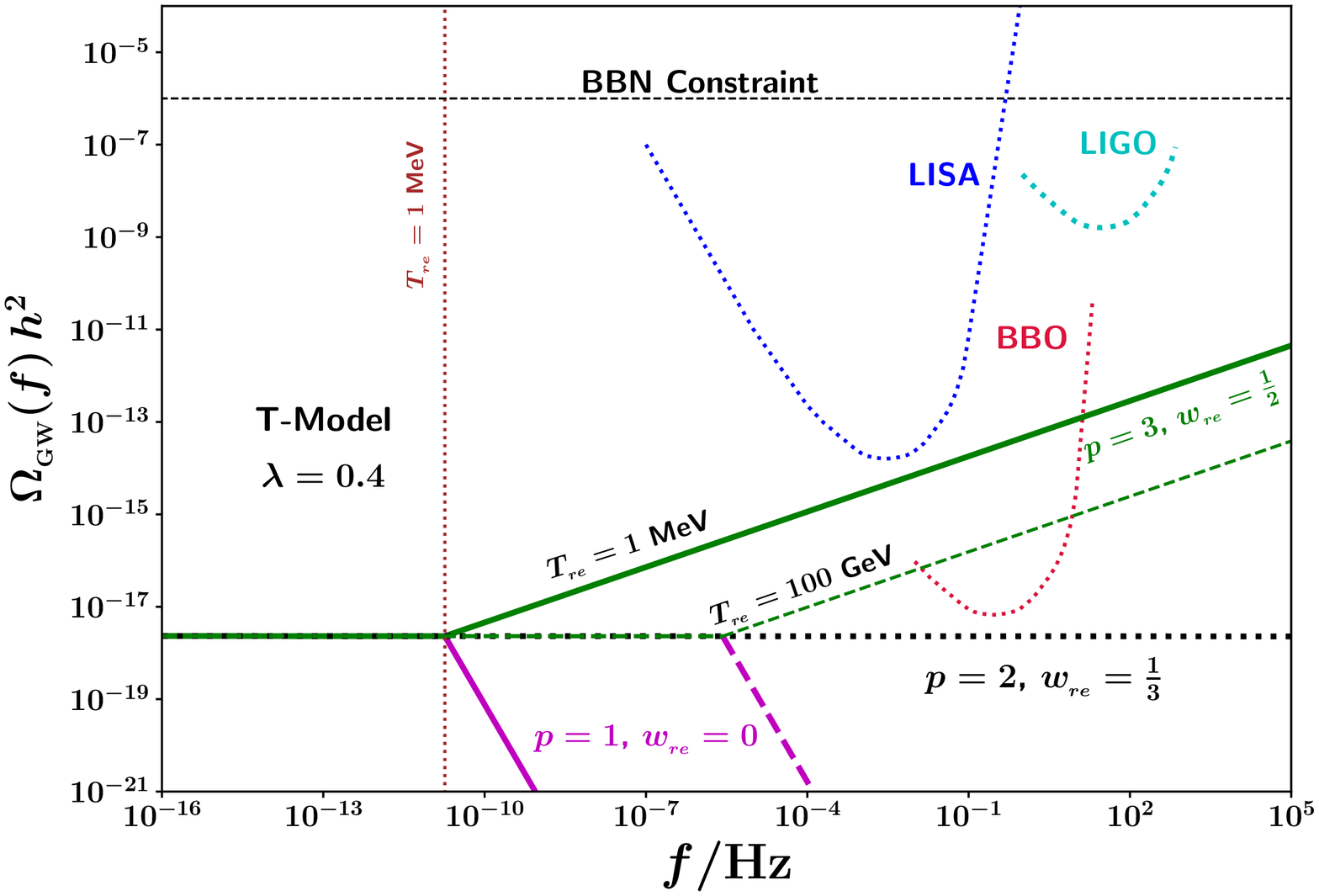}
\caption{The spectrum of relic gravitational waves is shown for the T-model $\alpha$-attractor potential (\ref{eq:pot_Tmodel}) for $\lambda=0.4$. The dotted black curve corresponds to $p=2$ for which the post-inflationary EOS is 
radiation-like, with $\wre=\l<w_{_{\phi}}\r>=1/3$, see (\ref{eq:EOS}), and the GW spectrum is flat. 
The solid and dashed green curves correspond to $p=3$ and reheating 
temperatures $\Tre = 1~{\rm MeV}$ and $100~{\rm GeV}$ respectively, for which
 the  post-inflationary EOS is $\wre=\l<w_{_{\phi}}\r>=1/2$ and the resulting GW spectrum has a blue tilt. 
One finds that in this case relic gravity waves
can be detected by future GW observatories such as BBO, for a range of reheating temperature  $\Tre\leq 10^6$ GeV.  
The solid and dashed purple curves correspond to reheating temperatures 
$\Tre = 1~{\rm MeV}$ and $100~{\rm GeV}$ respectively, and to 
a matter-like post inflationary EOS $\wre=\l<w_{_{\phi}}\r>=0$ which arises for $p=1$.
Note that in this case GWs have a  red tilt and their amplitude is suppressed relative to $p=2,3$.}
\label{fig:GW_Tmodel}
\end{center}
\end{figure}

\n
$\bullet$ ~ During the pre-radiation epoch the GW spectrum depends upon the EOS during reheating.
In the context of perturbative reheating, which is relevant for this work,   one finds
\ber
\ng(k)\bigg\vert_{\rm OSC} &=& 2\left (\frac{p-2}{2p-1}\right ) ~~~~
{\rm canonical ~oscillatory ~epoch }\nonumber\\
\ng(k)\bigg\vert^{NC}_{\rm OSC} &=& 2\left (2-3\alpha\right ) ~~~~~~ {\rm noncanonical ~oscillatory
 ~ epoch}
\eer
 where $p$ refers to the exponent in the inflationary potentials (\ref{eq:pot_Tmodel}) \& (\ref{eq:pot_Emodel}).
Canonical oscillatory epoch refers to post-inflationary oscillations of a
{\em canonical} scalar field with $p \geq 1$, while
noncanonical oscillatory epoch refers to post-inflationary oscillations by a {\em non-canonical} scalar field
with $\alpha \geq 1$ in (\ref{eq:lag_nc}). Note that
$\nt - \ng(k) \leq 2$ during canonical scalar field oscillations
whereas $\nt - \ng(k) \geq 2$ during non-canonical scalar field oscillations.
Thus the post-inflationary tensor spectra are distinctly different in the two cases, as shown in 
figure  \ref{fig:GW_can_nc}.

Since the blue tilted GWs, corresponding to $p>2$, are potentially important from the observational prospective, we discuss their implications in light of the ongoing and  near future GW observatories in figure  \ref{fig:GW_p}. The top panel shows the spectrum of blue tilted relic GWs corresponding  to $p=3,~4,~5$, plotted in solid, dashed and dotted green curves respectively, for  fixed reheating temperature $\Tre=1$ MeV and tensor-to-scalar ratio $r\simeq 0.001$. The bottom panel depicts the same with reheating temperature  fixed to $\Tre=100$ GeV. From figure  \ref{fig:GW_p}, we conclude that
\begin{itemize}
\item Relic GWs can be observed by LISA in the case of  low  reheating temperature, $\Tre \sim ~1-100$ MeV and  stiff enough reheating EOS $\wre>0.5$, corresponding to $p>3$. 

\item However for $\wre>2/3$, corresponding to $p>5$, with a low reheating temperature $\Tre<100$ MeV, the spectrum of relic GWs would violate the BBN constraint $\Og\, h^2 \leq 10^{-6}$ (as indicated by the dotted green curve in the top panel of figure  \ref{fig:GW_p}). Hence the corresponding parameter space of $\lbrace p, \Tre \rbrace$ is ruled out by the BBN constraint, even though it would have been possible to detect the signal by the advanced LIGO detectors. Similar conclusions were also drawn  in \cite{dany_19}.

\item The  blue-tilted  relic GW spectrum can be detected by  the BBO, for a range of reheating temperatures $\Tre\leq 10^6$ GeV.
\end{itemize}

In marked contrast to perturbative reheating, in models with non-perturbative reheating the 
reheating/preheating epoch can be a complex affair with explosive (resonant) particle production, backreaction
and non-equilibrium field theory all playing a significant role until thermalization is finally reached. For 
simplicity this epoch is usually characterised (see \cite{kamion_1,Creminelli,kamion_2,cook15}) by a constant effective EOS parameter, $\wre$, so that the general
formulae (\ref{eq:GW_spectrum_2a}) --  (\ref{eq:index1})    
also have bearing on this scenario.

\begin{figure}[t]
\begin{center}
\includegraphics[width=0.85\textwidth]{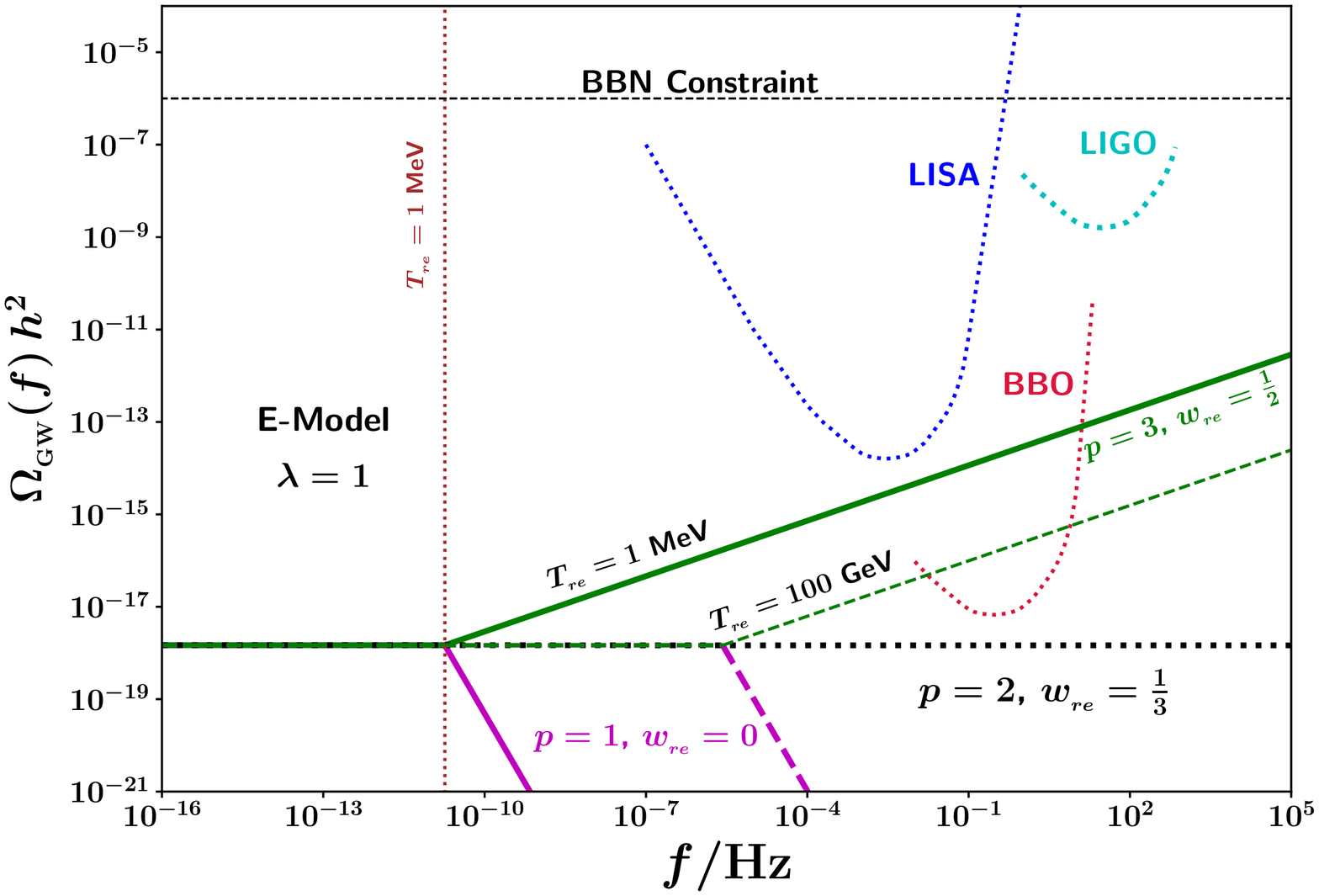}
\caption{
The spectrum of relic gravitational waves is shown for the E-model $\alpha$-attractor potential (\ref{eq:pot_Emodel}) for $\lambda=1$. The dotted black curve corresponds to $p=2$ for which the post-inflationary EOS is
radiation-like, with $\wre=\l<w_{_{\phi}}\r>=1/3$, see (\ref{eq:EOS}), and the GW spectrum is flat.
The solid and dashed green curves correspond to $p=3$ and reheating
temperatures $\Tre = 1~{\rm MeV}$ and $100~{\rm GeV}$ respectively, for which
 the  post-inflationary EOS is $\wre=\l<w_{_{\phi}}\r>=1/2$ and the resulting GW spectrum has a blue tilt.
One finds that in this case relic gravity waves
can be detected by future GW observatories such as BBO, for a range of reheating temperature  $\Tre\leq 10^6$ GeV.
The solid and dashed purple curves correspond to reheating temperatures
$\Tre = 1~{\rm MeV}$ and $100~{\rm GeV}$ respectively, and to
a matter-like post inflationary EOS $\wre=\l<w_{_{\phi}}\r>=0$ which arises for $p=1$.
Note that in this case GWs have a  red tilt and their amplitude is suppressed relative to $p=2,3$.}
\label{fig:GW_Emodel}
\end{center}
\end{figure}

In figures \ref{fig:deg_Tmodel} and \ref{fig:deg_Emodel} 
 we showed that the T and E model $\alpha$-attractors
 exhibited a degeneracy since, for $\lambda \ggeq 0.2$,
different values of $p$ in the T and E model potentials (\ref{eq:pot_Tmodel}) \&
(\ref{eq:pot_Emodel}) gave rise to identical values of the CMB parameters $\lbrace n_{_S}, r \rbrace$.
In section \ref{sec:reheat_Tmodel} we demonstrated that this degeneracy was 
easily broken if one took into account the reheating predictions encoded in the parameters $\Nre$, $\Tre$.
We now show that the degeneracy in $\lbrace n_{_S}, r \rbrace$ 
can also be broken by the GW spectrum since the latter depends explicitly on the reheating
EOS, which, in turn, depends upon the value of $p$ through (\ref{eq:GW_spectrum_2b}) and (\ref{eq:spectral_index}). 

This can be seen from figure  \ref{fig:GW_Tmodel} for the T-model and from figure  \ref{fig:GW_Emodel} for the E-model.
In both models one notices that for $p=2 \Rightarrow \wre=\l<w_{_{\phi}}\r>=1/3$
(dotted black curve), the relic GW spectrum is almost scale invariant, as suggested by (\ref{eq:index1}).
By contrast, for larger values of $p$ such as $p=3 \Rightarrow \wre=\l<w_{_{\phi}}\r>=1/2$ (solid and dashed green),
the relic GW spectrum is blue-tilted and can be detected by future GW observatories, such as the BBO, for a range of reheating temperatures. The GW spectrum for $p=1 \Rightarrow \wre=\l<w_{_{\phi}}\r>=0$, is red-tilted with
$\ng(k)\simeq -2$,  and is suppressed  relative to the other two cases.

Note that the
gravity wave spectrum for the T-model in figure  \ref{fig:GW_Tmodel} is shown with the value of the inflationary
parameter $\lambda$ in (\ref{eq:pot_Tmodel}) set at $\lambda = 0.4$. 
For the E-model in figure  \ref{fig:GW_Emodel}, on the other hand, we have chosen $\lambda = 1$.
Our choice for $\lambda$ is motivated by the following considerations:

\begin{enumerate}

\item These values of $\lambda$ correspond to the degeneracy regions in which different values of $p$
result in the same values of $\lbrace n_{_S}, r \rbrace$; see figures \ref{fig:deg_Tmodel} \& \ref{fig:deg_Emodel}.

\item Moreover, as shown in the right panels of figures \ref{fig:deg_Tmodel} \& \ref{fig:deg_Emodel},
a larger value of $\lambda$ corresponds to a smaller value of the tensor-to-scalar
ratio $r$. Our choice of $\lambda = 0.4$ (T-model) and $\lambda = 1$ (E-model)
corresponds to $r \sim 10^{-3}$ which lies within the observable range of upcoming  CMB missions such as the CMB-S4
\cite{CMB_S4} and the Simons Observatory \cite{Simons}.

\end{enumerate}

\section{\ Discussion}
\label{sec:dis}
The inflationary paradigm often exhibits  degeneracies, with two (or more) models
predicting essentially the same values of $\lbrace n_{_S}, r \rbrace$, leading to the existence of `cosmological attractors' or `universality classes' of inflation \cite{linde1,linde2,Roest1}. Such degeneracies render difficulties for the CMB observations alone to constitute a unique probe of the inflationary dynamics. Such degeneracies usually emerge either because multiple inflationary potentials
make similar predictions for  the scalar spectral index $n_{_S}$ and the tensor-to-scalar
ratio $r$, or because within the same model, the predicted  values of $\lbrace n_{_S}, r \rbrace$
are insensitive to some of the model parameters in the potential. In this work, we have demonstrated the existence of inflationary degeneracies in two classes of $\alpha$-attractor inflationary models, namely the T-model and E-model 
\cite{linde1,linde2}  discussed in section \ref{sec:alphainf}.
Inflationary degeneracies have also been shown to exist in the non-canonical framework of inflation \cite{non-can2,non-can3} discussed in section \ref{sec:non_can_inf}. 

In the context of the $\alpha$-attractors, we have shown that the scalar spectral index $\ns$  becomes insensitive to the potential parameter $\lambda$ (related to the  curvature of the superconformal K$\ddot{\rm a}$hler metric \cite{linde2}) as well as the exponent $p$, for $\lambda > \mathcal{O}(0.1)$. 
The tensor-to-scalar ratio $r$, decreases with an increase in $\lambda$, and becomes insensitive to the exponent 
$p$ for $\lambda > \mathcal{O}(0.1)$. A similar degeneracy also exists 
with respect to the non-canonical parameter $\alpha$ in (\ref{eq:lag_nc}) in
non-canonical inflation, as demonstrated in figure  \ref{fig:ns_r_nc}. 

In section \ref{sec:reheat} we provided an introduction to the kinematics of reheating 
in terms of the  reheating parameters $\lbrace \wre,\Nre,\Tre \rbrace$, and  spelled out our strategy (developed along the lines of  \cite{kamion_1,kamion_2,cook15}) for yielding tighter constraints on the CMB observables by taking into account 
reheating constraints developed  for the case of perturbative reheating.  In sections \ref{sec:reheat_Tmodel} and \ref{sec:reheat_Emodel}, we demonstrated that the inflationary degeneracies of the T-model  and the E-model $\alpha$-attractors can be easily broken by noting that the reheating EOS is very sensitive to the parameter $p$. 
In particular, we showed that imposing  the liberal reheating 
constraints $\Nre>0$ and $\Tre \in [1 \, {\rm MeV}, 10^{16}\, {\rm GeV}]$ on the  $\alpha$-attractor potentials, the CMB predictions for $\lbrace n_{_S}, r \rbrace$ got segregated into different regions of space,
 as illustrated  in figure  \ref{fig:T_model_deg_break} and \ref{fig:E_model_deg_break}.  

However for the case of quadratic potential in the non-canonical framework, we found
 that  reheating constraints are not able to break the degeneracy in $\lbrace n_{_S}, r \rbrace$ appreciably, owing to the fact that the non-canonical EOS, obeying  $\l<w^{^{\rm NC}}_\phi\r> < 0$, does not include the critical point of segregation $\wre=1/3$. 

In section \ref{sec:GW}, we showed that the spectrum of relic GWs  could easily distinguish between canonical 
and non-canonical inflation. We also demonstrated that by considering reheating consistent  tightened constraints 
on $\lbrace n_{_S}, r \rbrace$, the corresponding spectra of relic GWs could help distinguish between  different values 
of $p$ in the  T-model  and the E-model $\alpha$-attractors. In particular, as illustrated in figure  \ref{fig:GW_Tmodel} and \ref{fig:GW_Emodel}, we found that the relic GW background generated for a stiffer reheating equation of state 
could be detectable by the future space based GWs observatory  BBO. We also concluded that for $p=3$, the GW spectra 
do not possess enough power  to reach LISA sensitivity, in agreement with the conclusions drawn in \cite{dany_19}. 
We plan to investigate the existence of degeneracies in a wider class of inflationary model, including the universality class corresponding to  the non-minimally coupled power-law potentials \cite{linde_un}  in a future work\footnote{See \cite{Gong:2015qha} for an analysis of the dependence of reheating temperature on $\lbrace n_{_S}, r \rbrace$ in the case of non-minimally coupled scalar fields.}. Probing thermal history of the universe by combining  CMB and GW observations was  discussed in \cite{Durrer:2011bi,Kuroyanagi:2013ns}. 

In a scenario in which inflation is followed by a long duration of reheating ($\Nre\geq 10$),
 the reheating temperature can be much lower than the energy scale of inflation, as demonstrated by
 equation (\ref{eq:CMB_reheat_Tre_SR}). In this case, a relatively stiffer post inflationary EOS $\wre = \langle w_\phi \rangle > 1/3$   exhibits interesting observational signatures in the form of blue tilted GWs, as seen from  
(\ref{eq:GW_ng_cases}). However \cite{amin1,amin2} noted that a sufficiently long period of reheating with an 
equation of state $\wre \geq 0$ might lead to scalar field fragmentation due to self resonance for potentials, such as  (\ref{eq:pot_Tmodel}) and (\ref{eq:pot_Emodel}), which behave like $V(\phi) \propto \phi^{2p}$ during reheating while possessing  asymptotically flat wing/wings for larger field values. In the context of $\alpha$-attractors, the regime of scalar field fragmentation corresponds to $\lambda\gg 1$ in (\ref{eq:pot_Tmodel}) and (\ref{eq:pot_Emodel}). For $p=1$, the effective equation of state remains $\wre = \langle w_\phi \rangle =0$, independently of whether fragmentation occurs or not \cite{amin2}. 
For $p=2$, fragmentation occurs for $\lambda\gg 1$ and within a few $e$-folds, the effective EOS becomes $\wre\simeq 1/3$ which is the same as $\langle w_\phi \rangle$. So our analysis for $p=1,\,2$ remains unaltered independently
 of the value of $\lambda$. Coming to the case of  $p=3$, fragmentation occurs for $\lambda\gg 1$ and  the 
effective EOS, within a few $e$-folds, becomes $\wre\simeq 1/3$  which is different from $\langle w_\phi \rangle=1/2$. 
However fragmentation is effective only for  $\lambda\gg 1$, for which the tensor to scalar ratio is extremely small, 
namely $r \ll 10^{-3}$, rendering it undetectable by planned CMB missions. We therefore conclude that,
 as long as $\lambda$ is not too large, our analysis remains robust even for the case of $p=3$.

Our analysis in this paper was carried out within the framework of the perturbative theory of reheating.
 In the case of non-perturbative reheating, the physics of the reheating
epoch can be quite complex. In this case, one usually assumes the effective $\wre$ to be a constant for the sake of simplicity \cite{cook15,kamion_1,kamion_2}. During the initial stage of {\em preheating}, particle production  occurs
 in a rapid and explosive manner due to parametric resonance. However the backreaction of created particles usually leads
 to the termination of the resonance after which the universe reheats via the slow perturbative decay of the inflaton. As long as the duration of preheating is short, $\Delta N \leq 1$, and the inflaton dominates the energy budget of the universe,  our analysis 
may also be extended to the case of non-perturbative reheating by assuming that  a fraction `q' of the  energy density $\rho_e$ remains  
in the inflaton after the termination of the resonance. Accordingly the modified form of  equation (\ref{eq:CMB_reheat_Nre_SR}) can be written as 

\beq
\Nre = \f{4}{1-3\wre}\l[61.55 -N_k^{\rm inf} - \log{\l(\f{(q\,V_e)^{\f{1}{4}}}{H_k^{\rm inf}}\r)}  \r]~.
\label{eq:np_reheat_Nre_SR}
\eeq
This and related issues will be discussed in greater detail in a companion paper.

\section{Acknowledgements} 
S.S.M. thanks the Council of Scientific and Industrial Research (CSIR), India, for financial support as senior research fellow.
Varun Sahni was partially supported by the J.~C.~Bose Fellowship of Department of Science and Technology, Government of India. 
A.A.S. was partially supported by the Russian Foundation for Basic Research grant No. 20-02-00411.

\appendix
\section{Kinematics during reheating}
\label{app:A}
From entropy conservation in the universe during  the post reheating radiation dominated hot Big Bang epoch   we can relate the temperature  $T$ at any epoch to  the scale factor $a$ (and hence redshift $z$)  of the universe, through the known temperature $T_{\rm eq}$ and scale factor $a_{\rm eq}$ at the matter-radiation equality in the following way
\beq
a^3 g_T^s T^3 = a_{\rm eq}^3 g_{\rm eq}^s T_{\rm eq}^3~, 
\label{eq:A_entropy}
\eeq
where $g_T^s$ and $g_{\rm eq}^s$ are the effective relativistic degrees of freedom in entropy. Hence
\beq
T = \l(\f{g_{\rm eq}^s}{g_T^s}\r)^{\f{1}{3}}\l(\f{a_{\rm eq}}{a}\r)T_{\rm eq}~.
\label{eq:A_T_a}
\eeq
From the causal diagram in figure  \ref{fig:RH_causal}, the epoch at which  an observable CMB scale makes its Hubble re-entry is giving by
\beq
 k = a(z)H(z)~,
\label{eq:A_scale_k}
\eeq
where the  Hubble parameter is given in terms of the radiation density $\rho$ by 
\beq
 H^2 = \f{1}{3\, m_p^2} \rho = \f{1}{3\, m_p^2} \f{\pi^2}{30}T^4~.
\label{eq:A_H_T}
\eeq

Incorporating (\ref{eq:A_T_a}) into the above equation and substituting the subsequent  expression of Hubble parameter in (\ref{eq:A_scale_k}), we can obtain  the value of a mode $k$ which made its Hubble re-entry at an epoch $z$ with a temperature $T$. For example using the known values of $z_{\rm eq}$ and $T_{\rm eq}$, we obtain the value of the CMB scale that made its Hubble re-entry during the matter radiation equality to be 
\beq
k_{\rm eq} = 0.013~ {\rm Mpc}^{-1}~.
\label{eq:A_k_eq}
\eeq
It is important to know that the given value `$x$' of a CMB scale $k$, including the pivot scale $k_*$, should strictly be expressed in the form of $k = x\, a_0 \,{\rm Mpc}^{-1}$. However, following the standard convention in the literature, assuming $a_0=1$ implicitly, we will continue expressing  $k = x\,{\rm Mpc}^{-1}$.
Similarly the scale corresponding to the  Hubble radius at the present epoch is given by
\beq
k_0 = 2.25\times 10^{-4}~ {\rm Mpc}^{-1}~.
\label{eq:A_k_0}
\eeq

The epoch $z_p$ at which the CMB pivot scale $k=k_*=0.05~ {\rm Mpc}^{-1}$ makes its Hubble re-entry is obtained to be 

\beq
1+z_p \simeq 2.62\times 10^4~,
\label{eq:A_zp}
\eeq
which implies that the CMB pivot scale, which satisfies  $k_*=0.05~ {\rm Mpc}^{-1} > k_{\rm eq}$, consequently  became sub-Hubble during the radiation dominated epoch prior to the matter-radiation equality.

In order to obtain expressions for the duration of reheating $\Nre=\log{\l(a_{\rm re}/a_e\r)}$ as well as the temperature $\Tre$ at the end of reheating, we begin by matching the comoving Hubble radius  at  the Hubble exit of  the CMB pivot scale during inflation  (see figure  \ref{fig:RH_causal}) 

\beq
 k = a_k H_k^{\rm inf} = 0.05~ {\rm Mpc}^{-1}~,
\label{eq:A_scale_k_match}
\eeq
$$\Rightarrow \f{a_k H_k^{\rm inf}}{a_0 H_0} = \f{k}{a_0 H_0}~,$$
$$\Rightarrow \l(\f{a_k}{a_e}\r)\l(\f{a_e}{a_{\rm re}}\r)\l(\f{a_{\rm re}}{a_{\rm eq}}\r)\l(\f{a_{\rm eq}}{a_0}\r)\l(\f{H_k^{\rm inf}}{H_0}\r) = \l(\f{k}{a_0 H_0}\r)~.$$
Taking logarithm of the above expression, we obtain
\beq
\log\l(\f{k}{a_0 H_0}\r)  = -N_k^{\rm inf} - \Nre - \Nrd - \log\l(1+z_{\rm eq}\r) +\log\l(\f{H_k^{\rm inf}}{H_0}\r)~.
\label{eq:A_scale_k_reheat}
\eeq
Assuming the effective reheating equation of state $\wre$ to be a constant, we obtain the following expression by matching the density at the end of reheating to the density at the end of inflation.
 \beq
\Rre = \rho_e \l(\f{a_e}{a_{\rm re}} \r)^{3(1+\wre)}~,
\label{eq:A_reheat_rho}
\eeq
from which we can obtain  the expression for the duration of reheating to be 

\beq
\Nre \equiv \log{\l(\f{a_{\rm re}}{a_e}\r)} = \f{1}{3(1+\wre)}\log{\l(\f{\rho_e}{\Rre}\r)}~.
\label{eq:A_reheat_Nre}
\eeq

In the radiation dominated epoch,  since $\Rre = \f{\pi^2}{30}g_{\rm re}\Tre^4$, equation (\ref{eq:A_reheat_Nre}) becomes 
\beq
\Nre = \f{1}{3(1+\wre)}\log{\l(\f{\rho_e}{\f{\pi^2}{30}g_{\rm re}\Tre^4}\r)}~,
\label{eq:A_reheat_Nre_Tre}
\eeq
 where $g_{\rm re} \equiv g(\Tre)$ is the effective number of relativistic degrees of freedom in energy density at the end of reheating. From the entropy conservation, using (\ref{eq:A_T_a}), we get
 \beq
\Tre = \l(\f{g_{\rm eq}^s}{g_{\rm re}^s}\r)^{\f{1}{3}} \l(\f{a_{\rm eq}}{a_{\rm re}}\r)T_{\rm eq}~.
\label{eq:A_reheat_Tre}
\eeq
 Using (\ref{eq:A_reheat_Tre}) in (\ref{eq:A_reheat_Nre_Tre}), we get 
 
\beq
\Nre = \f{1}{3(1+\wre)} \log{\l[\l(\f{30}{\pi^2 g_{\rm re}}\r)\l(\f{\rho_e^{\f{1}{4}}}{T_{\rm eq}}\r)^4\l(\f{a_{\rm re}}{a_{\rm eq}}\r)^4\l(\f{g_{\rm re}^s}{g_{\rm eq}^s}\r)^{\f{4}{3}} \r]}~,
\label{eq:A_reheat_Nre_0}
\eeq
which becomes 
\beq
\Nre = \f{4}{3(1+\wre)}\l[ \f{1}{4} \log{\l(\f{30}{\pi^2 g_{\rm re}}\r)} + \f{1}{3}\log{\l(\f{g_{\rm re}^s}{g_{\rm eq}^s}\r)} + \log{\l(\f{\rho_e^{\f{1}{4}}}{T_{\rm eq}}\r)}-\Nrd\r]~.
\label{eq:A_reheat_Nre_1}
\eeq
Substituting the expression for $\Nrd$ from (\ref{eq:A_scale_k_reheat}) in (\ref{eq:A_reheat_Nre_1}), we obtain 
\beq
\Nre = \f{-4}{3(1+\wre)}\l[ \f{1}{4} \log{\l(\f{30}{\pi^2 g_{\rm re}}\r)} + \f{1}{3}\log{\l(\f{g_{\rm re}^s}{g_{\rm eq}^s}\r)} + \log{\l(\f{\rho_e^{\f{1}{4}}}{H_k^{\rm inf}}\r)} + \log{\l(\f{k}{a_0 T_{\rm eq}}\r)}+N_k^{\rm inf}+\log\l(1+z_{\rm eq}\r) \r]~.
\label{eq:A_reheat_Nre_1a}
\eeq
Using the relation $T_{\rm eq} = \l(1+z_{\rm eq}\r)T_0$, we get
\beq
\Nre = \f{-4}{3(1+\wre)}\l[ \f{1}{4} \log{\l(\f{30}{\pi^2 g_{\rm re}}\r)} + \f{1}{3}\log{\l(\f{g_{\rm re}^s}{g_0^s}\r)} + \log{\l(\f{\rho_e^{\f{1}{4}}}{H_k^{\rm inf}}\r)} + \log{\l(\f{k}{a_0 T_0}\r)}+N_k^{\rm inf}+\Nre \r]~.
\label{eq:A_reheat_Nre_2}
\eeq
Assuming $w\neq 1/3$ and bringing the term involving $\Nre$ on the right hand side of the above equation to the left hand side, we obtain the following expression 
\beq
\Nre = \f{-4}{1-3\wre}\l[N_k^{\rm inf} + \log{\l(\f{\rho_e^{\f{1}{4}}}{H_k^{\rm inf}}\r)}  + \log{\l(\f{k}{a_0 T_0}\r) + \f{1}{4} \log{\l(\f{30}{\pi^2 g_{\rm re}}\r)} + \f{1}{3}\log{\l(\f{g_{\rm re}^s}{g_0^s}\r)} } \r]~.
\label{eq:A_reheat_Nre_3}
\eeq

For slow-roll inflation, using the fact that $\rho_e = \f{3}{2}V_e$ and using $k=k_*=0.05~{\rm Mpc}^{-1}$, $T_0 = 2.7255$ K, $g_{\rm re}=g_{\rm re}^s = 106.75$ and $g_0^s=3.94$, we obtain the following master formula for the duration of reheating in terms of $\wre$, $N_k^{\rm inf}$ and the inflationary parameters $H_k^{\rm inf}$, $V_e$.

\beq
\Nre = \f{4}{1-3\wre}\l[61.55 -N_k^{\rm inf} - \log{\l(\f{V_e^{\f{1}{4}}}{H_k^{\rm inf}}\r)}  \r]~, ~~~~\wre \neq 1/3~.
\label{eq:A_reheat_Nre_SR}
\eeq
Consequently, from equation (\ref{eq:A_reheat_Nre_Tre}),  the expression for the reheating temperature $\Tre$ becomes
\beq
\Tre =  \l(\f{45}{\pi^2 g_{\rm re}}\r)^\f{1}{4} V_e^{\f{1}{4}}\, e^{-\f{3}{4}\l(1+\wre\r)\Nre}.
\label{eq:A_reheat_Tre_SR}
\eeq

Note that for $\Tre<100$ GeV, the effective number of relativistic degrees of freedom $g_{\rm re},~g_{\rm re}^s$ are smaller than 106.75 and vary with temperature. At the beginning of Big Bang Nucleosynthesis, $T_{\rm BBN} \simeq 1~{\rm MeV}$ which corresponds to $g_{\rm re}=g_{\rm re}^s \simeq 10.75$. For $\Tre\in (1\,{\rm MeV},\, 100\,{\rm GeV})$, the variation of   $g_{\rm re}$ and $g_{\rm re}^s$  with temperature  can be incorporated by using lattice QCD calculations. However the variation has a small effect on $\Nre$ and hence we ignore it in our analysis.

\section{CMB constraints on inflation}
\label{app:B}
Consider the case of a canonical scalar field  minimally coupled to gravity with potential
\beq
V(\phi)=V_0 \, f\l(\f{\phi}{m_p}\r)~.
\label{eq:B_pot}
\eeq
The potential slow-roll parameters are given by 
\ber
\epsilon_{_V} = \f{\mpl^2}{2}\l(\f{f'}{f}\r)^2~,\label{eq:B_eV} \\
\eta_{_V} = \mpl^2\l(\f{f''}{f}\r)~.\label{eq:B_etaV}
\eer
In the  slow-roll limit $\epsilon_{_V}\, ,\eta_{_V} \ll 1 $, the scalar  power spectrum is given by \cite{baumann07}
\beq
P_{\cal R}(k) = A_{_S}\l(\f{k}{k_*}\r)^{\ns - 1},\label{eq:B_Ps}
\eeq
with the amplitude of scalar power spectrum at the CMB pivot scale $k\equiv k_*=0.05~{\rm Mpc}^{-1}$ is given by\cite{Planck1}
\beq
A_{_S}\equiv P_{\cal R}(k_*) = \f{1}{24\pi^2}\f{V_0}{\mpl^4}\f{f\l(\phi_k\r)}{\epsilon_{_V}(\phi_k)}\bigg\vert_{k=k_*}~,
\label{eq:B_As}
\eeq
and the scalar spectral index (with negligible running) is given by
\beq
\ns = 1 + 2 \, \eta_{_V}(\phi_*) - 6 \, \epsilon_{_V}(\phi_*) ~,
\label{eq:B_ns}
\eeq
where $\phi_*$ is the value of the inflaton field at the Hubble exit of CMB pivot scale $k_*$. Similarly the tensor power spectrum, in the slow-roll limit,  is given by
\beq
P_{T}(k) = A_{_T}\l(\f{k}{k_*}\r)^{\nt},\label{eq:B_Pt}
\eeq
with the amplitude of tensor power spectrum at the CMB pivot scale  is given by
\beq
A_{_T}\equiv P_T(k_*) = \f{2}{\pi^2}\l(\frac{H_k^{\rm inf}}{\mpl}\r)^2\bigg\vert_{k=k_*} \simeq  \f{2}{3\pi^2}\f{V_0}{\mpl^4}f\l(\phi_k\r)\bigg\vert_{k=k_*}~,
\label{eq:B_At}
\eeq
and the tensor spectral index (with negligible running) is given by
\beq
\nt = -2 \, \epsilon_{_V}(\phi_*)~.
\label{eq:B_nt}
\eeq
The tensor-to-scalar ratio $r$ is defined by
\beq
r \equiv \f{A_{_T}}{A_{_S}} = 16\, \epsilon_{_V}(\phi_*)~,
\label{eq:B_r}
\eeq
yielding the single field consistency relation 
\beq
r = -8\, \nt~.
\label{eq:B_consis}
\eeq
From the CMB observations of Planck 2018 \cite{Planck1}, we have  
\beq
A_{_S} = 2.1\times 10^{-9}~,
\label{eq:B_As_CMB}
\eeq
while the  $1\sigma$ constraint on the scalar spectral index is given by
\beq
n_{_S} = 0.9649 \pm 0.0042~.
\label{eq:B_ns_CMB}
\eeq
Similarly constraint on the tensor-to-scalar ratio $r$, from the combined observations of Planck 2018 \cite{Planck1} and BICEPII/Keck \cite{bicep2}, is given by
\beq
r \leq 0.06~,
\label{eq:B_r_CMB}
\eeq
which translates to the fact that $A_{_T}\leq 6\times 10^{-2}\, A_{_S}$, putting an upper bound on the inflationary Hubble scale $H_k^{\rm inf}$ from equation (\ref{eq:B_At}) as well as the energy scale during inflation $T_{\inf}$, given by 
\ber
H_k^{\rm inf} \leq 2.5\times 10^{-5}~\mpl = 6.1\times 10^{13}~{\rm GeV}~,\label{eq:B_H_k_CMB} \\
T_{\inf}\equiv \l(3\, m_p^2 \l(H_k^{\rm inf}\r)^2\r)^{1/4} \leq 1.6\times 10^{16}~{\rm GeV}~~.\label{eq:B_T_inf_CMB}
\eer
Similarly the CMB bound on $r$ translates directly to an  upper bound on the first slow-roll parameter
 \beq
\epsilon_{_V} \leq 0.00375~,
\label{eq:B_eV_CMB}
\eeq
 rendering the  tensor tilt from equation (\ref{eq:B_nt}) to be negligibly small
\beq
| \nt| \leq 0.0075~.
\label{eq:B_nt_CMB}
\eeq
Given the upper limit on $\epsilon_{_V}$, using the CMB bound on $\ns$ from (\ref{eq:B_ns_CMB}) in (\ref{eq:B_ns}), we infer that the  second slow-roll parameter is negative and  obtain interesting upper and lower limits on its magnitude, given by
\beq
0.0042 \leq \vert \eta_{_V} \vert \leq 0.0197~.
\label{eq:B_etaV_CMB}
\eeq
 The EOS $w_\phi$ of  the inflaton field is given by  
\beq
w_\phi = \f{\f{1}{2}\dot{\phi}^2 - V(\phi)}{\f{1}{2}\dot{\phi}^2 + V(\phi)}=-1+\f{2}{3}\epsilon_{_V}(\phi)
\label{eq:B_w_inf}
\eeq
An interesting consequence of (\ref{eq:B_eV_CMB}) is the fact that, around the pivot scale, the EOS during inflation is constrained to be
\beq
w_\phi \leq -0.9975~,
\label{eq:B_w_inf_CMB}
\eeq
implying that the expansion of the universe during inflation was near exponential (quasi-de Sitter like). End of inflation is marked by $w_\phi=-1/3$ which corresponds to $\dot{\phi}^2 = V(\phi)$. Hence the energy density of the inflaton at the end of inflation is given by 
 
\beq
\rho_e \equiv \rho(\phi_e) = \f{1}{2}\dot{\phi}^2+V(\phi) \bigg \vert_{\phi = \phi_e}=\f{1}{2}V(\phi_e)+V(\phi_e) = \f{3}{2}V(\phi_e)
\label{eq:B_rho_e}
\eeq


\begin{thebibliography}{69}

\bibitem{inf_star80}
A.~A.~Starobinsky, \plb {\bf 91}, 99  (1980).
\bibitem{inf_guth81}
A. H. Guth, \prd {\bf 23}, 347 (1981).
\bibitem{inf_linde82}
A. D. Linde, \plb {\bf 108}, 389 (1982). 
\bibitem{inf_alstein82}
A.~Albrecht and P.~J.~Steinhardt, \prl {\bf 48}, 1220 (1982).

\bibitem{linde90}
A.~D.~Linde, {\em Particle Physics and Inflationary Cosmology}, Harwood, Chur, Switzerland (1990).

\bibitem{baumann07}
D.~Baumann, TASI Lectures on Inflation, [arXiv:0907.5424].

\bibitem{star79}
 A.~A.~Starobinsky, JETP Lett. {\bf 30}, 682 (1979).

\bibitem{grish75}
 L.~P.~Grishchuk, Sov. Phys. JETP {\bf 40}, 409 (1975). 

\bibitem{st81} 
 A.~A.~Starobinsky, JETP Lett. {\bf 34}, 438 (1981).
 

\bibitem{sahni90}
V.~Sahni, Phys. Rev. D {\bf 42}, 453 (1990).


\bibitem{kofman96a}
L.  Kofman,  in
Relativistic  Astrophysics:  A  Conference  in
Honor of Igor Novikov’s 60th Birthday
, Proceedings, Copen-
hagen,  Denmark,  1996,  edited  by  B.  Jones  and  D.  Marcovic
~
Cambridge  University  Press,  Cambridge,  England  
[astro-ph/9605155]

\bibitem{ema17}
Y.~Ema, R.~Jinno, K.~Mukaida and K.~Nakayama, JCAP {\bf 1702} (2017) 045 [arXiv:1609.05209 [hep-ph]].

\bibitem{he19} 
M.~He, R.~Jinno, K.~Kamada, S.~C.~Park, A.~A.~Starobinsky and J.~Yokoyama, Phys. Lett. B {\bf 791}, 
36 (2019) [arXiv:1812.10099 [hep-ph]].

\bibitem{kofman94}
L.~A.~Kofman, A.~D.~Linde and A.~A.~Starobinsky, Phys. Rev. Lett. {\bf 73}, 3195 (1994).

\bibitem{yuri95}
Y.~Shtanov, J.~Traschen and R.~Brandenberger, Phys. Rev. D {\bf 51}, 5438 (1995);
also see  Y.~Shtanov,  Ukr.  Fiz.  Zh.  {\bf 38}, 1425 (1993)
(in Russian).

\bibitem{kofman96}
L.~A.~Kofman, A.~D.~Linde and A.~A.~Starobinsky, Phys. Rev. Lett. {\bf 76}, 1011 (1996).

\bibitem{kofman97}
L.~A.~Kofman, A.~D.~Linde and A.~A.~Starobinsky, Phys. Rev. D {\bf 56}, 3258 (1997).

\bibitem{liddle_lyth}
A.~R.~Liddle and D.~H.~Lyth, Cosmological Inflation and Large Scale Structure,
Cambridge University Press, 2000.

\bibitem{linde1}
R. Kallosh and A. Linde, JCAP07 (2013) 002 [arXiv:1306.5220].

\bibitem{linde2}
R. Kallosh, A. Linde and D. Roest, JHEP {\bf 1311} (2013) 198 [arXiv:1311.0472].

\bibitem{Roest1}
D.~Roest,
JCAP \textbf{01} (2014) 007
[arXiv:1309.1285 [hep-th]].


\bibitem{Planck1}
Planck collaboration: P.~A.~R.~Ade \etal, Planck 2018 results. X. Constraints on inflation
A\&A 641, A10 (2020)
[arXiv:1807.06211]


\bibitem{bicep2}
P.~A.~R.~Ade \textit{et al.} [BICEP2 and Keck Array],
Phys. Rev. Lett. \textbf{121} (2018) 221301
[arXiv:1810.05216 [astro-ph.CO]].

\bibitem{Martin:2010kz}
J.~Martin and C.~Ringeval,
Phys. Rev. D \textbf{82} (2010), 023511
[arXiv:1004.5525 [astro-ph.CO]].

\bibitem{Easson:2010zy}
D.~A.~Easson and B.~A.~Powell,
Phys. Rev. Lett. \textbf{106} (2011), 191302
[arXiv:1009.3741 [astro-ph.CO]]. 

\bibitem{kamion_1}
L.~Dai, M.~Kamionkowski and J.~Wang,
Phys. Rev. Lett. \textbf{113} (2014), 041302
[arXiv:1404.6704 [astro-ph.CO]].

\bibitem{Creminelli}
P.~Creminelli, D.~L\'opez Nacir, M.~Simonovi\'c, G.~Trevisan and M.~Zaldarriaga,
Phys. Rev. D \textbf{90} (2014) no.8, 083513
[arXiv:1405.6264 [astro-ph.CO]].

\bibitem{Martin:2014nya}
J.~Martin, C.~Ringeval and V.~Vennin,
Phys. Rev. Lett. \textbf{114} (2015) no.8, 081303
[arXiv:1410.7958 [astro-ph.CO]].

\bibitem{kamion_2}
J.~B.~Munoz and M.~Kamionkowski,
Phys. Rev. D \textbf{91} (2015) no.4, 043521
[arXiv:1412.0656 [astro-ph.CO]].

\bibitem{cook15}
J.~A.~Cook, E.~Dimastrogiovanni, D.~A.~Easson and L~.A.~Krauss, 
JCAP {1504} (2015) 047 [arXiv:1502.04673].

\bibitem{German:2020iwg}
G.~German,
JCAP \textbf{11} (2020), 006
[arXiv:2003.09420 [astro-ph.CO]].

\bibitem{German:2020cbw}
G.~German,
[arXiv:2010.09795 [astro-ph.CO]].

\bibitem{turner83}
M.~S.~Turner, Phys. Rev. D {\bf 28}, 1243 (1983).

\bibitem{st78}
 A.~A.~Starobinsky, Sov. Astron. Lett. {\bf 4}, 82 (1978).

\bibitem{non-can1}
V.~Mukhanov and A.~Vikman, JCAP {\bf 0602} (2006) 004 [astro-ph/0512066].

\bibitem{non-can2}
S.~Unnikrishnan, V.~Sahni and A.~Toporensky, JCAP
{\bf 1208}
(2012) 018 [arXiv:1205.0786].

\bibitem{non-can3}
S.~Unnikrishnan and V.~Sahni, JCAP {\bf 1310}
(2013) 063 [arXiv:1305.5260].

\bibitem{Starobinsky:1981vz}
A.~A.~Starobinsky, in: Proc. of the Second Seminar "Quantum Theory of Gravity"
(Moscow, 13-15 October 1981), INR Press, Moscow, 1982, p. 58-72; \\
reprinted in: Quantum Gravity, eds.. M.~A.~Markov, P.~C.~West, Plenum Publ. Co.,
New York, 1984, p. 103-128.

\bibitem{DeFelice:2010aj}
A.~De~Felice and S.~Tsujikawa, Living Rev. Rel. {\bf 13}, 3 (2010)
[arXiv:1002.4928].

\bibitem{albrecht82}
A.~Albrecht, P.~J.~Steinhardt, M.~S.~Turner and F. Wilczek,
Phys. Rev. Lett. {\bf 48} 1437 (1982).

\bibitem{Kolb_Turner}
E.~Kolb and M.~Turner, 
{\em The Early Universe},
Addison-Wesley, Redwood (1990).

\bibitem{tkachev96}
S.~Khlebnikov and I.~Tkachev, Phys. Lett. B
{\bf 390}, 80 (1977).

\bibitem{allen88}
B.~Allen, Phys. Rev. D {\bf 37}, 2078 (1988).

 \bibitem{inf_encyclo}
J.~Martin, C.~Ringeval and V.~Vennin,
Phys. Dark Univ. \textbf{5-6} (2014), 75-235
[arXiv:1303.3787 [astro-ph.CO]].

\bibitem{dany_18}
C.~Caprini and D.~G.~Figueroa,
Class. Quant. Grav. \textbf{35} (2018) no.16, 163001
[arXiv:1801.04268 [astro-ph.CO]].

\bibitem{dany_19}
D.~G.~Figueroa and E.~H.~Tanin,
JCAP \textbf{1908} (2019), 011
[arXiv:1905.11960 [astro-ph.CO]].

\bibitem{Bernal:2019lpc}
N.~Bernal and F.~Hajkarim,
Phys. Rev. D \textbf{100} (2019) no.6, 063502
[arXiv:1905.10410 [astro-ph.CO]].


\bibitem{sami2002}
V. Sahni,  M. Sami and T. Souradeep,  Phys. Rev. D {\bf 65}, 023518 (2002) [arXiv:gr-qc/0105121].



\bibitem{Ema_2020}
Y.~Ema, R.~Jinno and K.~Nakayama,
JCAP \textbf{09} (2020), 015
[arXiv:2006.09972 [astro-ph.CO]].

\bibitem{LISA}
J.~Baker, J.~Bellovary, P.~L.~Bender, E.~Berti, R.~Caldwell, J.~Camp, J.~W.~Conklin, N.~Cornish, C.~Cutler and R.~DeRosa, \textit{et al.},
arXiv:1907.06482 [astro-ph.IM].

\bibitem{BBO}
W.~T.~Ni,
Int. J. Mod. Phys. D \textbf{25} (2016) 1630001
[arXiv:1610.01148 [astro-ph.IM]].

\bibitem{QI}
 P.~J.~Peebles and A.~Vilenkin, Phys. Rev. D
59
, 063505 (1999)

\bibitem{instant_preheat}
 G.~Felder, L.~Kofman and A.~D.~Linde, Phys.Rev. D
{\bf 59}, 123523 (1999) [hep-ph/9812289];
G.~Felder, L.~Kofman and A.~D.~Linde, Phys.Rev. D
{\bf 60}, 103505 (1999) [hep-ph/9903350].

\bibitem{geng17}
C-Q.~Geng, C-C.~Lee, M.~Sami, E.~N.~Saridakis and 
A.~A.~Starobinsky,
JCAP {\bf 1706} (2017) 011 [arXiv:1705.01329 [gr-qc]].

\bibitem{souradeep}
 T.~Souradeep and V.~Sahni, Mod. Phys. Lett. A
{\bf 7}, 3541 (1992) [hep-ph/9208217].

\bibitem{giovani}
 M.~Giovannini, Phys. Rev. D {\bf 58}, 083504 (1998); Phys.
Rev. D {\bf 60}, 123511 (1999).

\bibitem{CMB_S4}
K.~N.~Abazajian \textit{et al.} [CMB-S4],
arXiv:1610.02743 [astro-ph.CO].

\bibitem{Simons}
A.~Suzuki \textit{et al.} [POLARBEAR],
J. Low Temp. Phys. \textbf{184} (2016) no.3-4, 805-810
[arXiv:1512.07299 [astro-ph.IM]].



\bibitem{linde_un}
R.~Kallosh, A.~Linde and D.~Roest,
Phys. Rev. Lett. \textbf{112} (2014) 011303
[arXiv:1310.3950 [hep-th]].

\bibitem{Durrer:2011bi}
R.~Durrer and J.~Hasenkamp,
Phys. Rev. D \textbf{84} (2011), 064027
[arXiv:1105.5283 [gr-qc]].

\bibitem{Kuroyanagi:2013ns}
S.~Kuroyanagi, C.~Ringeval and T.~Takahashi,
Phys. Rev. D \textbf{87} (2013) no.8, 083502
[arXiv:1301.1778 [astro-ph.CO]].

\bibitem{Gong:2015qha}
J.~O.~Gong, S.~Pi and G.~Leung,
JCAP \textbf{05} (2015), 027
[arXiv:1501.03604 [hep-ph]].

\bibitem{amin1}
K.~D.~Lozanov and M.~A.~Amin,
Phys. Rev. Lett. \textbf{119} (2017) no.6, 061301
[arXiv:1608.01213 [astro-ph.CO]].

\bibitem{amin2}
K.~D.~Lozanov and M.~A.~Amin,
Phys. Rev. D \textbf{97} (2018) no.2, 023533
[arXiv:1710.06851 [astro-ph.CO]].


\end{thebibliography}
\end{document}